\documentclass[11pt,a4paper]{article}
\pdfoutput=1
\usepackage{a4wide}
\usepackage{amsfonts}
\usepackage{amssymb}
\usepackage{amsmath}
\usepackage{graphicx}
\usepackage{booktabs}
\usepackage{array}
\usepackage{color}
\usepackage{ulem}
\usepackage[small,bf]{caption}
\usepackage{cite}
\usepackage[usenames,dvipsnames]{xcolor}
\usepackage{subfigure}
\usepackage{verbatim}
\usepackage{dsfont}
\def\1{\mathbf{1}}
\def\3{\mathbf{3}}
\def\2{\mathbf{2}}

\def\gtap{\ \raisebox{-.4ex}{\rlap{$\sim$}} \raisebox{.4ex}{$>$}\ }
\def\ltap{\ \raisebox{-.4ex}{\rlap{$\sim$}} \raisebox{.4ex}{$<$}\ }
\allowdisplaybreaks

\numberwithin{equation}{section}

\newcounter{mysubequation}[equation]

\definecolor{pink}{rgb}{1.,.2,.8}

\begin{document}
\begin{titlepage}

\vspace*{-15mm}
\begin{flushright}
SISSA 48/2016/FISI\\
IPMU16-0125\\
TTP16-035
\end{flushright}
\vspace*{0.7cm}

\begin{center}
{
\bf\LARGE Renormalisation Group Corrections to\\[0.3em] Neutrino Mixing Sum Rules
}
\\[8mm]
J.~Gehrlein$^{\, a,}$ \footnote{E-mail: 
\texttt{julia.gehrlein@student.kit.edu}},
S.~T.~Petcov$^{\, b,c,}$ \footnote{Also at:
 Institute of Nuclear Research and Nuclear Energy,
  Bulgarian Academy of Sciences, 1784 Sofia, Bulgaria.},
M.~Spinrath$^{\, a,}$ \footnote{E-mail: \texttt{martin.spinrath@kit.edu}},
A.~V.~Titov$^{\, b,}$ \footnote{E-mail: \texttt{arsenii.titov@sissa.it}}
\\[1mm]
\end{center}
\vspace*{0.50cm}
\centerline{$^{a}$ \it Institut f\"ur Theoretische Teilchenphysik, Karlsruhe Institute of Technology,}
\centerline{\it Engesserstra\ss{}e 7, D-76131 Karlsruhe, Germany}
\vspace*{0.2cm}
\centerline{$^{b}$ \it SISSA/INFN, Via Bonomea 265, I-34136 Trieste, Italy }
\vspace*{0.2cm}
\centerline{$^{c}$ \it Kavli IPMU (WPI), University of Tokyo,} 
\centerline{\it 5-1-5 Kashiwanoha, 277-8583 Kashiwa, Japan}
\vspace*{1.20cm}

\begin{abstract}
\noindent
Neutrino mixing sum rules are common to a large class
of models based on the (discrete) symmetry approach 
to lepton flavour. In this approach the neutrino mixing matrix 
$U$ is assumed to have an underlying approximate symmetry 
form $\tilde{U}_{\nu}$, which is dictated by, or associated with, 
the employed (discrete) symmetry. In such a setup the cosine of 
the Dirac CP-violating phase $\delta$ can be related to
the three neutrino mixing angles in terms of a sum rule 
which depends on the symmetry form of $\tilde{U}_{\nu}$. 
We consider five extensively  discussed 
possible symmetry forms of $\tilde{U}_{\nu}$:
i) bimaximal (BM) and ii) tri-bimaximal (TBM) forms, 
the forms corresponding to
iii) golden ratio type A (GRA) mixing, 
iv) golden ratio type B (GRB) mixing, 
and v) hexagonal (HG) mixing. 
For each of these forms we investigate the renormalisation 
group corrections to the sum rule predictions 
for $\delta$ in the cases of neutrino Majorana mass 
term generated by the Weinberg (dimension 5) operator added to  
i) the Standard Model, and ii) the minimal SUSY extension of the 
Standard Model.
\end{abstract}

\end{titlepage}
\setcounter{footnote}{0}

\section{Introduction}

Understanding the observed pattern of neutrino mixing and establishing 
the status of leptonic CP violation are among the ``big'' open
questions in particle physics. 
Considerable efforts have been made in the past years trying to answer these 
fundamental questions. 
In particular, the approach based on a discrete 
non-Abelian family symmetry in the lepton sector, 
assumed to be existing at some high-energy scale, 
has been widely studied in the literature (for reviews on the subject see 
\cite{Ishimori:2010au,Altarelli:2010gt,King:2013eh, King:2014nza}).
In this approach the family symmetry has necessarily 
to be broken at low energies 
to some residual symmetries of the charged lepton and neutrino 
mass matrices. These residual symmetries constrain the form of the 
matrices which diagonalise the charged lepton and neutrino mass matrices, 
and hence the form of the Pontecorvo, Maki, Nakagawa, Sakata (PMNS) 
neutrino mixing matrix. 

 In the three neutrino mixing case (see, e.g., \cite{PDG2016})
the $3\times3$ unitary PMNS matrix can be parametrised in 
terms of three mixing angles, 
$\theta_{12}$, $\theta_{13}$, $\theta_{23}$, 
one Dirac phase $\delta$ and, if the massive neutrinos are Majorana particles, 
two Majorana phases \cite{Bilenky:1980cx}. 
The Dirac and Majorana phases are responsible for CP violation 
in the lepton sector. 
The neutrino mixing parameters $\sin^2 \theta_{12}$, $\sin^2 \theta_{13}$ 
and $\sin^2 \theta_{23}$ have been determined with a relatively high precision 
in the recent global analyses 
\cite{Capozzi:2016rtj,Capozzi:2013csa,Gonzalez-Garcia:2014bfa}. 
These analyses provided only a hint so far that  $\delta \approx 3\pi/2$.
In Table~\ref{tab:exp_parameters} we summarise the best fit values, 
$1\sigma$ and $3\sigma$ allowed ranges of the mixing parameters and the mass 
squared differences $\Delta m_{21}^2$ and $\Delta m_{31}^2$ ($\Delta m_{23}^2$), 
with $\Delta m_{ij}^2 \equiv m_i^2 - m_j^2$,  
$m_{1,2,3}$ being the neutrino masses, found in ref. \cite{Capozzi:2016rtj}
for the neutrino mass spectrum with normal (inverted) ordering 
(denoted further as the NO (IO) spectrum). 
We will use the results given in Table~\ref{tab:exp_parameters} in 
our numerical analyses. 
%%%%%%%%%%%%%%
\begin{table}
\centering
\begin{tabular}{lccc} 
\toprule
Parameter & Best fit & $1\sigma$ range & $3\sigma$ range\\ 
\midrule 
$\sin^2\theta_{12}/10^{-1}$ & $2.97$ & $2.81\rightarrow 3.14$ & $2.50\rightarrow 3.54$\\[0.5 pc]
$\sin^2\theta_{13}/10^{-2}$~(NO) & $2.14$ & $2.05\rightarrow 2.25$ & $1.85\rightarrow 2.46$\\[0.5 pc]
$\sin^2\theta_{13}/10^{-2}$~(IO) & $2.18$ & $2.06\rightarrow 2.27$ & $1.86\rightarrow 2.48$\\[0.5 pc]
$\sin^2\theta_{23}/10^{-1}$~(NO) & $4.37$ & $4.17\rightarrow 4.70$ & $3.79\rightarrow 6.16$\\[0.5 pc]
$\sin^2\theta_{23}/10^{-1}$~(IO) & $5.69$ & $4.28\rightarrow 4.91 \oplus 5.18\rightarrow 5.97$ & $3.83\rightarrow 6.37$\\[0.5 pc]
$\delta/\pi$~(NO) & $1.35$ & $1.13\rightarrow 1.64$ & $0\rightarrow 2$\\[0.5 pc]
$\delta/\pi$~(IO) & $1.32$ & $1.07\rightarrow 1.67$ & $0\rightarrow 2$\\
\midrule
$\Delta m_{21}^{2}/10^{-5}$~eV$^2$ & $7.37$ & $7.21\rightarrow 7.54$ & $6.93\rightarrow 7.97$\\[0,5 pc]
$\Delta m_{31}^{2}/10^{-3}$~eV$^2$~(NO) & $2.54$ & $2.50\rightarrow 2.58$ & $2.40\rightarrow 2.67$\\[0,5 pc]
$\Delta m_{23}^{2}/10^{-3}$~eV$^2$~(IO) & $2.50$ & $2.46\rightarrow 2.55$ & $2.36\rightarrow 2.64$\\
\bottomrule
\end{tabular}
\caption{The best fit values, $1\sigma$ and 3$\sigma$ ranges of the 
neutrino oscillation parameters taken from~\cite{Capozzi:2016rtj}. 
}
\label{tab:exp_parameters}
\end{table}
%%%%%%%%%%%%%%%%%%%%%%%%%%%%%%%%%
%

 In the discrete symmetry approach specific correlations 
between the mixing angles 
and the CP-violating (CPV) phases occur. 
These correlations are usually 
referred to as neutrino mixing sum rules (see, e.g.,
\cite{King:2013eh,King:2014nza,Hanlon:2013ska,Marzocca:2013cr,Petcov:2014laa,
Girardi:2014faa,Girardi:2015zva,Girardi:2015vha,Girardi:2015rwa,Girardi:2016zwz}).
\footnote{In flavour models there exists another type of correlations 
which hold  between the neutrino masses and the Majorana phases.
These correlations are called neutrino mass sum rules (for recent extensive studies,
see, e.g., \cite{King:2013psa, Gehrlein:2015ena}).}
Since mixing sum rules are concrete relations 
between different observables, i.e., 
the neutrino mixing angles and the Dirac phase, 
they can be tested experimentally. 
Thus, via sum rules, one can examine 
the current phenomenologically viable
flavour models based on different discrete symmetries.

 In \cite{Petcov:2014laa,Girardi:2015vha} different mixing sum rules 
have been derived and in \cite{Girardi:2014faa,Girardi:2015vha, Girardi:2015zva} the phenomenological
consequences of these sum rules have been studied.
In \cite{Girardi:2015rwa} sum rules and predictions for
$\cos\delta$ have been obtained from different types 
of residual symmetries in the charged lepton and neutrino sectors.
In these studies it was assumed that the sum rule 
is exactly realised at low energy. 
However, as every quantity in quantum field theory, the mixing parameters
get affected by renormalisation group (RG) running.
Similar to the study of renormalisation group corrections 
to neutrino mass sum rules
in \cite{Gehrlein:2015ena},
we investigate in the present article the impact 
of corrections from the renormalisation group equations (RGEs) on the mixing sum rule 
predictions for the Dirac phase $\delta$.
The main question we want to address 
is how stable the predictions for 
$\delta$ are under RG corrections which
under certain conditions 
can be expected to 
be quite sizeable \cite{Antusch:2003kp}. 

In the literature RG corrections to certain type of mixing 
sum rules have been studied before.
The first attempt to study RG corrections to mixing angle sum rules, 
to our knowledge, has been made in \cite{Schmidt:2006rb} 
for the quark-lepton complementarity relations, 
$\theta_{12} + \theta_C \cong \pi/4$ 
and $\theta_{23} + \arcsin\,V_{cb} \cong \pi/4$, 
$\theta_C$ and $V_{cb}$ being the Cabibbo angle and an element of the 
Cabibbo, Kobayashi, Maskawa (CKM) quark mixing matrix.
In \cite{Boudjemaa:2008jf} the RG corrections 
for the sum rule 
relating the element $U_{\tau e}$ of the PMNS matrix to the 
element $V^{\rm TBM}_{\tau e} = -\,1/\sqrt{6}$
of the tri-bimaximal mixing matrix, $|U_{\tau e}| = 1/\sqrt{6}$, 
and for the leading order in $\theta_{13}$ versions of this sum rule,
have been investigated.
In refs.~\cite{Schmidt:2006rb} and  \cite{Boudjemaa:2008jf} 
the bimaximal (BM) mixing \cite{Petcov:1982ya} scheme 
and the tri-bimaximal (TBM) mixing scheme \cite{Xing:2002sw} (see also \cite{Wolfenstein:1978uw}) , 
respectively, were analysed.
In \cite{Antusch:2008yc} the study of RG perturbations  
was done for an approximate (leading order) 
mixing sum rule and  for normal hierarchical neutrino mass spectrum, 
$m_1 \ll m_2 < m_3$,  
neglecting terms of order $\mathcal{O}(m_1/m_2)$ and  $\mathcal{O}(m_1/m_3)$.
The authors of 
\cite{Antusch:2008yc} extended their analysis 
to incorporate canonical normalisation effects
besides RG corrections. Both type of corrections were  
assumed to be dominated by the third family effects.
The authors of \cite{Ballett:2014dua} estimated
the size of RG corrections to the sum rules we 
will be considering in the present study by taking into account  
{\it only} the RG correction to $\theta_{12}$.

In the present article we go beyond these previous works 
i) by considering the exact form of the general mixing sum rules  
derived in \cite{Petcov:2014laa}, 
ii) by taking into account the RG corrections not only to 
the angle $\theta_{12}$, but to all three neutrino mixing angles 
$\theta_{12}$, $\theta_{23}$, $\theta_{13}$ and the CPV phases,
iii) discussing not only the cases of BM or TBM mixing schemes, 
but also the cases of golden ratio type A (GRA) \cite{Kajiyama:2007gx}, 
golden ratio type B (GRB) \cite{Rodejohann:2008ir}
and hexagonal (HG) \cite{Albright:2010ap} mixing schemes, 
and iv) by considering both the cases of NO and IO neutrino mass spectra. 
We perform the analysis assuming that the neutrino Majorana mass matrix 
is generated by the Weinberg (dimension 5) operator. The RG corrections 
to the sum rules of interest are calculated in  
the Standard Model as well as in the minimal supersymmetric
extension of the Standard Model (MSSM).

Our study goes also beyond  \cite{Zhang:2016djh} where only the GRA, 
BM and TBM mixing schemes were analysed.
We discuss different forms of the charged lepton mixing matrix and 
present a significantly larger number of results. 
In particular, we derive values of the neutrino mass scale and 
$\tan \beta$ for which the various mixing schemes are still viable.
We make a thorough numerical analysis from which we derive 
likelihood functions for the value of the Dirac CPV phase 
$\delta$ at low energies if the
specified mixing sum rule holds at high energies.

The paper is organised as follows: after a short review of 
the framework for mixing sum rules
in Section~\ref{sec:framework}, we present analytical 
estimates for the allowed parameter regions
for $\delta$ taking RG corrections into account 
in Section~\ref{sec:analytical}. In
Section~\ref{sec:numerical_results} we present the 
numerical results for the different
mixing schemes. Finally, we summarise and conclude 
in Section~\ref{sec:summary} and present in the appendix
plots for the likelihoods in terms of $\cos \delta$ for
better comparison with previous literature.

\section{Mixing Sum Rules}
\label{sec:framework}

In this section we briefly review the framework in which mixing sum rules 
are obtained and fix notation and conventions.
In the most general case the PMNS matrix $U$ 
can be parametrised as \cite{Frampton:2004ud}
%%%%%%%
\begin{align}
U = U_e^\dagger U_\nu = (\tilde{U}_e)^\dagger \Psi \tilde{U}_\nu Q_0\,.
\end{align}
%%%%%%%
Here $U_e$ and $U_\nu$ are $3\times3$ unitary matrices, 
which diagonalise, respectively, the charged lepton and neutrino 
mass matrices. 
$\tilde{U}_e$ and $\tilde{U}_\nu$ are CKM-like $3\times3$ unitary matrices,
and $\Psi$ and $Q_0$ are diagonal phase matrices: 
%%%%%%%
\begin{align}
\Psi&=\text{diag}\left(1,\text{e}^{-\text{i}\, \psi},\text{e}^{-\text{i}\, \omega}\right)\,,\\
Q_0&=\text{diag}\left(1,\text{e}^{\text{i}\tfrac{\xi_{21}}{2}},\text{e}^{\text{i}\tfrac{\xi_{31}}{2}}\right)\,.
\end{align}
%%%%%%%
%
The phases in $Q_0$ contribute to the Majorana phases in the PMNS matrix. 

 Similar to what has been done in \cite{Petcov:2014laa,Girardi:2014faa,Girardi:2015zva,Girardi:2015vha}
we will consider the cases when $\tilde{U}_\nu$  has the BM, 
TBM, GRA, GRB and HG forms. For all these forms $\tilde U_\nu$ 
can be expressed as a product of $3\times3$ orthogonal matrices 
$R_{23}$ and $R_{12}$ describing rotations in the 2-3 and 1-2 planes, i.e.,
%%%%%%%
\begin{align}
\tilde{U}_\nu=R_{23}(\theta_{23}^\nu)R_{12}(\theta_{12}^\nu)\,,
\label{eq:tildeU}
\end{align}
%%%%%%%
with $\theta_{23}^\nu=-\pi/4$ and $\theta_{12}^{\nu}=\pi/4$ (BM); 
$\theta_{12}^{\nu}=\arcsin (1/\sqrt{3})$ (TBM); 
$\theta_{12}^{\nu}= \arctan (1/\phi)$ (GRA), 
$\phi=(1+\sqrt{5})/2$ being the golden ratio;
$\theta_{12}^{\nu}=\arccos (\phi/2)$ (GRB); 
$\theta_{12}^{\nu}=\pi/6$ (HG).
For convenience, in another convention the same list reads
$\sin^2\theta_{23}^\nu= 1/2$ and $\sin^2 \theta_{12}^{\nu}=1/2$ (BM); 
$\sin^2\theta_{12}^{\nu}=1/3$ (TBM); 
$\sin^2\theta_{12}^{\nu}= (5 - \sqrt{5})/10$ (GRA);
$\sin^2\theta_{12}^{\nu}= (5 - \sqrt{5} )/8$ (GRB); 
$\sin^2\theta_{12}^{\nu}=1/4$ (HG).

 For the matrix $\tilde{U}_e$, following \cite{Petcov:2014laa}, 
we will consider 
two different forms both of which correspond to negligible $\theta^e_{13}$. 
They are realised in a class of flavour models based on a GUT and/or 
a discrete symmetry 
(see, e.g., \cite{Gehrlein:2014wda,Meroni:2012ty,Marzocca:2011dh,
Antusch:2012fb,Girardi:2013sza,Antusch:2013wn,Antusch:2010es}). 
The first form is characterised also by zero $\theta^e_{23}$, i.e.,
%%%%%%%
\begin{align}
\tilde{U}_e = R_{12}^{-1}(\theta^e_{12})\,.
\label{eq:Ue12}
\end{align} 
%%%%%%% 
%
In this case there is a correlation between the values of
$\sin^2\theta_{23}$ and $\sin^2\theta_{13}$:  
%%%%%%%
\begin{align}
\sin^2\theta_{23} = \frac{\sin^2\theta^\nu_{23} - \sin^2\theta_{13}}
{1 - \sin^2\theta_{13}}\,,
\end{align}
%%%%%%%
%
which for all the symmetry forms of $\tilde U_\nu$ introduced above leads to
%%%%%%%
\begin{align}
\sin^2\theta_{23} = \frac{1 - 2 \sin^2\theta_{13}}
{2\,(1 - \sin^2\theta_{13})}
= \frac{1}{2} - \frac{1}{2} \sin^2\theta_{13} + \mathcal{O}(\sin^4 \theta_{13})\,.
\label{eq:ssth23}
\end{align}
%%%%%%%
%
This implies in turn that $\theta_{23}$ cannot 
deviate significantly from $\pi/4$.
The second form of $\tilde U_e$ corresponds to non-zero $\theta^e_{12}$ and
$\theta^e_{23}$, i.e., 
%%%%%%%
\begin{align}
\tilde{U}_e = R_{23}^{-1}(\theta_{23}^e) R_{12}^{-1}(\theta_{12}^e)\,.
\label{eq:Ue2312}
\end{align}
%%%%%%%
%
This matrix provides the corrections to $\tilde{U}_\nu$ necessary to 
reproduce the current best fit values of all the three neutrino mixing 
angles $\theta_{12}$, $\theta_{13}$ and $\theta_{23}$ in the 
PMNS matrix $U$ without any further contributions like RG or other corrections.

It was shown in \cite{Petcov:2014laa} that for $\tilde{U}_\nu$ given
in eq.~\eqref{eq:tildeU} 
and $\tilde{U}_e$ determined in eqs.~\eqref{eq:Ue12} or \eqref{eq:Ue2312}, 
the Dirac phase $\delta$ present in the PMNS matrix satisfies a sum rule 
which reads 
%%%%%%%
\begin{align}
\cos \delta=\frac{\tan\theta_{23}}{\sin 2\theta_{12}\sin\theta_{13}}
\left[\cos2 \theta_{12}^{\nu}+\left(\sin^2\theta_{12}-\cos^2\theta_{12}^{\nu}\right)
\left(1-\text{cot}^2\theta_{23}\sin^2\theta_{13}\right)\right]\,.
\label{eq:delta}
\end{align}
%%%%%%%
%
Additionally, in the case of $\tilde U_e$ given in eq.~\eqref{eq:Ue12}, 
the correlation between $\theta_{23}$ and $\theta_{13}$, eq.~\eqref{eq:ssth23},  
has to be respected. The sum rule, eq.~\eqref{eq:delta}, in this case 
reduces to \cite{Petcov:2014laa}
%%%%%%%%%%%%%%%%%%%%%%%
\begin{align} 
\cos\delta = \frac{(1 - 2 \sin^2\theta_{13})^{\frac{1}{2}}}
{\sin2\theta_{12} \sin\theta_{13}}\,
\bigg[ \cos2\theta^{\nu}_{12} + \left (\sin^2\theta_{12} - \cos^2\theta^{\nu}_{12}\right )\,
\frac{1 - 3 \sin^2\theta_{13}}{1 - 2 \sin^2\theta_{13}} \bigg]\,.
\label{eq:delta12e}
\end{align}
%%%%%%%%%%%%%%%%%%%%%%%

 In the following we will refer to the case with $\tilde U_e$ given in 
eq.~\eqref{eq:Ue12} (eq.~\eqref{eq:Ue2312}) as to the case of 
zero (non-zero) $\theta^e_{23}$. In this article we will study the impact 
of the RG corrections on the mixing sum rules in 
eqs.~\eqref{eq:delta} and \eqref{eq:delta12e}, and the angle 
sum rule in eq.~\eqref{eq:ssth23}, which are assumed to hold at
some high-energy scale specified later.

 In \cite{Girardi:2015vha} other forms of the matrices 
$\tilde U_e$ and $\tilde U_\nu$ corresponding to different rotations 
and leading to sum rules for $\cos\delta$ of the type of 
eqs.~\eqref{eq:delta} and \eqref{eq:delta12e} 
have been investigated. 
The RG corrections to them, however, are expected 
to be similar to the ones which take place for the sum rules 
described above. 
For this reason we will not consider them in the present study. 

\section{Analytical Estimates}
\label{sec:analytical}
Before we present our numerical results in the next section, we give 
in this section analytical estimates 
of the effect of radiative corrections 
on the mixing sum rules. We discuss how we obtain 
constraints on the mass scale
and on $\tan \beta$ (in the MSSM) from the requirement that the mixing
sum rule has to be fulfilled at the high scale. 

\subsection{General Effects of Radiative Corrections}

The running of the mixing parameters is already known for quite some time, see,
e.g.,~\cite{Antusch:2003kp}. One might wonder if  RG corrections have 
a large impact on the predicted value for $\delta $ 
from the sum rule in eq.~\eqref{eq:delta}.
Indeed, we expect large RG corrections for a 
large Yukawa coupling (large $\tan \beta$) and a heavy 
neutrino mass scale.
To be more precise, the $\beta$-functions of the mixing angles, 
in the leading order in $\theta_{13}$ and neglecting 
the electron and muon Yukawa couplings 
in comparison to the tau one,
depend on the tau Yukawa coupling,
the absolute neutrino mass scale (or ${\rm min}(m_j)$, $j=1,2,3$),
the mixing angles, the type of spectrum~--~normal or inverted ordering~--~% 
the neutrino masses obey,
on the Majorana phases 
$\alpha_1$ and $\alpha_2$~%
\footnote{The Majorana phases $\alpha_1$ and $\alpha_2$ are related to those 
of the standard parametrisation of the PMNS matrix \cite{PDG2016}, 
 $\alpha_{21}$ and $\alpha_{31}$, as follows:
$\alpha_{21} = \alpha_1 - \alpha_2$ and 
$\alpha_{31} = \alpha_1$.
}, and in the MSSM~--~on $\tan \beta$. 
In the leading order in $\theta_{13}$ only the $\beta$-function for $\theta_{13}$ 
depends on $\delta$.  
The $\beta$-functions read up to $\mathcal{O}(\theta_{13})$ \cite{Antusch:2003kp}:
%%%%%%%%%%%%%%%%%%%%%%%%%%%%%%%%%%%%
\begin{align}
\frac{\text{d}\, \theta_{12}}{\text{d}\, \ln (\mu/\mu_0)} &=-\frac{C y_{\tau}^{2}}{32\pi^{2}}\sin 2\theta_{12} s^{2}_{23}\frac{\left|m_{1}\text{e}^{\text{i}\alpha_{1}}+m_{2}\text{e}^{\text{i}\alpha_{2}}\right|^{2}}{\Delta m_{21}^{2}}+\mathcal O(\theta_{13}) \;,
\label{eq:theta_12_r} \\
\frac{\text{d}\, \theta_{13}}{\text{d}\, \ln (\mu/\mu_0)} &=\frac{C y_{\tau}^{2}}{32\pi^{2}}\sin 2\theta_{12} \sin 2\theta_{23} \frac{m_{3}}{\Delta m_{32}^{2}(1+\zeta)} \nonumber\\
&\nonumber \times
\left[m_{1}\cos(\alpha_{1}-\delta)-(1+\zeta)m_{2}\cos(\alpha_{2}-\delta)-\zeta m_{3}\cos\delta \right]\\
&+\mathcal O(\theta_{13}) \;, \label{eq:theta_13_r} \\
\frac{\text{d}\, \theta_{23}}{\text{d}\, \ln (\mu/\mu_0)}&=-\frac{C y_{\tau}^{2}}{32\pi^{2}}\sin 2\theta_{23} \frac{1}{\Delta m_{32}^{2}}\left[c_{12}^{2}\left|m_{2}\text{e}^{\text{i}\alpha_{2}}+m_{3}\right|^{2}+s_{12}^{2}\frac{\left|m_{1}\text{e}^{\text{i}\alpha_{1}}+m_{3}\right|^{2}}{1+\zeta}\right]\\&+\nonumber\mathcal O(\theta_{13}) \;, \label{eq:theta23_running}
\end{align}
%%%%%%%%%%%%%%%%%%%%%%%%%%%%%%%%%%%%
%
with $\mu$ being the renormalisation scale, $\zeta=\frac{\Delta m_{21}^2}{\Delta m_{32}^2}$ and
$\frac{C y_\tau^2}{32 \pi^2}\approx 0.3 \cdot 10^{-6} (1+\tan^2 \beta) $ in
the MSSM and $\frac{C y_\tau^2}{32 \pi^2}\approx -0.5\cdot 10^{-6}$ in the SM.
In the SM there is no $\tan \beta$ enhancement and hence 
the effects are usually relatively small.

We would like to note at this point that we consider here only minimal scenarios, namely
the SM and the MSSM augmented with Majorana neutrino masses. In standard seesaw
scenarios it would be correct to integrate out the additional heavy states at their respective
mass scale which would change the $\beta$-functions and the running. Nevertheless, we want
to assume the heavy masses all to be roughly of the same order, so that it is a good approximation
to impose the sum rules at the high scale and use the minimal $\beta$-functions for the running.
For low scale seesaw mechanisms this would certainly be a bad  approximation, 
but there the sum rule should be realised at the
low scale as well and running effects can be more generally expected to be small.

%%%%%%%%%%%%%%%%%%%%%%%%%%
\begin{figure}
\begin{center}
\includegraphics[width=0.45\textwidth]{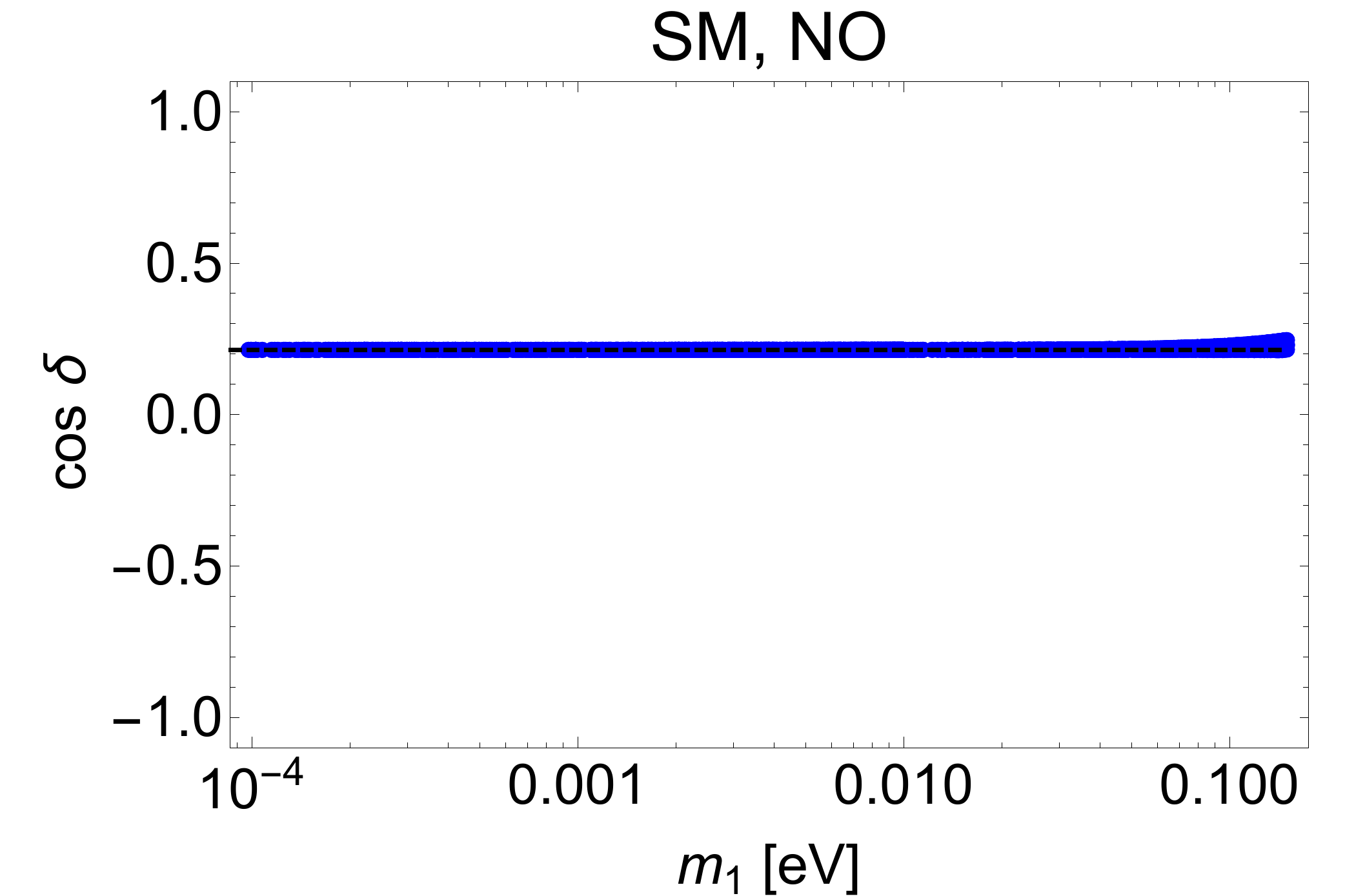} \hfill  
\includegraphics[width=0.45\textwidth]{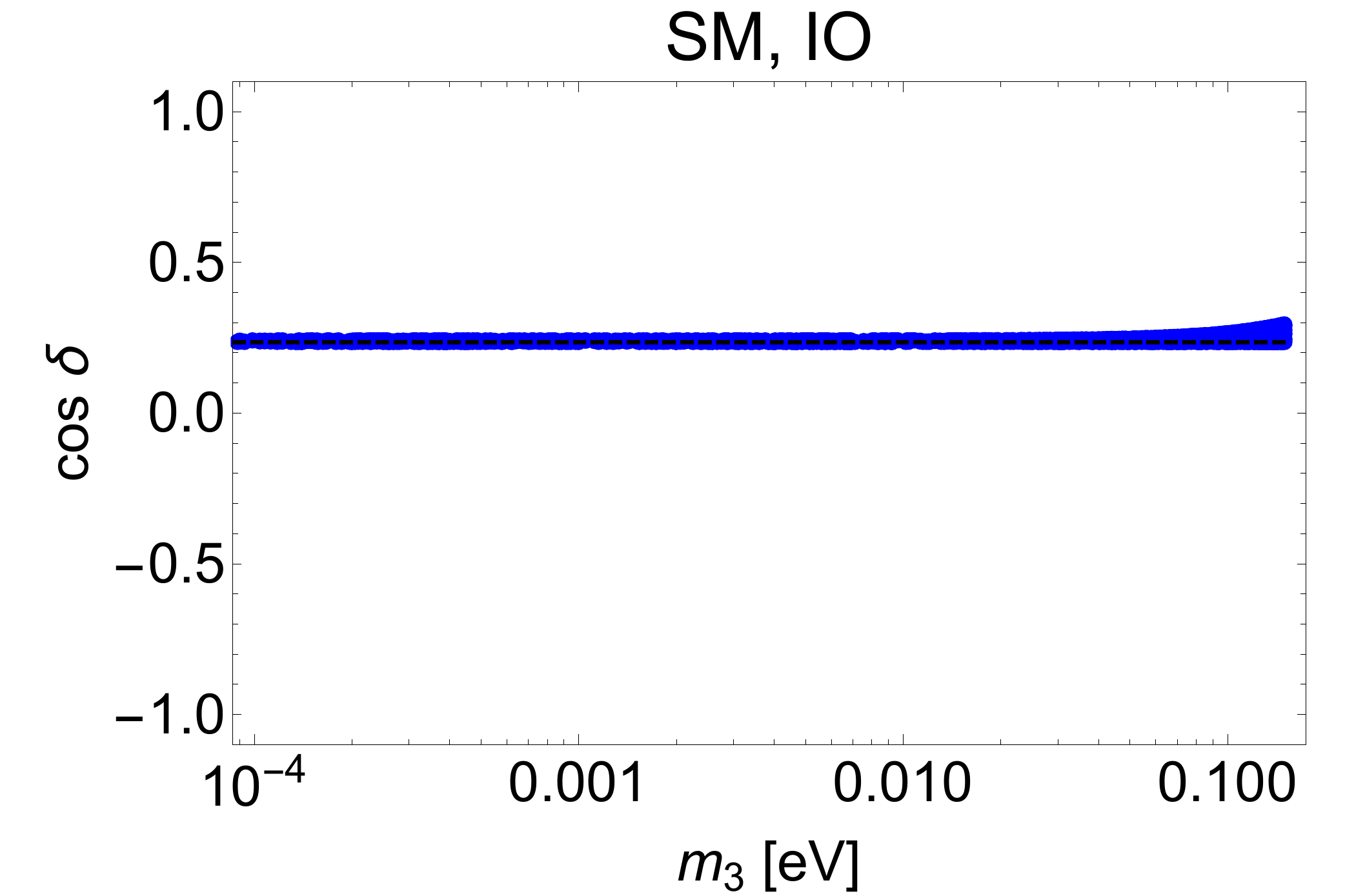} \\[2ex]
\includegraphics[width=0.45\textwidth]{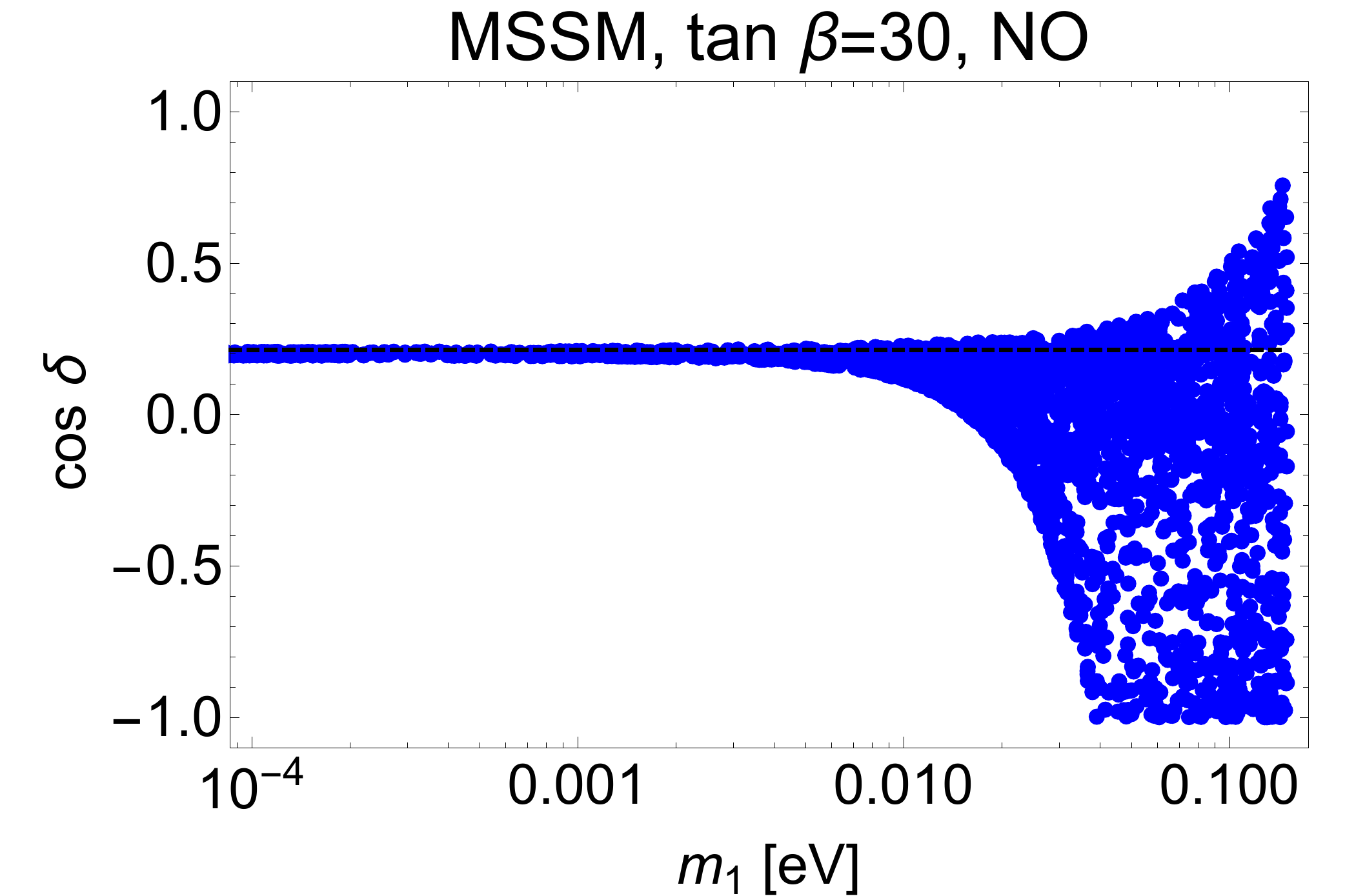} \hfill
\includegraphics[width=0.45\textwidth]{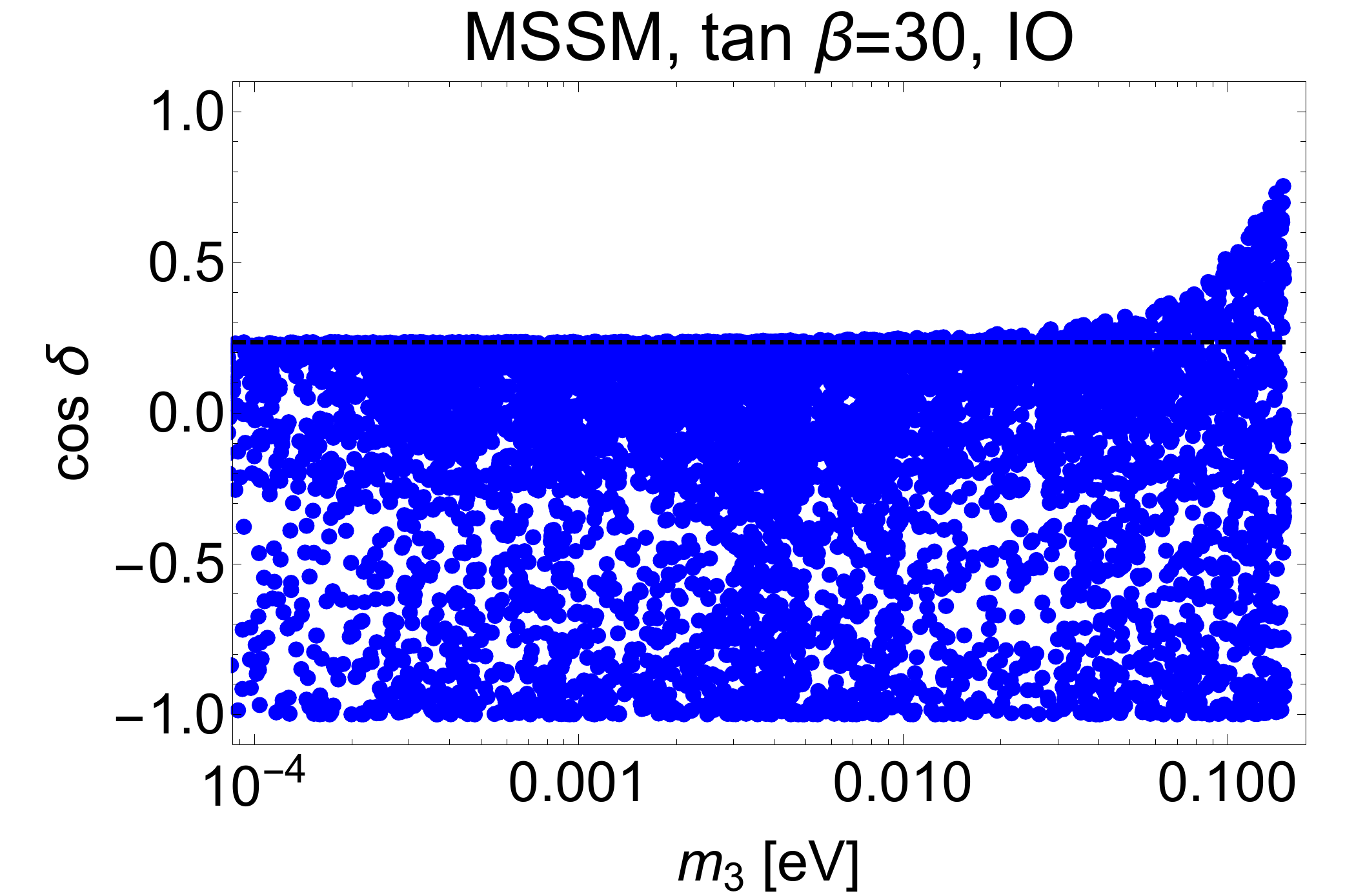} \\[2ex]
\includegraphics[width=0.45\textwidth]{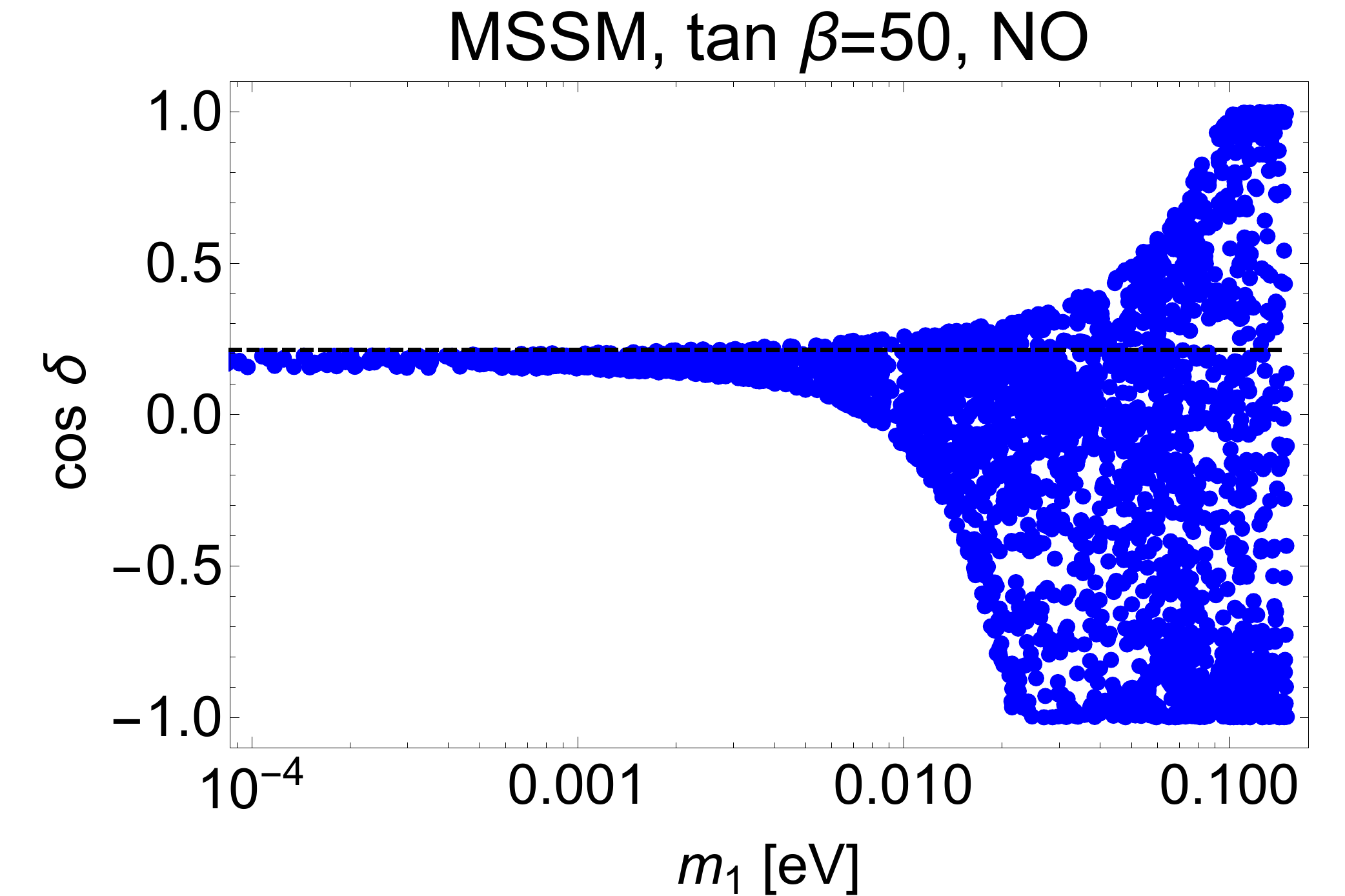} \hfill
\includegraphics[width=0.45\textwidth]{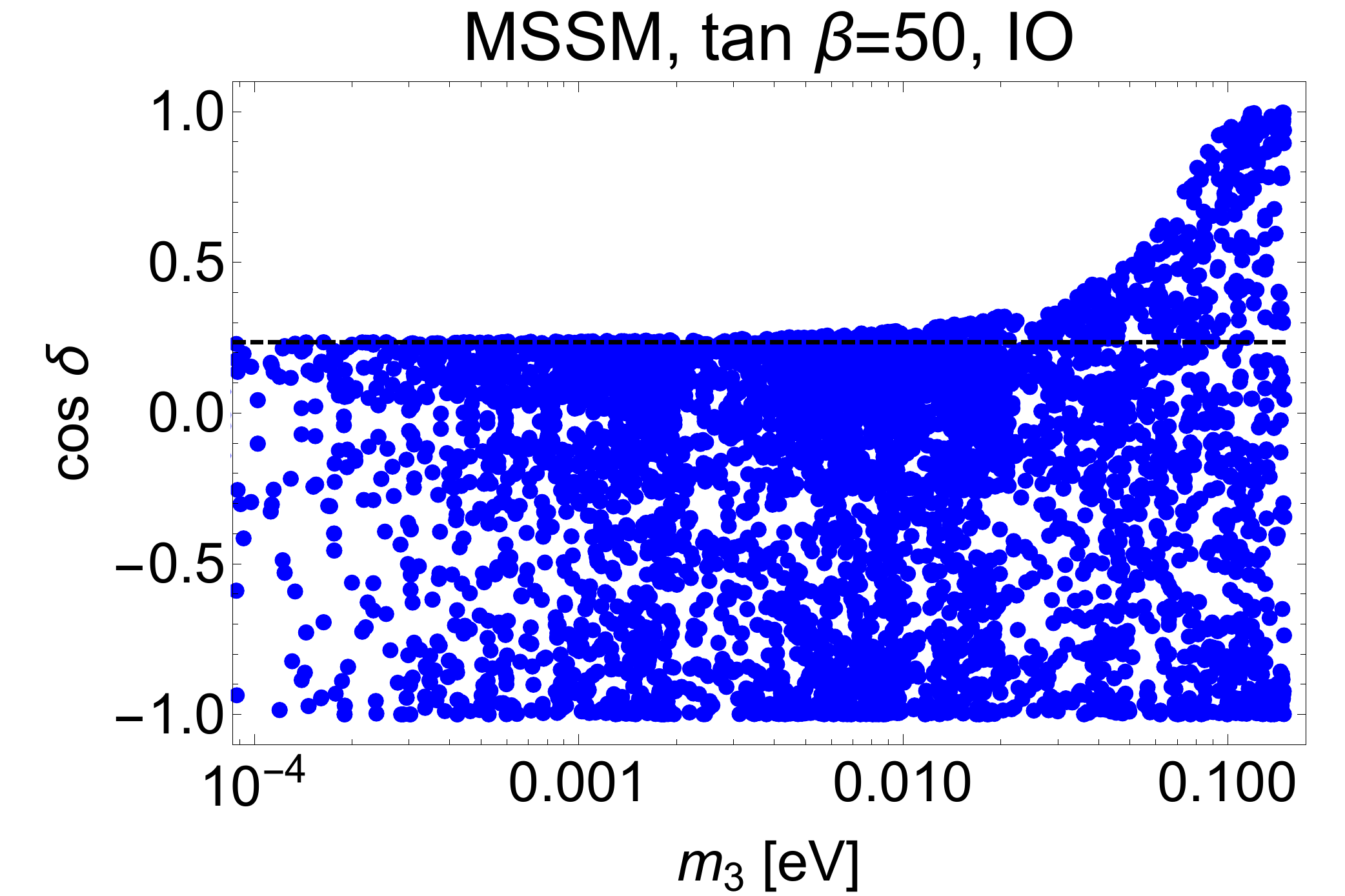} 
\end{center}
\caption{
Results for the predicted value of $\cos\delta$ from the sum rule 
in eq.~\eqref{eq:delta} for the GRA mixing scheme in 
the case where 
$\theta_{12}^e\neq 0$, $\theta_{23}^e\neq0$ and $\theta_{13}^e=0$. 
The black dashed lines represent the tree level result. 
The blue points are our scan points.
For the angles and the mass squared differences 
we took the best fit values from Table~\ref{tab:exp_parameters}.
We let the parameters run between the high-scale $M_S\approx 10^{13}$ GeV and the low-scale $M_Z$. The Majorana
phases are chosen randomly between 0 and $2 \pi$.  
The plots on the left (right) side correspond to
normal (inverted) mass ordering.
\label{Fig:FirstImpression}}
\end{figure}
%%%%%%%%%%%%%%%%%%%%%%%%%%%%%
%

To give an idea about the size of the effect of interest 
we  show in Fig.~\ref{Fig:FirstImpression}
results for  $\cos\delta$ as derived
from the sum rule in eq.~\eqref{eq:delta} for the GRA mixing scheme. 
We used the \texttt{REAP} package~\cite{Antusch:2005gp} 
to solve the RGEs
for the mixing parameters between the low-energy scale $M_Z$ and the 
high-energy scale which we have set equal to the seesaw 
scale $M_S\approx 10^{13}$ GeV.
We only consider the case with $\theta_{12}^e\neq 0$, $\theta_{23}^e\neq 0$
and $\theta_{13}^e=0$.
We have set all mass squared differences and angles to their best fit values 
given in Table~\ref{tab:exp_parameters},
scanned over the lightest neutrino mass and chose random values for the low
energy Majorana phases.
For the SM case we see no effect, while
for $\tan \beta = 30$ and 50, the RG effects 
are significant.
Even for a moderate $\tan \beta$ in the MSSM and a 
relatively small mass scale $m_\text{lightest}\approx 0.04$ eV 
the effect is non-negligible.
Since the running of the angles is stronger with an inverted mass ordering, 
the effect for the prediction of $\cos \delta$ is larger in the IO case.
For that case it is furthermore in particular remarkable that the corrections
do not go to zero for $m_3$ going to zero. This is due to the well-known
fact, cf.~\cite{Antusch:2003kp}, that the $\beta$-functions
for $\delta$ and $\theta_{12}$ are in this limit enhanced by a factor of
$\Delta m_{23}^2/\Delta m_{21}^2$. Together with the $\tan \beta$ enhancement
this leads to quite sizeable effects for all relevant neutrino mass scales.

\subsection{Allowed Parameter Regions with RG Corrections}
\label{subsec:allowedregions}

In this subsection we derive constraints  
on  $\tan \beta$ (in the case of the MSSM) and
the mass of the lightest neutrino, $m_{\text{lightest}}$,  
by imposing the mixing sum rule
at the high scale
and by requiring that 
$\cos \delta\in[-1,1]$ at the high scale.
We have chosen the high-scale to be equal to 
the seesaw scale $M_S\approx 10^{13}$~GeV.
The BM mixing scheme is 
strongly disfavoured for the current 
best fit values of the neutrino mixing angles
without taking the RG corrections into account. 
Thus, one of the  questions we are interested in
is whether the corrections can
reconstitute the validity of the BM scheme even for
the best fit values of the angles.

We give first analytical estimates of the RG effects on
eq.~\eqref{eq:delta}.
At the high scale we can write, for instance, for the mixing
angles
\begin{equation}
\theta_{ij} (M_S) = \theta_{ij}(M_Z) + \delta \theta_{ij} \equiv \theta_{ij} + \delta \theta_{ij} \;,
\end{equation}
where $ \delta \theta_{ij}$ is the RG correction or the difference between
the high-scale and low-scale values of the mixing angle $\theta_{ij}$. Since the
RG corrections are small we can expand the mixing sum
rule at the high-scale in the small quantities and find:
%%%%%%%%%%%%%%%%%%%%%%%%%%%%%%%%%%%%%%%
\begin{align}
\cos \delta(M_S) \approx & \cos\delta(M_Z)+\delta(\cos\delta) \nonumber\\
&\nonumber = \frac{\tan\theta_{23}}{\sin 2\theta_{12}\sin\theta_{13}}(\cos 2 \theta_{12}^{\nu}+(\sin^2\theta_{12}-\cos^2\theta_{12}^{\nu})
\nonumber (1-\text{cot}^2\theta_{23}\sin^2\theta_{13})) \\
&+\nonumber f_{13}(\theta_{13},\theta_{12},\theta_{23},\theta_{12}^{\nu}) \, 
\delta \theta_{13}\\ 
&\nonumber+f_{23}(\theta_{13},\theta_{12},\theta_{23},\theta_{12}^{\nu}) \,
\delta \theta_{23}\\
&+ f_{12}(\theta_{13},\theta_{12},\theta_{23},\theta_{12}^{\nu}) \, 
\delta \theta_{12}~,
\end{align}
%%%%%%%%%%%%%%%%%%%%%%%%%%%%%%%
%
where the $f_{ij}$ are prefactors from the expansion. For the angles
and mass squared differences at the low scale we use the best fit values.
Note that the Dirac phase $\delta$ appears in the $\beta$-function for the
mixing angles. Here, we use the approximation $\delta(M_Z) \approx \delta(M_S)$
and evaluate the value from the sum rule neglecting RG corrections.
This is formally correct since their inclusion would be a two-loop correction.
The Majorana phases are free parameters.

For the best fit values of the angles the function $f_{12}$  
is always positive independent of the value of $\theta_{12}^{\nu}$.
Since the sign of $\delta{\theta}_{12}$ is always negative  
to leading order in $\theta_{13}$,
the correction to $\cos\delta (M_Z)$ due to the running 
of $\theta_{12}$ has a fixed negative sign 
in this approximation.
The sign of the correction due to the running of $\theta_{23}$ depends 
on $\theta_{12}^{\nu}$ and the mass ordering:  
$\delta{\theta}_{23}$  is positive for inverted ordering 
and negative for normal ordering and
$f_{23}$ is negative for $\theta_{12}^{\nu} \gtrsim 33^{\circ}$.
The sign of the correction due to the running of 
$\theta_{13}$ depends on the CPV phases and $\theta_{12}^{\nu}$.

For BM mixing the function $f_{13}$ dominates in $\delta(\cos\delta)$, 
in contrast to  the other
mixing patterns for which $f_{12}$ has the largest influence.
This means that the contribution in TBM, GRA, GRB and HG mixings 
due to the running of $\theta_{12}$, 
which is larger than the contributions due to the
running of the other angles
(except for the case of a parametric suppression of the 
$\beta$-function which  will be discussed later),
is additionally enhanced by the large prefactor $f_{12}$ 
making the $\delta \theta_{12}$ even more important.

Since the running depends also on the unknown Majorana phases 
we will vary them 
and give in the rest of the subsection the results for 
minimal or maximal corrections. Note that minimal 
corrections can also correspond to negative values
of $\delta(\cos\delta)$.

\begin{figure}
\begin{center}
\vspace{-10mm}
\includegraphics[scale=0.5]{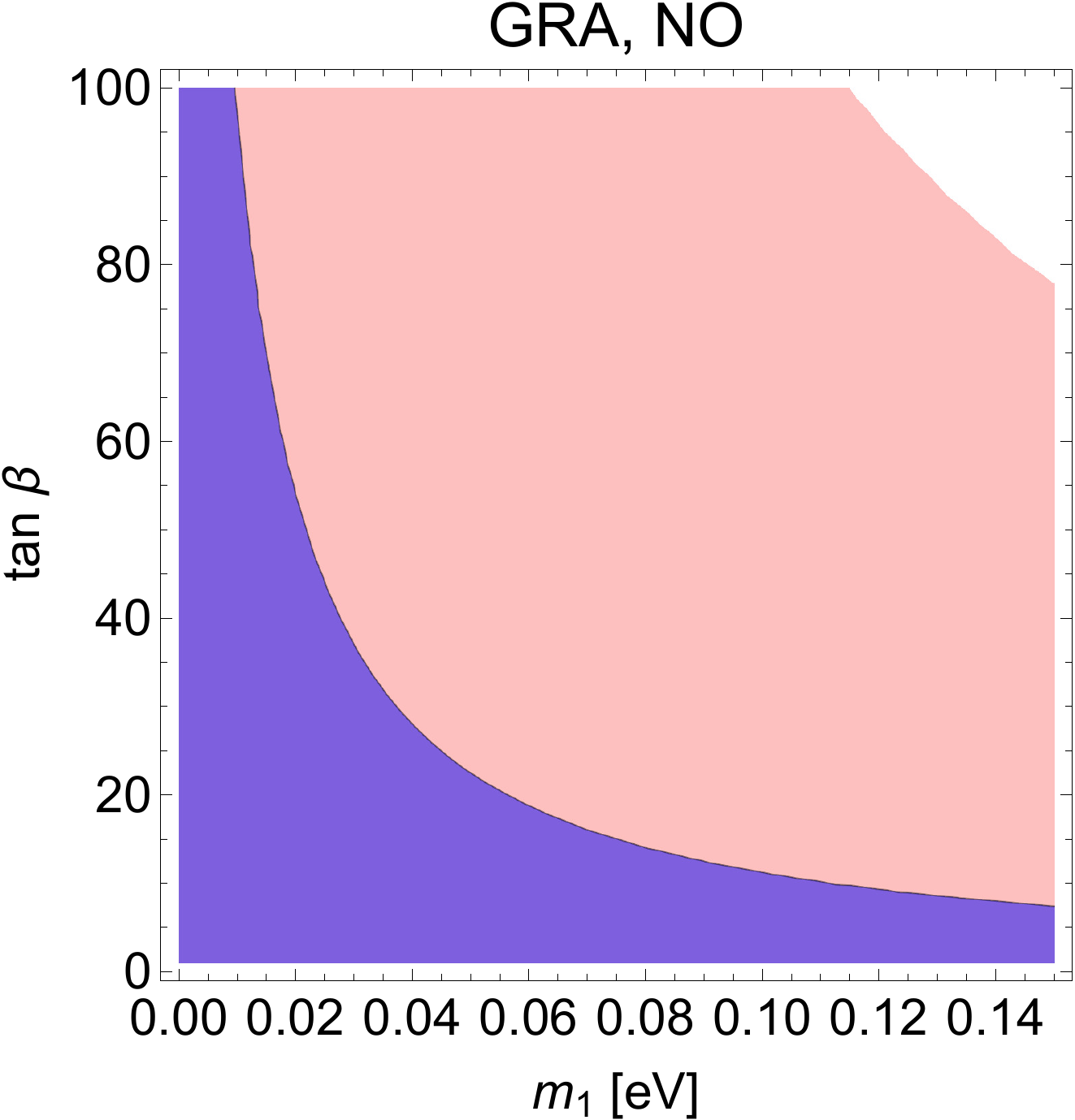}\hspace{0.4cm}
\includegraphics[scale=0.5]{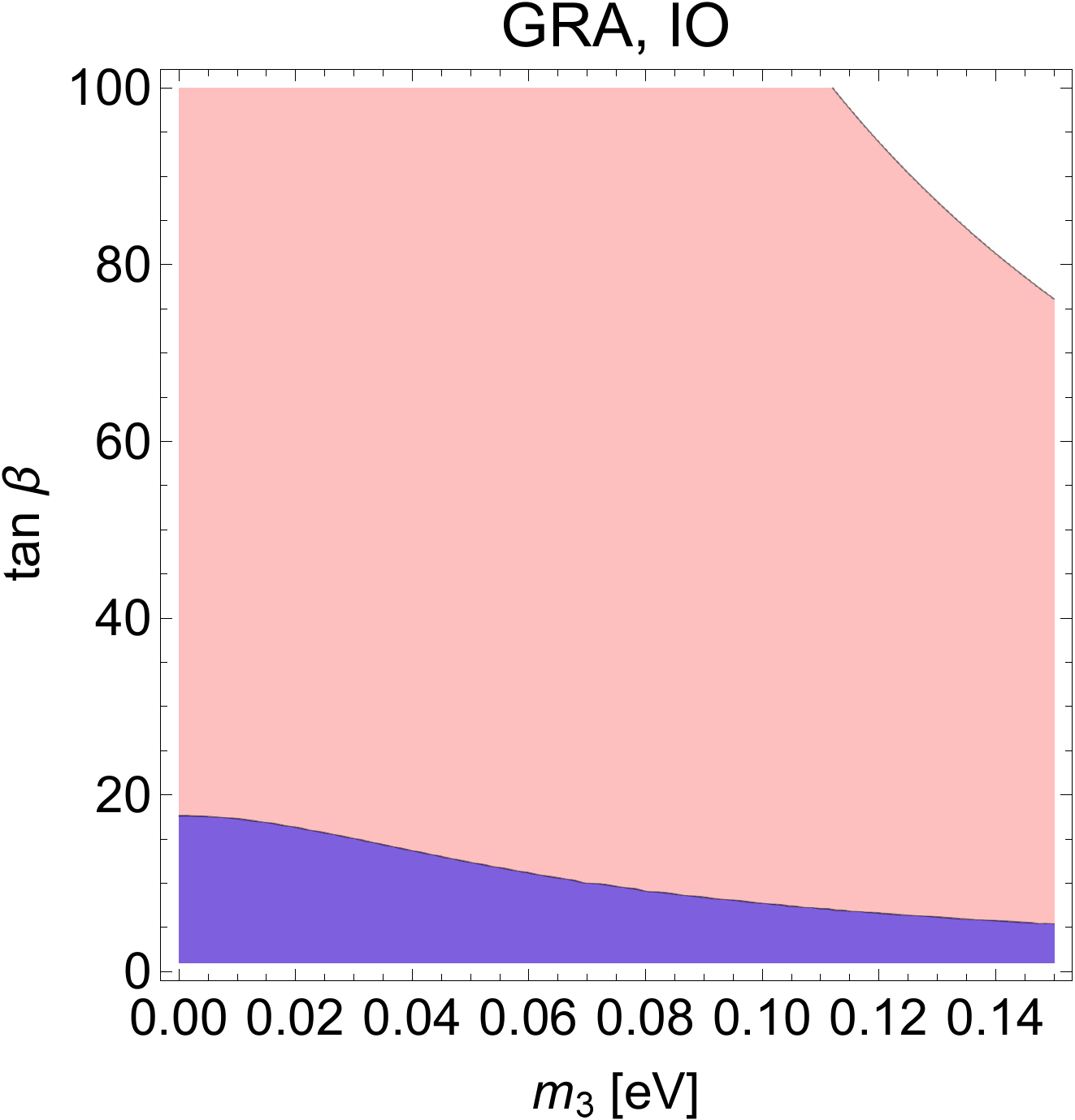} \vspace{4mm}\\ 
\includegraphics[scale=0.5]{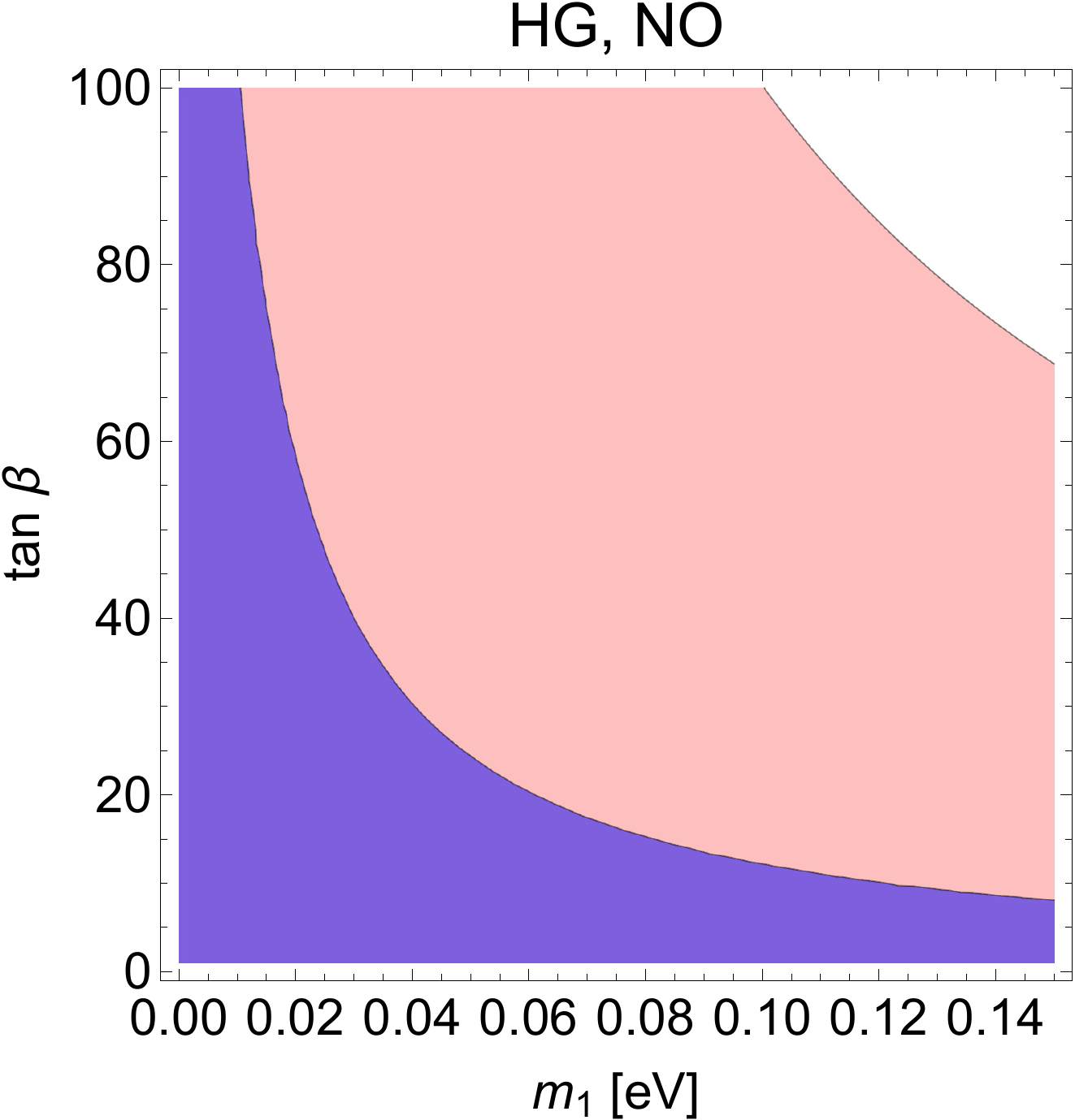}\hspace{0.4cm}
\includegraphics[scale=0.5]{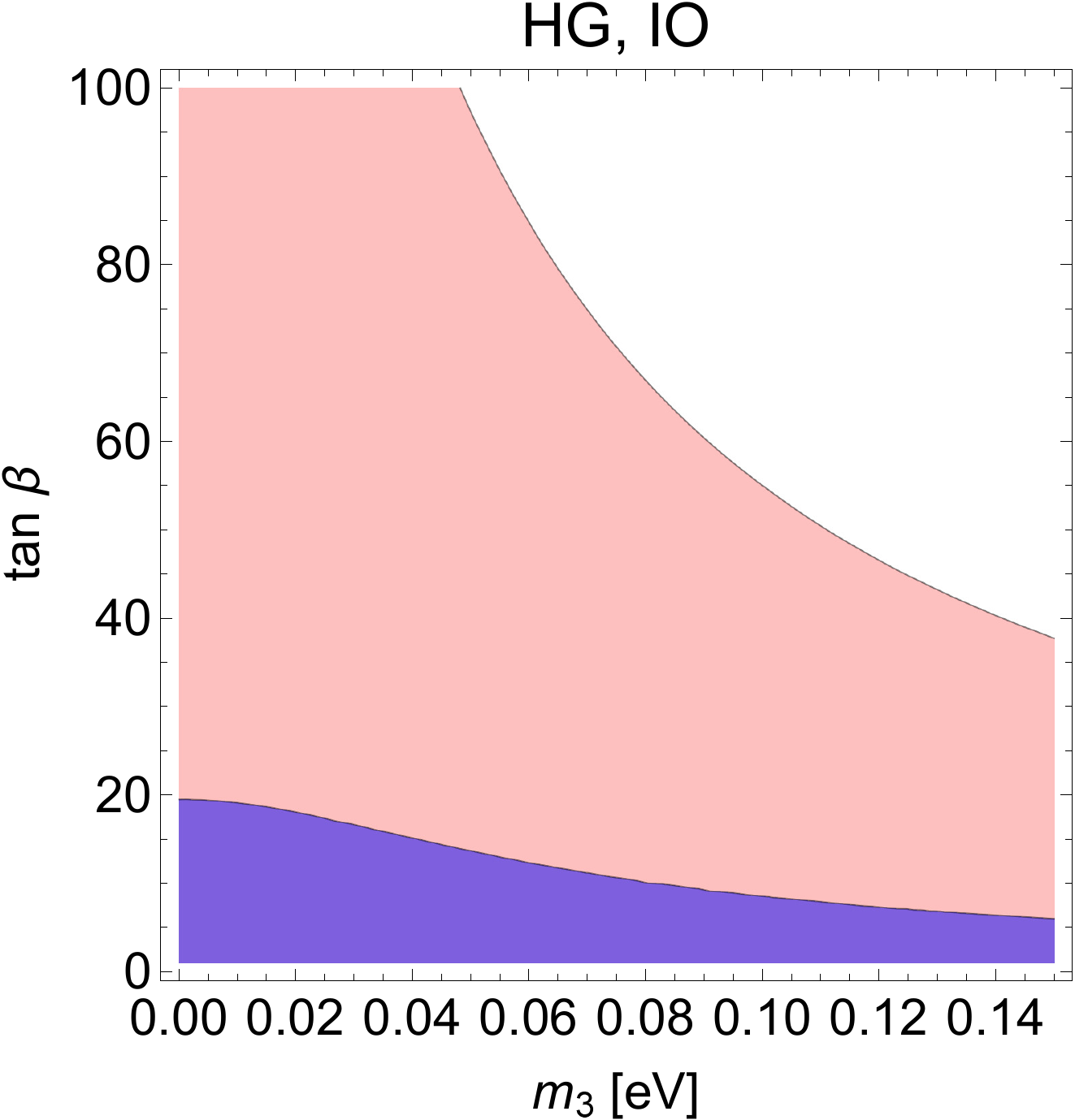} \vspace{4mm}\\ 
\end{center}\vspace{-4mm}
\caption{Allowed regions for $\tan\beta$ and $m_{\text{lightest}}$ 
for the NO and IO spectra in the cases of minimal (blue) and 
maximal (pink) corrections 
for $\cos \delta$ in the GRA mixing scheme (upper plots)
and the HG mixing scheme (lower plots).
We used the best fit values for the mixing angles.
The high-energy scale is set to $10^{13}$ GeV.}
\label{fig:runningGR}
\end{figure}

%%%%%%%%%%%%%%%%%%%%%%%%%%%%%%%%%%%
\begin{figure}
\begin{center}
\vspace{-10mm}
\includegraphics[scale=0.5]{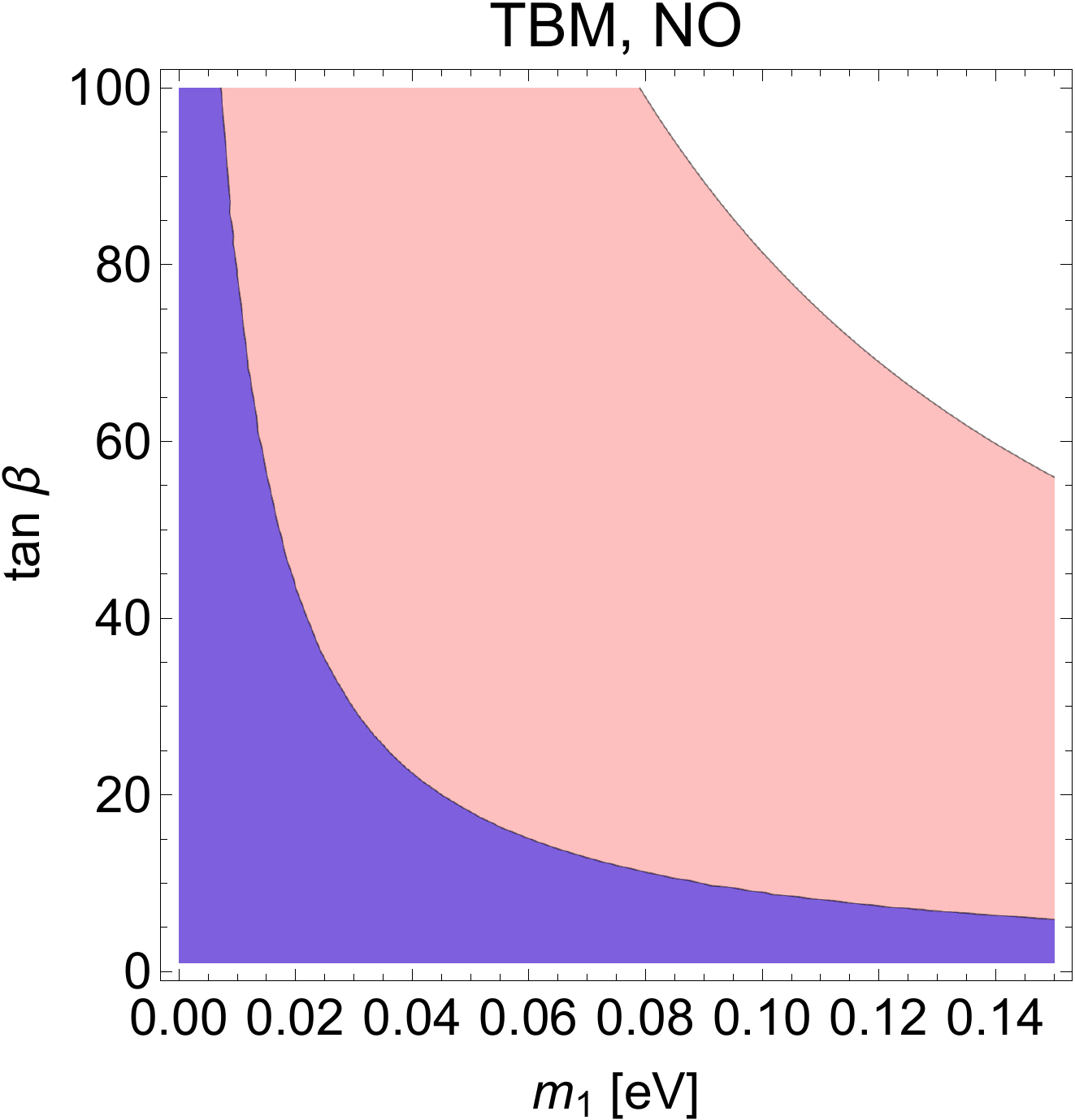}\hspace{0.4cm}
\includegraphics[scale=0.5]{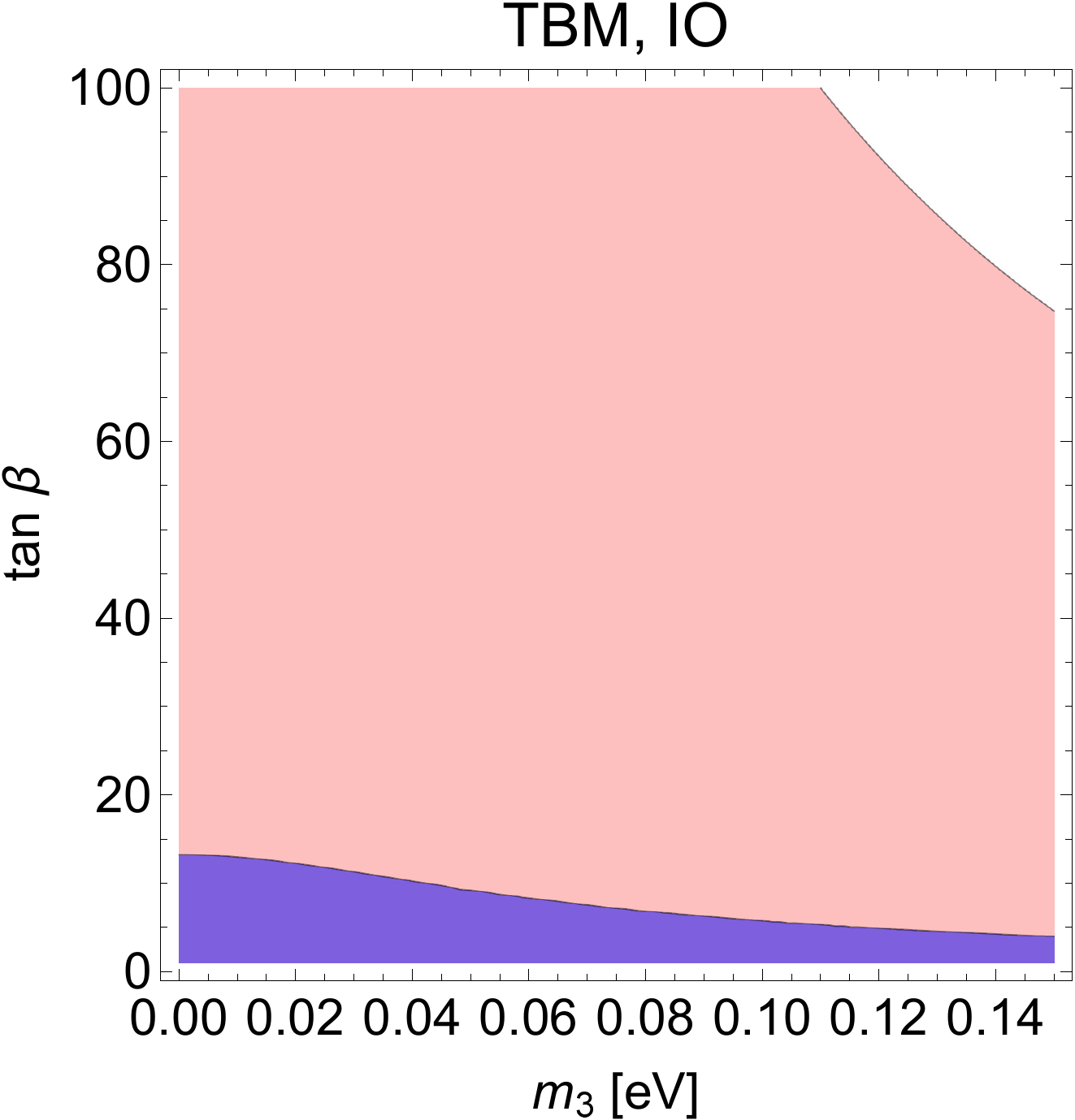} \vspace{4mm}\\ 
\includegraphics[scale=0.5]{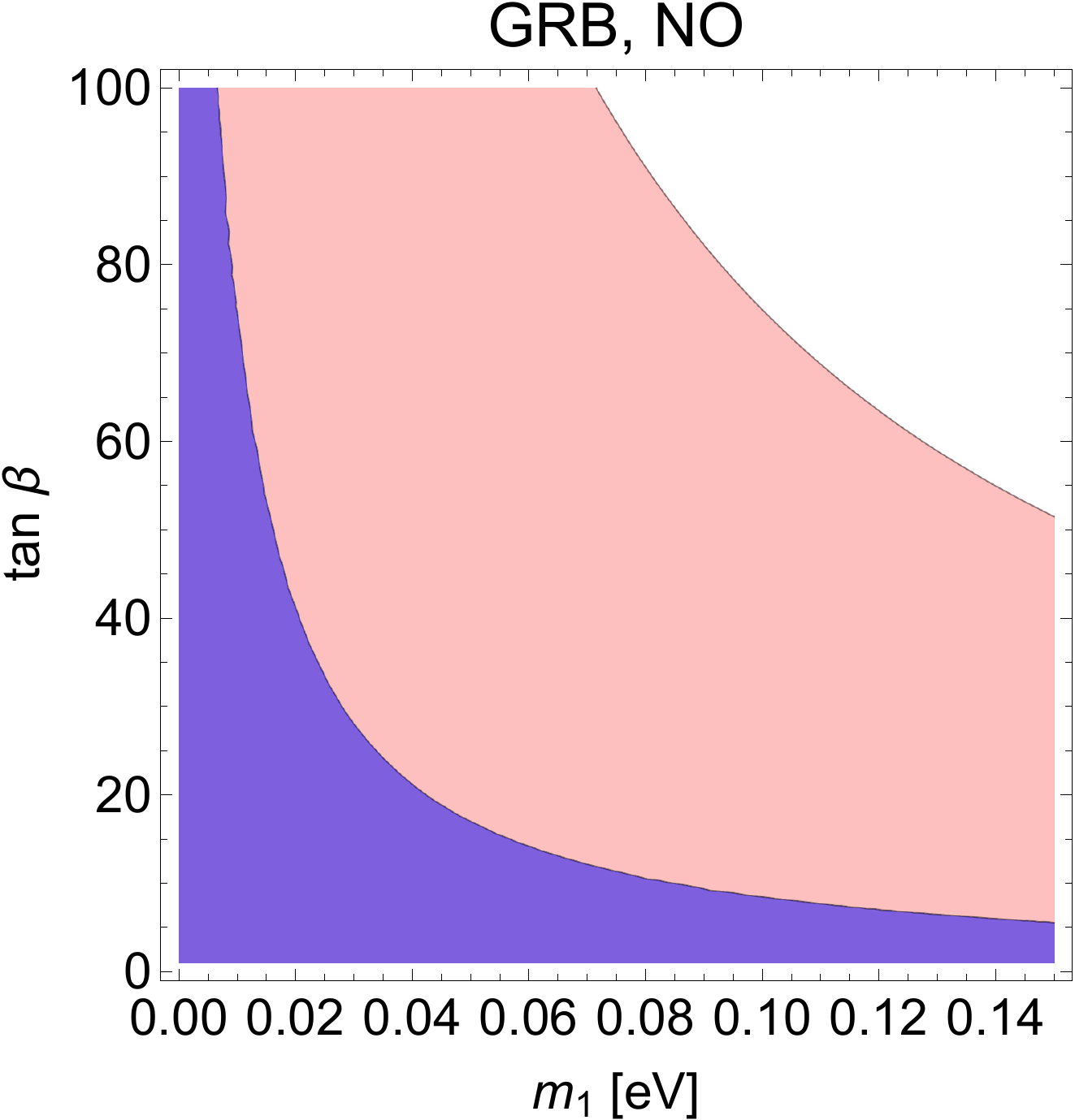}\hspace{0.4cm}
\includegraphics[scale=0.5]{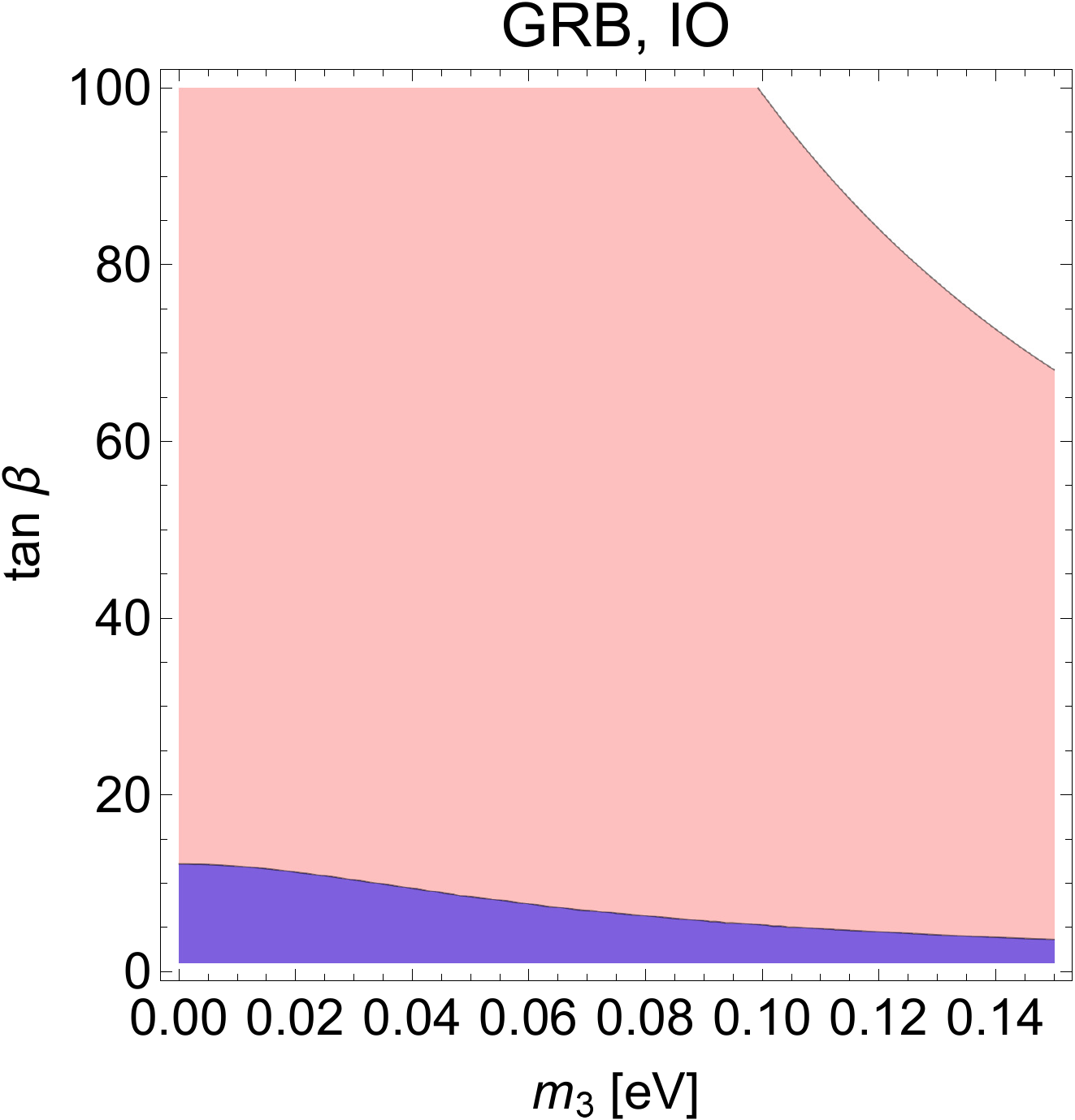} \vspace{4mm}\\ 
\end{center}\vspace{-4mm}
\caption{Allowed regions for $\tan\beta$ and $m_{\text{lightest}}$ 
for the NO and IO spectra in the cases of minimal (blue) and maximal 
(pink) corrections 
for $\cos \delta$ in the TBM mixing scheme (upper plots)
and the GRB mixing scheme (lower plots). 
We used the best fit values for the mixing angles.
The high-energy scale is set to $10^{13}$ GeV.
}
\label{fig:runningTBM}
\end{figure}
%%%%%%%%%%%%%%%%%%%%%%%%%%%%%

The allowed parameter regions in the $m_{\text{lightest}}$-$\tan \beta$ 
plane for the GRA and HG
cases are shown in Fig.~\ref{fig:runningGR}.
For minimal corrections the parameter regions get severely 
constrained, $\tan \beta>20$
is incompatible with $\cos\delta (M_S) \in[-1,1]$ for IO spectrum; 
for NO spectrum it is incompatible with $\cos\delta (M_S) \in[-1,1]$ for 
$m_1 \gtap 0.06$ eV.
This can be understood since
$\cos \delta(M_Z)$ is positive for GRA mixing
and the dominant contribution to $\delta(\cos\delta)$ 
comes from the correction due to $\delta \theta_{12}$, which is negative.
A similar argument holds also for HG mixing. 

For TBM and GRB
$\cos \delta(M_Z)$ is negative and the 
corrections further decrease the value. The plots for 
the allowed parameter regions can be found in Fig.~\ref{fig:runningTBM}.

%%%%%%%%%%%%%%%%%%%%%%%%%%%%
\begin{figure}
\begin{center}
\includegraphics[scale=0.5]{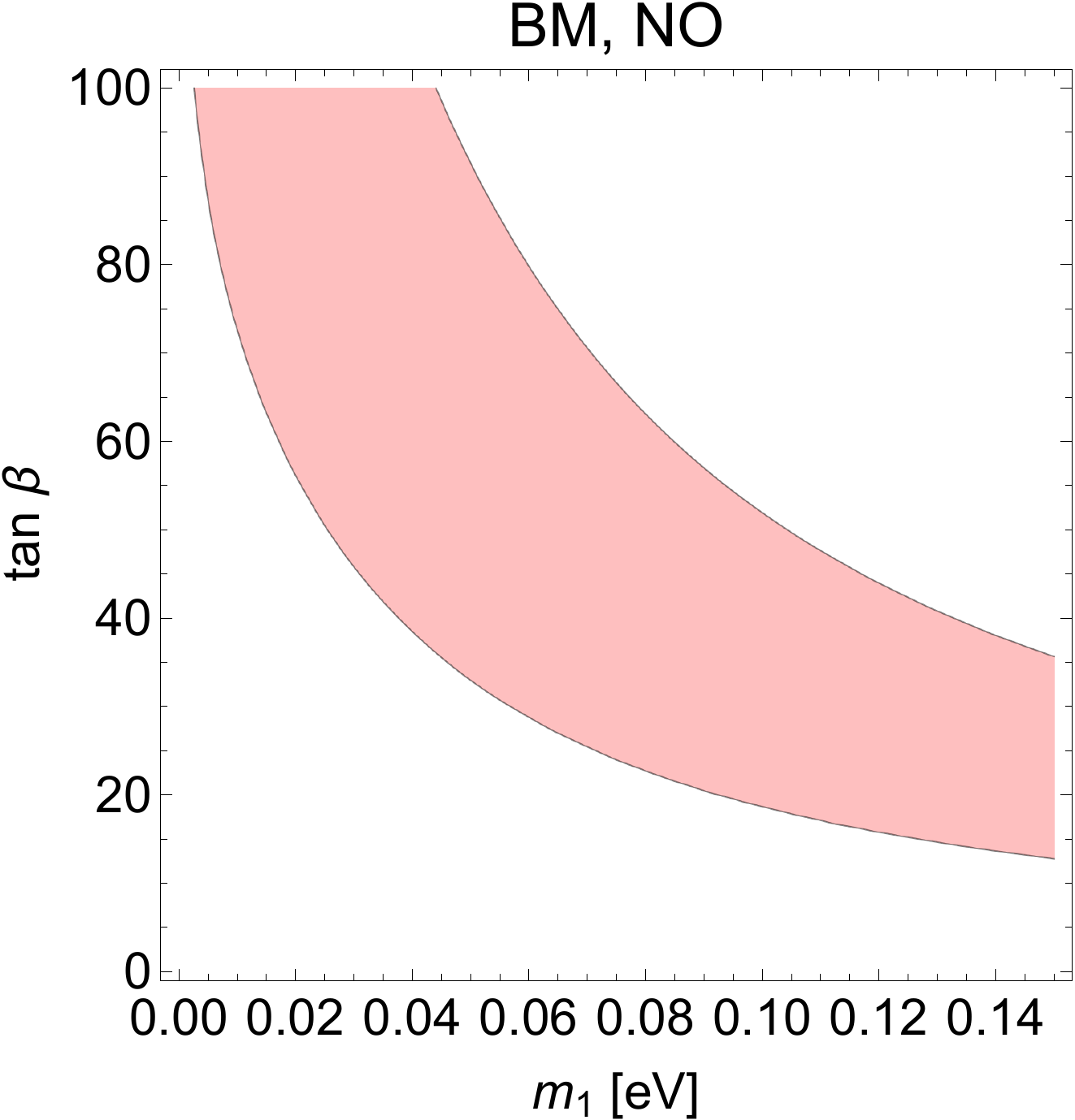}\hspace{0.4cm}
\includegraphics[scale=0.5]{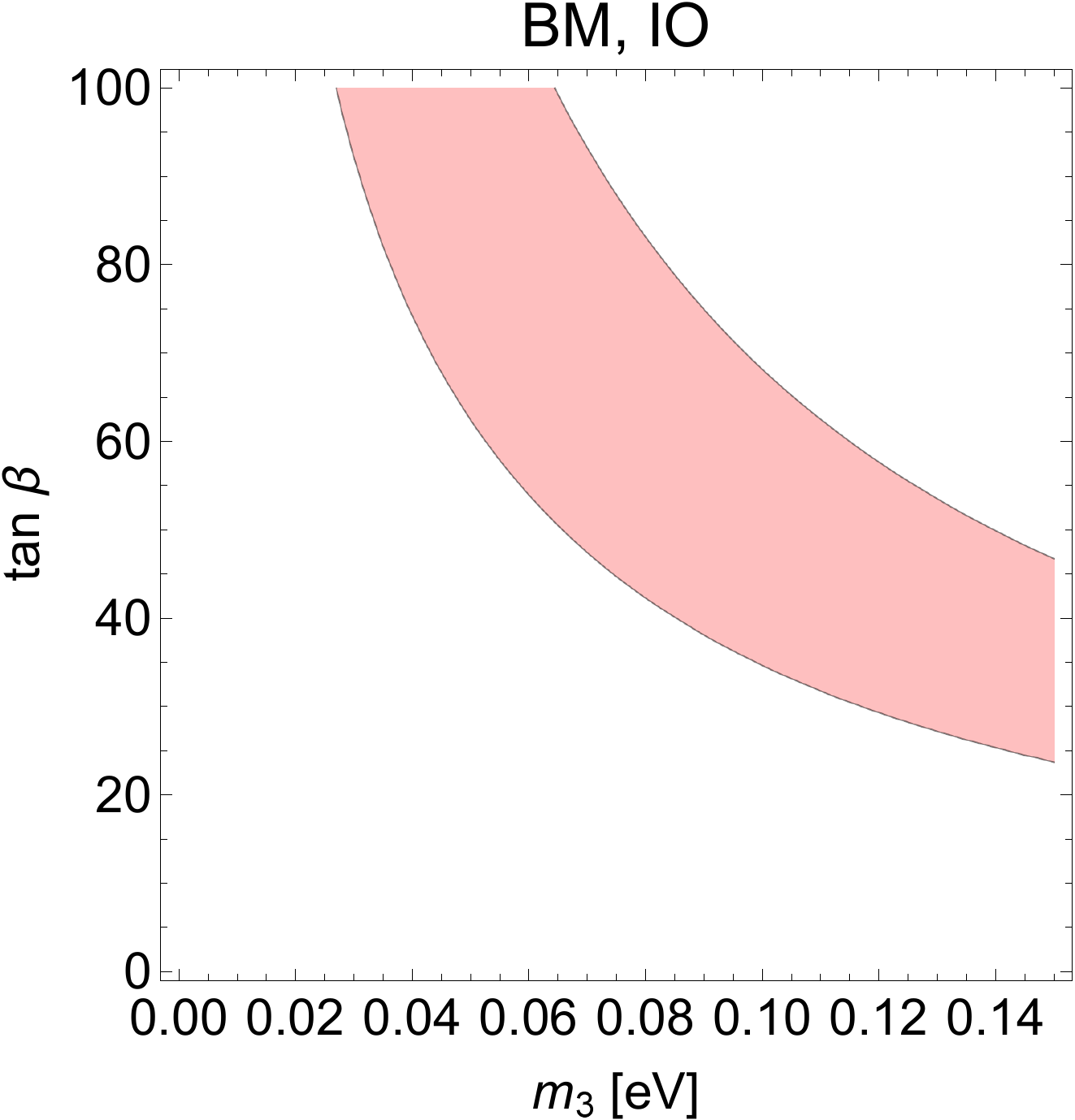}
\end{center}
\caption{Allowed regions for $\tan\beta$ and $m_{\text{lightest}}$ 
for the NO and IO spectra in the case of maximal corrections 
for $\cos \delta$ in the BM mixing scheme. We used the best fit values 
of the mixing angles. For the minimal corrections 
there is no allowed parameter region 
which is compatible with $|\cos \delta|\leq 1$. 
We set the high-energy scale to $10^{13}$ GeV.
}
\label{fig:runningBM}
\end{figure}
%%%%%%%%%%%%%%%%%%%%%%%%%%%%

For BM mixing
$\cos \delta(M_Z) < -1$
for the best fit values of the angles, which is ruled out.
As best approximation for the value of $\delta$ in the
$\beta$-functions we use then $\delta(M_Z) = \pi$.
The dominant contribution to 
the correction is due to $\delta\theta_{13}$, which is positive for 
the  maximal correction. Since $f_{13}$ is also positive 
in BM mixing, the value of  $\cos \delta(M_S)$ increases. 
Hence, the RG corrections have shifted  $\cos \delta(M_S)$ to allowed
values, but for too large values of $\tan\beta$ the corrections overshoot 
$\cos \delta(M_S) = 1$ and the points are excluded.
The allowed banana-shaped parameter regions are 
displayed in Fig.~\ref{fig:runningBM}.

Note that in this example we have only employed the 
constraint on $\delta$ from eq.~\eqref{eq:delta} 
at the high-energy scale. This corresponds to the scheme 
where $\theta_{23}^e\neq 0$. 
To fulfil the sum rule, $\theta_{12}$ is allowed to run  weakly.
In the case of the SM running, the RG effects are already small. 
In the case of the MSSM running,  
they are relatively small if the Majorana phases satisfy the
relation $\alpha_2\approx \alpha_1+\pi$. 
The restrictions on the Majorana phases
in the case of $\theta_{23}^e=0$ from eq.~\eqref{eq:ssth23} are rather weak.

\subsection{Implications of $\boldsymbol{\alpha_2 - \alpha_1 = 0}$ and $\boldsymbol{\pi}$ and Small $\boldsymbol{\tan\beta}$}

In this subsection we show how the specific values 
of the difference of the Majorana phases, namely, 
$\alpha_2 - \alpha_1 = 0$ and $\pi$, contribute to 
the total likelihood profile obtained after the RG 
corrections are taken into account.
These values might seem to be very special
at a first glance but in fact many symmetric matrices
belong at leading order to one of the two cases.
The CP-violating effects of 
the requisite corrections from $\tilde{U}_e$
then might be controlled using, for instance, spontaneous
CP violation with the discrete vacuum alignment method
proposed in \cite{Antusch:2011sx}.

These two cases are also interesting because they correspond
to extremal values of the neutrinoless double beta decay
observable~--~the effective Majorana mass, $|m_{ee}|$, 
in the cases of neutrino mass spectrum with IO or of 
quasi-degenerate type (see, e.g., \cite{PDG2016,bb0nu}). 
For $\alpha_2 - \alpha_1 = 0$, $|m_{ee}|$ is maximal in the two cases,
while if $\alpha_2 - \alpha_1 = \pi$,  $|m_{ee}|$ has a minimal value 
for both types of spectrum. In the case of IO spectrum 
and $m_3 \ll m_{1,2}$, for example, 
$|m_{ee}| \cong \sqrt{\Delta m^2_{23} + m^2_3}\cos^2\theta_{13} 
\cong 4.7\times 10^{-2}$ eV if  $\alpha_2 - \alpha_1 = 0$, 
while for  $\alpha_2 - \alpha_1 = \pi$ we have 
$|m_{ee}| \cong \sqrt{\Delta m^2_{23} + m^2_3}\cos^2\theta_{13}\cos2\theta_{12} 
\gtap 0.014$ eV, where we have used the 
$3\sigma$ allowed ranges of  $\Delta m^2_{23}$, $\sin^2\theta_{13}$ and 
$\sin^2\theta_{12}$ (for the IO spectrum) from Table~\ref{tab:exp_parameters}.

 As can be understood from eq.~\eqref{eq:theta_12_r}, 
in the case of equal Majorana phases, the running of $\theta_{12}$ 
is maximal, while for $\alpha_2 - \alpha_1 = \pi$ it is maximally suppressed. 
Since for the TBM, GRA, GRB and HG symmetry forms 
the correction to the tree-level value of $\cos\delta$ 
is dominated by the running of $\theta_{12}$ 
(see subsection~\ref{subsec:allowedregions}), 
we consider as example the case of TBM and $\theta^e_{23} \neq 0$ 
with the values of $\alpha_2 - \alpha_1$  specified above.
The results we obtain in the  GRA, GRB and HG cases 
are very similar.

It is interesting to see, in particular, what is the quantitative relation 
between the corrections obtained in the setup with 
relatively large $\tan\beta$, e.g., $\tan\beta = 30$, and suppression 
of $\theta_{12}$ running due to $\alpha_2 - \alpha_1 = \pi$, and 
the setup with relatively small $\tan\beta$, e.g., $\tan\beta = 5$ or $10$, 
but enhancement due to $\alpha_2 = \alpha_1$.  

To answer this question, we employ a simplified 
one-step integration procedure (linearised running), in which
the high-energy values of the mixing parameters entering the sum rule
are obtained using one-step integration of 
the exact one-loop beta functions for the mixing parameters 
from \cite{Antusch:2003kp}. 
We set $\theta_{13}$, $\theta_{23}$, 
$\Delta m_{21}^2$, $\Delta m_{31 (23)}^2$ to their 
best fit values and impose 
i)~$\alpha_2 = \alpha_1$, and ii)~$\alpha_2 = \alpha_1 + \pi$. 
For each set of these low-energy values, 
we solve the high-energy sum rule for the low-energy value of $\theta_{12}$. 

 In order to perform a statistical analysis of the low-energy data 
after RG corrections 
we construct the $\chi^2$ function as 
\begin{equation}
\chi^2 (\vec{x}) = \sum_{i=1}^6 \chi^2_i (x_i)\,,
\label{eq:chisquare}
\end{equation}
%%%%%%%%%%%%%%%%%
%
where $\vec{x} = (\sin^2\theta_{12}, \sin^2\theta_{13}, \sin^2\theta_{23}, \delta, 
\Delta m_{21}^2, \Delta m_{31 (23)}^2)$ for the NO (IO) spectrum, and 
$\chi^2_i$ are one-dimensional projections taken 
from \cite{Capozzi:2016rtj}. 
In order to obtain the one-dimensional projection $\chi^2(\delta)$ 
from the constructed 
$\chi^2(\vec{x})$ function we need to minimise the latter 
with respect to all other parameters 
($\sin^2\theta_{ij}$, $\Delta m_{21}^2$ and $\Delta m_{31 (23)}^2$), i.e., 
we need to find a minimum of $\chi^2(\vec{x})$ for a fixed value of $\delta$:
%%%%%%%%%%%%%%%%%%%%%%%%%%%%%%
\begin{equation}
\chi^2 (\delta) = \min\left[\chi^2(\vec{x})|_{\delta = {\rm const}} \right]\,.
\label{eq:chisqcosdelta}
\end{equation}
%%%%%%%%%%%%%%%%%%%%%%%%%%%%%%%
%
The likelihood function $L$, which represents the most probable values 
of $\delta$ in each of the considered cases,  reads
%%%%%%%%%%%%%%%%%%%%
\begin{equation}
L (\delta)= \exp\left(-\frac{\chi^2(\delta)}{2}\right)\,.
\label{eq:likelihood}
\end{equation}
%%%%%%%%%%%%%%%%%%%%%%%%%%%%%
%
We will present the results  in terms of the likelihood functions, 
considering three values for the absolute mass scale, 
$m_\text{lightest}=0.005$, $0.01$ and $0.05$~eV, 
and four values of $\tan\beta = 5$, $10$, $30$ and $50$.

It is worth noting here that, 
as shown in ref.~\cite{Antusch:2003kp} (see eq.~(26) therein), 
for the running of the difference $\alpha_1 - \alpha_2$ 
we have  up to $\mathcal O(\theta_{13})$ terms:  
%%%%%%%%%%%%%%%%%%%%%%%%%%%%%%%%%%%
\begin{equation}
\frac{\text{d}}{\text{d}\, \ln (\mu/\mu_0)} (\alpha_1 - \alpha_2) \propto \sin(\alpha_1 - \alpha_2)\,.
\end{equation}
%%%%%%%%%%%%%%%%%%%%%%%%%%%%%%%%
%
This implies that if the phases are equal 
(different by $\pi$) at some scale
to a good approximation, 
they remain equal (differ by $\pi$) at another scale. 
Thus, the relation imposed by us at the low scale 
holds also at the high scale 
(up to $\mathcal O(\theta_{13})$ corrections).

%%%%%%%%%%%%%%%%%%%%%%%%%%%%%%%%%%%%%%%%
\begin{figure}
\centering
\includegraphics[width=\textwidth]{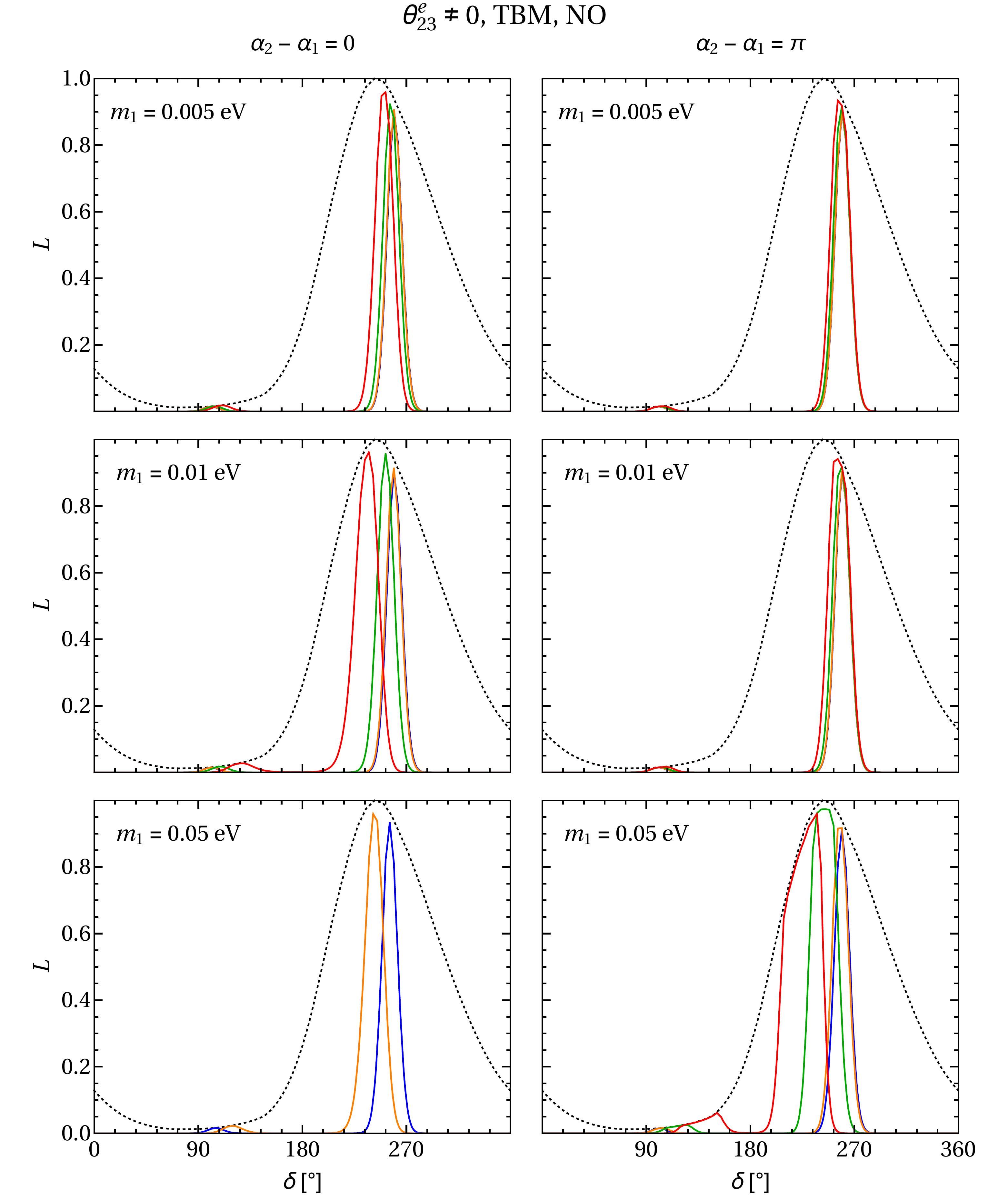}
\caption{Likelihood function vs. $\delta$ 
in the case of non-zero $\theta^e_{23}$ for the TBM symmetry form of 
the matrix $\tilde U_\nu$ and the NO spectrum. 
The dotted black line stands for likelihood extracted from the global analysis
in \cite{Capozzi:2016rtj}.
The blue, orange, green and red lines are for 
the running within MSSM with
$\tan\beta = 5$, $10$, $30$ and $50$, respectively.
The left panels correspond to $\alpha_2 = \alpha_1$,  
while the right panels are for $\alpha_2 = \alpha_1 + \pi$. 
}
\label{fig:likelihoodTBMNOparticphases}
\end{figure}

\begin{figure}
\centering
\includegraphics[width=\textwidth]{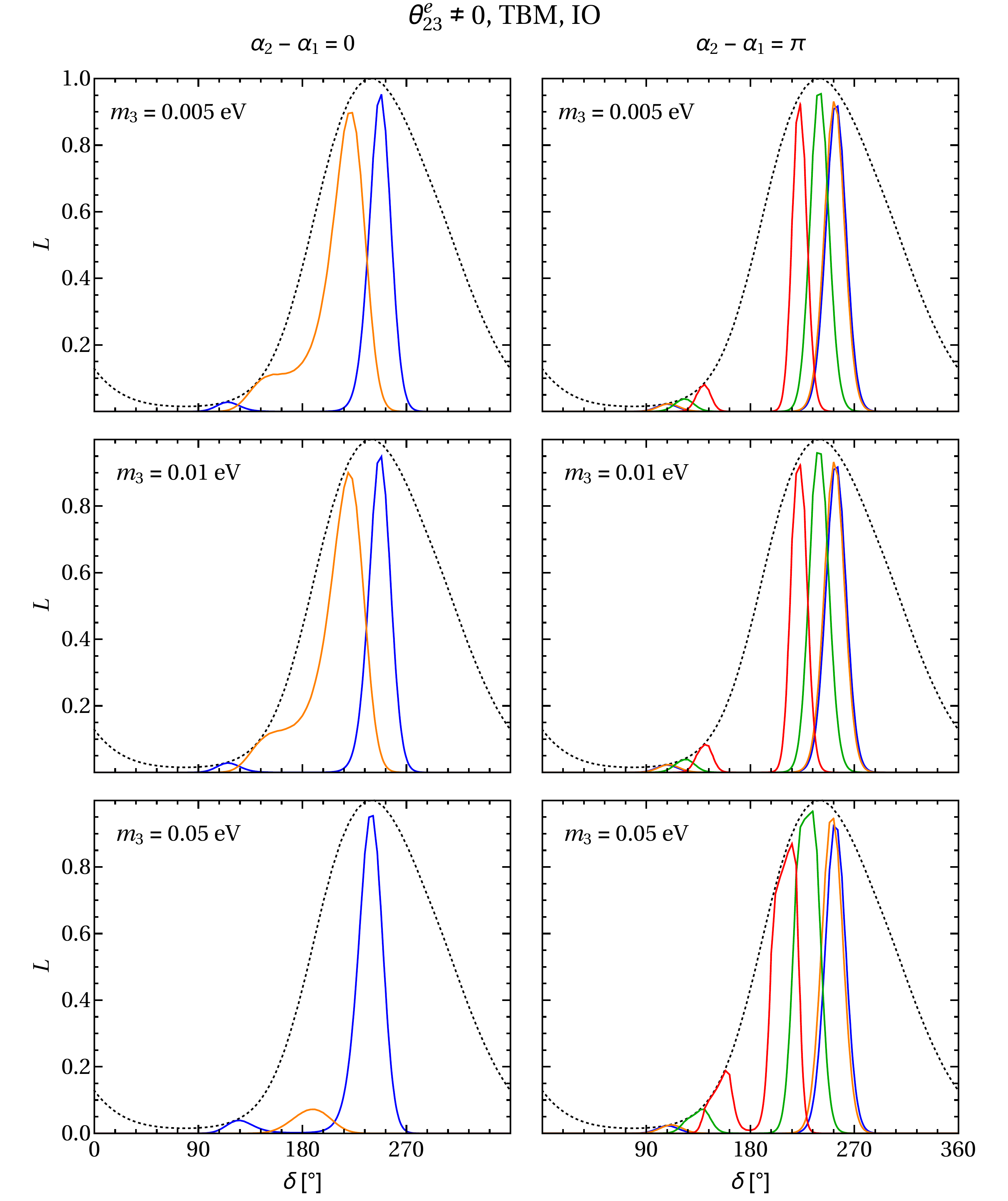}
\caption{Likelihood function vs. $\delta$ 
in the case of non-zero $\theta^e_{23}$ for the TBM symmetry form of 
the matrix $\tilde U_\nu$ and the IO spectrum. 
The dotted black line stands for likelihood extracted from the global analysis
in \cite{Capozzi:2016rtj}.
The blue, orange, green and red lines are for 
the running within MSSM with
$\tan\beta = 5$, $10$, $30$ and $50$, respectively.
The left panels correspond to $\alpha_2 = \alpha_1$, 
while the right panels are for $\alpha_2 = \alpha_1 + \pi$. 
}
\label{fig:likelihoodTBMIOparticphases}
\end{figure}

\begin{figure}
\centering
\includegraphics[width=\textwidth]{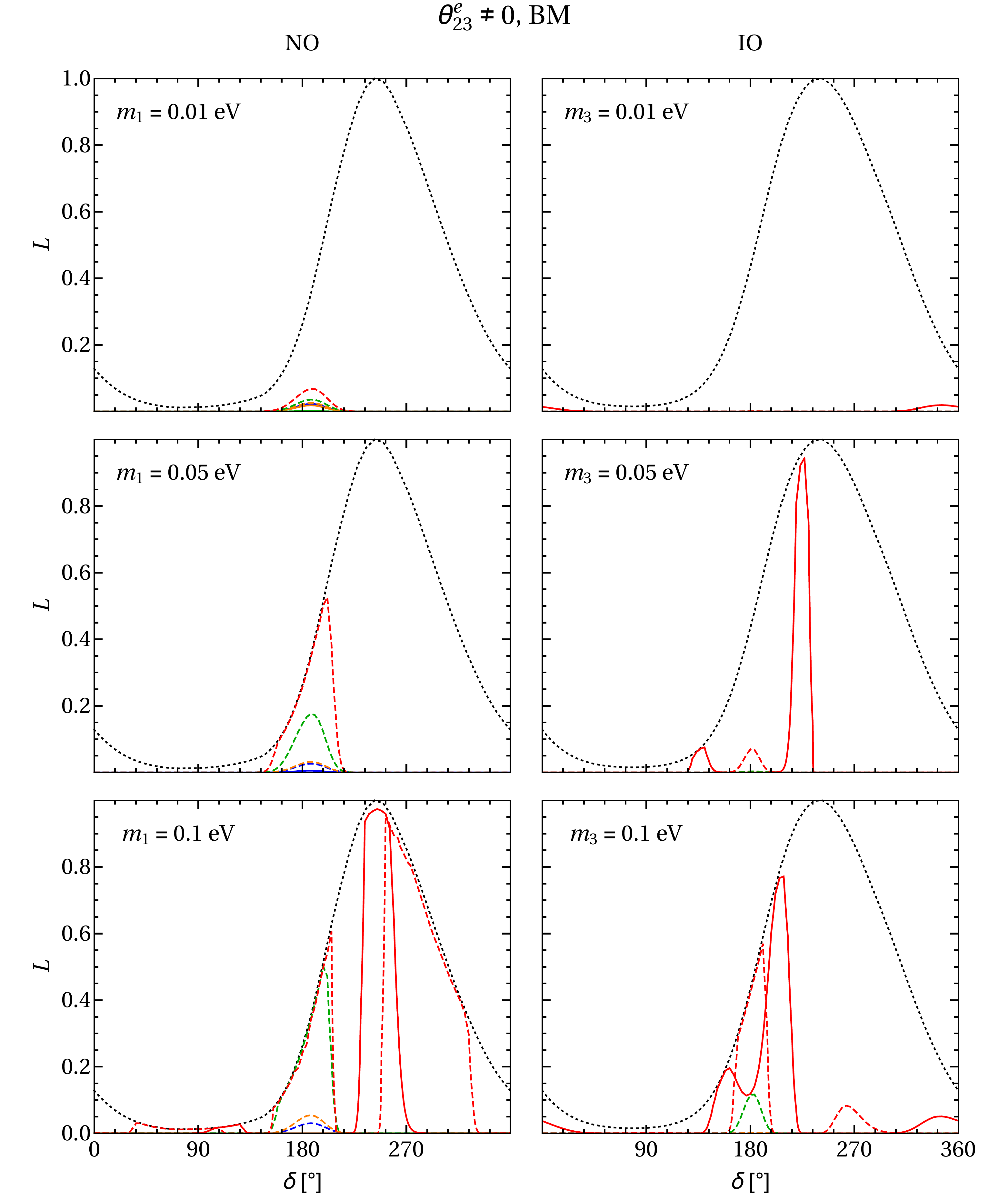}
\caption{Likelihood function vs. $\delta$ 
in the case of non-zero $\theta^e_{23}$ for the BM symmetry form of 
the matrix $\tilde U_\nu$. 
The dotted black line stands for likelihood extracted from the global analysis
in \cite{Capozzi:2016rtj}.
The blue, orange, green and red lines are for 
the running within MSSM with
$\tan\beta = 5$, $10$, $30$ and $50$, respectively.
The solid lines correspond to $\alpha_2 = \alpha_1$. 
The dashed lines correspond to $\alpha_2 = \alpha_1 + \pi$.
Note that the lines for $\tan\beta < 50$ are often
barely visible.
}
\label{fig:likelihoodBMparticphases}
\end{figure}
%%%%%%%%%%%%%%%%%%%%%%%%%%%%%%%%%%%%%%%%

 We present graphically the results obtained 
for the TBM symmetry form 
in Figs.~\ref{fig:likelihoodTBMNOparticphases} 
and~\ref{fig:likelihoodTBMIOparticphases} for the NO and IO 
neutrino mass spectra, respectively.
The dotted black line stands for likelihood extracted from 
the global analysis \cite{Capozzi:2016rtj}.
The blue, orange, green and red lines are for 
$\tan\beta = 5$, $10$, $30$ and $50$, 
respectively.
The left panels in each of the two figures 
correspond to $\alpha_2 = \alpha_1$,  
while the right panels are for $\alpha_2 = \alpha_1 + \pi$. 

Several comments are in order. As expected, the results for 
$\alpha_2 - \alpha_1 = \pi$ and
small $\tan\beta$, $\tan\beta = 5$ and $10$   
(blue and orange lines, respectively), 
are quantitatively very similar to the result 
without running (this is why we do not present the latter 
in the plots)
for all three mass scales considered and both orderings 
due to the suppression of the running of $\theta_{12}$ 
discussed above. However, this is not the case for
the large values of $\tan\beta = 30$ and $50$ 
(green and red lines, respectively)
and the NO spectrum with $m_1 = 0.05$~eV, 
and for all three values of $m_3$ considered in the case of 
the IO spectrum. Clearly, the enhancement due to $\tan\beta$ prevails over
the suppression due to the Majorana phases in these cases.

The next interesting point to note is that for the IO spectrum,
the corrections in the case of $\tan\beta = 5$ and $\alpha_2 = \alpha_1$ 
(blue line) are comparable 
with the corrections for $\tan\beta = 30$ and $\alpha_2 = \alpha_1 + \pi$ 
(green line) for all three mass scales considered. 
A similar observation holds also for the
NO spectrum if  $m_1 = 0.05$~eV: the corrections for  
$\tan\beta = 10$ and  $\alpha_2 = \alpha_1$  (orange line) are similar 
in magnitude to those for 
$\tan\beta = 30$ and $\alpha_2 - \alpha_1 = \pi$ (green line). 

Further, we note also that 
the absence of the green and red lines, corresponding to 
$\tan\beta = 30$ and $50$ and equal Majorana phases, 
in all cases, except for NO with $m_1 = 0.005$~eV and $m_1 = 0.01$~eV,
reflects the fact that the RG corrections  
lead, in particular, to a low-energy 
value of $\theta_{12}$, which is outside of the current $3\sigma$ range. 
For the IO spectrum with $m_3 = 0.05$~eV and $\alpha_2 = \alpha_1$, 
even for $\tan\beta = 10$ (orange line) the RG corrections are quite large, 
such that only a small region of values of $\delta$ around $\pi$ is allowed, 
with the likelihood of these values being suppressed.

For the BM symmetry form the results we obtain 
are quite different. 
In this case we consider  values of 
$m_\text{lightest}=0.01$, $0.05$ and $0.1$~eV, 
and  $\tan\beta = 5$, $10$, $30$ and $50$.
We find that the small values of  $\tan\beta$ 
considered,  $\tan\beta = 5$, 10,
cannot  provide the RG corrections 
which allow one to have $\cos\delta \in [-1,1]$ and low-energy 
values of the mixing angles compatible with the current data 
(except for the small range of values of 
$\delta$ close to $\pi$
allowed without running). 
For the large values of $\tan\beta$ and the NO spectrum,  
we get significant RG corrections compatible with all constraints,  
as can be seen from Fig.~\ref{fig:likelihoodBMparticphases}, 
i) for $\alpha_2 - \alpha_1 = \pi$ (dashed lines), provided 
$m_1\gtap 0.05$ eV, and  
ii) for  $\alpha_2 = \alpha_1$ (solid line) 
if  $m_1 \cong 0.10$ eV and $\tan\beta=50$.
For the IO spectrum and $m_3 \gtap 0.05$~eV, 
the predictions are compatible with the data for  
$\alpha_2 = \alpha_1$ provided  $\tan\beta=50$. 
If  $m_3 = 0.1$~eV, 
$\alpha_2 - \alpha_1 = \pi$  also contributes to the final 
likelihood profile for $\tan\beta=50$, 
although this contribution is less favoured.

As already discussed above, the running of $\theta_{12}$ is 
suppressed if the difference of the Majorana phases is equal 
to $\pi$, otherwise the running of $\theta_{12}$ is always the 
dominant correction to $\cos \delta$. If the running of $\theta_{12}$ 
is minimal, the running of $\theta_{23}$ and $\theta_{13}$ is dominant 
(for a maximal running of $\theta_{13}$ we need additionally 
to have $\delta=\alpha_2$). Then $\delta \theta_{13}$ and  
$\delta \theta_{23}$ are  roughly two orders of magnitude larger 
then  $\delta \theta_{12}$. This implies that the correction to 
$\cos\delta$ in the HG, GRA, GRB and TBM mixing schemes 
is not longer determined by the running of $\theta_{12}$ 
but by the running of  $\theta_{23}$ and $\theta_{13}$. 
For BM mixing the contribution of $\delta \theta_{13}$ is still dominant. 
The sign and size of the correction to $\cos \delta$ depends on 
$\delta$ because the size of $\delta \theta_{13}$ depends on $\delta$ 
and the contributions to $\delta(\cos\delta)$ by the running of  
$\theta_{23}$ and $\theta_{13}$ are approximately equal.

Finally, we would like to note that 
the cases studied in the present subsection 
were analysed rather qualitatively in \cite{Ballett:2014dua}, 
considering only the running of $\theta_{12}$. 
Our analysis goes beyond the discussion in \cite{Ballett:2014dua}, 
since we present explicitly in graphic form 
the impact of the RG effects on the likelihood functions 
(Figs.~\ref{fig:likelihoodTBMNOparticphases}\,--\,\ref{fig:likelihoodBMparticphases}).
In particular, as was discussed above, the results 
depend strongly on the symmetry form considered~--~%
the TBM, GRA, GRB and HG forms on the one hand 
and the BM form on the other~--~%  
and this distinction was not discussed in \cite{Ballett:2014dua}.
Furthermore, in our quantitative results we find a
region of parameter space where their conclusions are not fully
correct.
Although this region seems somewhat tuned,
it is actually motivated, as we mentioned above,
in setups with spontaneous CP violation.
We find that, e.g., in the case of the TBM symmetry form, 
for $m_3 = 0.01$~eV (IO), $\tan\beta = 30$ 
and $\alpha_2 - \alpha_1 = \pi$ 
(green line in the corresponding panel of 
Fig.~\ref{fig:likelihoodTBMIOparticphases}) 
the RG corrections are noticeable,
in contrast to the conclusion 
in \cite{Ballett:2014dua} that the RG corrections can
be neglected for  $\tan\beta \ltap 35$ if the spectrum
is not quasi-degenerate.

\subsection{Notes on the $\boldsymbol{\theta_{23}^e= 0}$ Case}
\label{sec:th23ecase}

Before we turn to the numerical results we want to 
make a few more remarks on the case of $\theta_{23}^e = 0$,  i.e., 
imposing also the sum rule from eq.~\eqref{eq:ssth23} 
at the high scale. This will help to understand the
numerical results in the next section.
In eq.~\eqref{eq:delta12e} we can replace $\theta_{12}(M_S)$ by $\theta_{12}(M_Z)$
plus the small RG correction $\delta \theta_{12}$ in which we expand.
Since $\theta_{13}$ and $\delta\theta_{13}$ are small we can neglect the latter
($\theta_{13}(M_S)\approx \theta_{13}(M_Z)$) and expand the correction in the first
to end up with
\begin{align}
\cos\delta(M_S) \approx \cos \delta(M_Z) +
 \frac{  1 - \cos 2 \theta_{12} \, \cos 2 \theta_{12}^\nu  }{ \theta_{13} \sin^2 2 \theta_{12}  } \delta \theta_{12} ~.
\label{eq:deltaapprox}
\end{align}
In the case of BM mixing $\cos \delta(M_Z)$ is smaller than $-1$ for the
best fit values of the angles and the correction is always negative since the
running of $\theta_{12}$ has a fixed sign. Note, that the value of $\cos \delta(M_Z)$ could
be adjusted by $\theta_{23}^e \neq 0$ to a value larger than $-1$, cf.~eq.~\eqref{eq:delta}.
So, from that estimate we expect the BM mixing scheme 
not to be valid in the case of $\theta_{23}^e = 0$. This is
confirmed in our extensive numerical scan, where we employed the exact sum rules
from eqs.~(\ref{eq:ssth23}, \ref{eq:delta12e}) and the full 1-loop $\beta$-functions
for all parameters but did not find any physically acceptable
points as well. Nevertheless, our
estimate is a bit rough and a numerical scan cannot 
cover the whole parameter space
such that a tiny, highly tuned region of parameter space might 
still be allowed. 

Let us now turn to the other mixing cases. There the absolute 
value of $\cos \delta(M_Z)$ in our estimate eq.~\eqref{eq:deltaapprox} is
always smaller than one.
For TBM and GRB it is still negative, but
for TBM mixing, for instance, we get 
\begin{align}
 \cos \delta(M_Z) \approx  -0.21 \;,
\end{align}
which allows for a sizeable correction of $\theta_{12}$ up to $-6.5^\circ$,
so that these two scenarios are not disfavoured by our estimate.
For GRA and HG mixing the first term is even positive such that we can
account for even more sizeable RG corrections in these cases.

\section{Numerical Results}
\label{sec:numerical_results}

In the present section we will first describe our numerical 
approach before we show the results we obtain 
for the $\delta$ likelihood functions in the TBM, GRA, GRB, BM and HG mixing 
schemes in the cases of  $\theta_{23}^e \neq 0$ and 
$\theta_{23}^e = 0$.

\subsection{Numerical Approach}
\label{subsec:numapproach}

To obtain the low-energy predictions for $\delta$ from the high-scale
mixing sum rule,  eq.~\eqref{eq:delta} in the case 
of $\theta^e_{23} \neq 0$ 
(eq.~\eqref{eq:delta12e} in the case of $\theta^e_{23} = 0$),
we employ the running of the parameters using the 
\texttt{REAP} package~\cite{Antusch:2005gp}. 
For the running we set the low-energy scale to be $M_Z$ 
and the high-energy scale to be equal
to the seesaw scale $M_S\approx 10^{13}$~GeV. Since the dependence on 
the scales is only logarithmic a mild change of the high-energy 
or low-energy scale would not change our results significantly.

In our scans we  present the results for the SM and MSSM extended minimally 
by the Weinberg operator. We have fixed the scale where we switch from 
the SM to MSSM RGEs to 1~TeV. 
Again the dependence on the scale is only logarithmic and hence weak. 
The exact supersymmetric (SUSY) particle spectrum plays only a minor role since we have
neglected the SUSY threshold corrections~\cite{SUSYthresholds}.

In the MSSM we consider as benchmarks $\tan \beta=30$ and $\tan \beta=50$.  
In the SM the running is relatively small and hence 
the results are very similar to the results without running. 
In fact the SM results look like
the results obtained in \cite{Girardi:2014faa} apart from 
relatively small changes due to the
different global fit results \cite{Capozzi:2013csa} 
used therein. 
For a given mass scale and a given model 
(SM or MSSM with a given $\tan \beta$), we 
employ the mixing sum rules at the high scale
to determine $\delta$ (and $\theta_{23}$ for
$\theta_{23}^e = 0$) at the low scale depending on the
other parameters.
For a given mass scale and a given model 
(SM or MSSM with a given $\tan \beta$), we determine the low-scale
parameters (the angles, mass squared differences 
and the Majorana phases) such that the  
 mixing sum rule eq.~\eqref{eq:delta} (and eq.~\eqref{eq:delta12e} for 
$\theta_{23}^e = 0$) at the high scale
is fulfilled and their likelihood function is maximal.
We choose a ``small'' neutrino mass scale, $m_\text{lightest}=0.01$~eV, 
a ``medium''  mass scale, $m_\text{lightest}=0.05$~eV, and a 
``large''  mass scale, $m_\text{lightest}=0.1$~eV. 
The ``large'' neutrino  mass scale 
is still compatible with the cosmological bound on the sum of 
the neutrino masses \cite{Planck:2015xua}
%%%%%%%%%%%%%%%%%%%%%%%%%
\begin{equation}
\sum m_{\nu} < 0.49\text{ eV.}
\label{eq:kosmoschranke}
\end{equation} 
%%%%%%%%%%%%%%%%%%%%%%%%%
%
Note that for very small neutrino mass scales,
$m_\text{lightest} \ll 0.01$~eV 
and sufficiently small $\tan \beta$, the RG effects are
negligibly small even in the MSSM.
We present the results for different cases considered 
in the present study in terms of the likelihood functions 
defined in eq.~\eqref{eq:likelihood}.

\subsection{Results for Different Mixing Schemes 
in the Case of Non-zero $\boldsymbol{\theta^e_{23}}$}

%%%%%%%%%%%%%%%%%%%%%%%%%%%%%%%%%%%%%
\begin{figure}
\centering
\includegraphics[width=\textwidth]{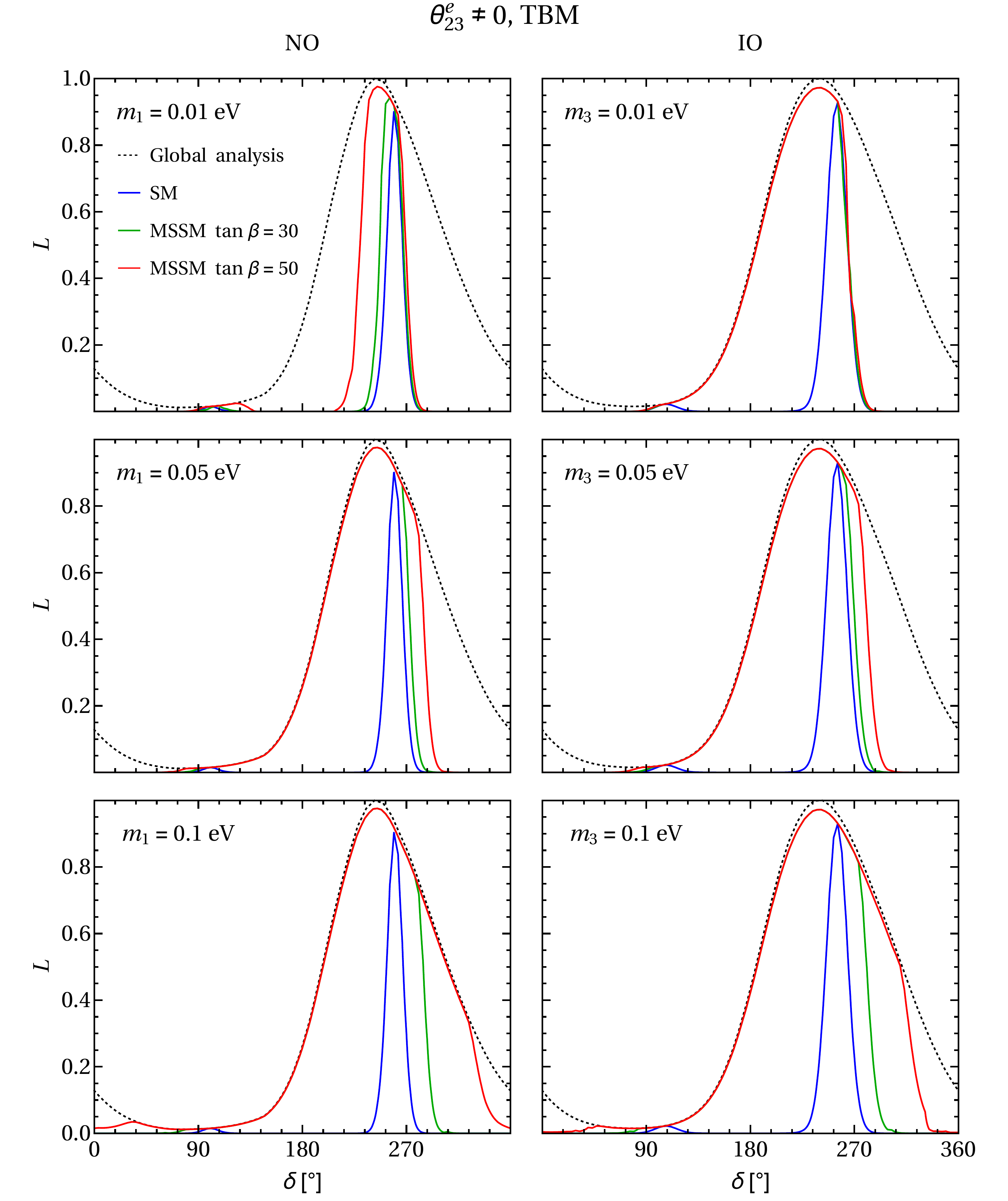}
\caption{Likelihood function vs. $\delta$ 
in the case of non-zero $\theta^e_{23}$ for the TBM symmetry form of 
the matrix $\tilde U_\nu$ in all the setups considered. 
The dotted line stands for likelihood extracted 
from the global analysis in \cite{Capozzi:2016rtj}.
The blue line is for the SM running, while 
the green and red lines are for the running within MSSM with 
$\tan\beta = 30$ and $\tan\beta = 50$, respectively.
}
\label{fig:likelihoodTBM}
\end{figure}

\begin{figure}
\centering
\includegraphics[width=\textwidth]{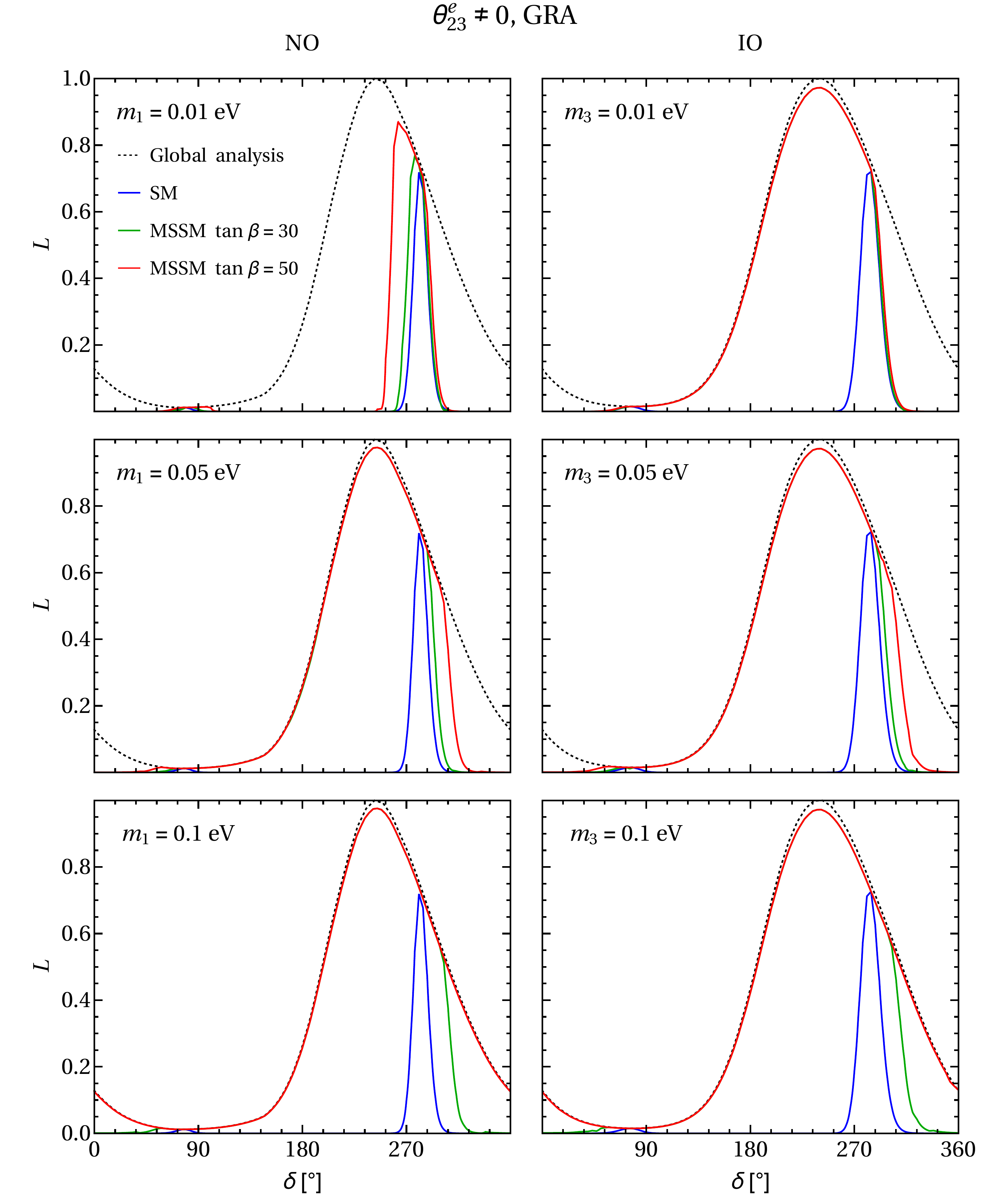}
\caption{Likelihood function vs. $\delta$ 
in the case of non-zero $\theta^e_{23}$ for the GRA symmetry form of 
the matrix $\tilde U_\nu$ in all the setups considered. 
The dotted line stands for likelihood extracted from the global analysis
in \cite{Capozzi:2016rtj}.
The blue line is for the SM running;
the green and red lines are for the running within MSSM with 
$\tan\beta = 30$ and $\tan\beta = 50$, respectively.}
\label{fig:likelihoodGRA}
\end{figure}

\begin{figure}
\centering
\includegraphics[width=\textwidth]{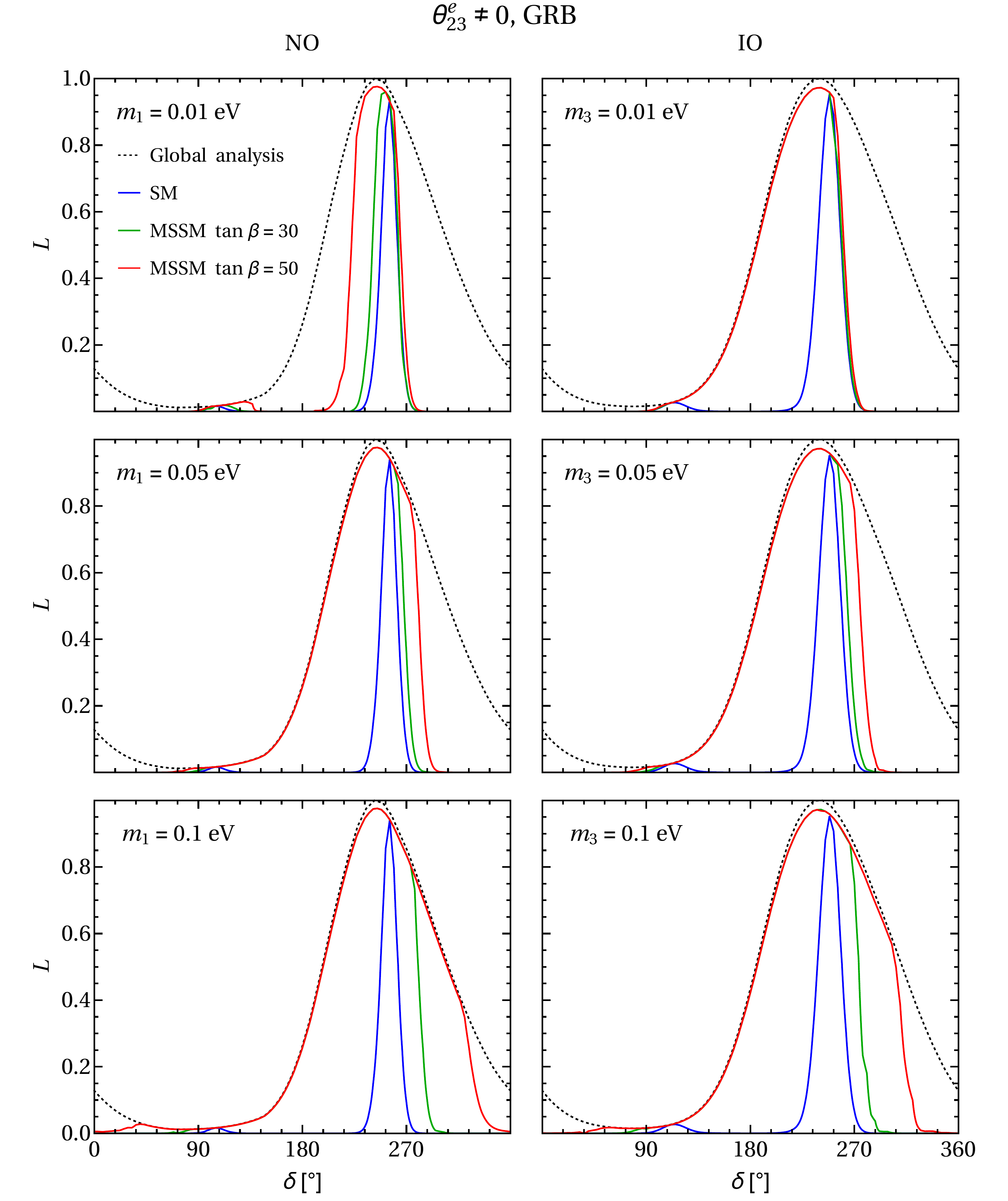}
\caption{Likelihood function vs. $\delta$ 
in the case of non-zero $\theta^e_{23}$ for the GRB symmetry form of 
the matrix $\tilde U_\nu$ in all the setups considered. 
The dotted line stands for likelihood extracted from the global analysis
in \cite{Capozzi:2016rtj}.
The blue line is for the SM running, while 
the green and red lines are for the running within MSSM with 
$\tan\beta = 30$ and $\tan\beta = 50$, respectively.}
\label{fig:likelihoodGRB}
\end{figure}

\begin{figure}
\centering
\includegraphics[width=\textwidth]{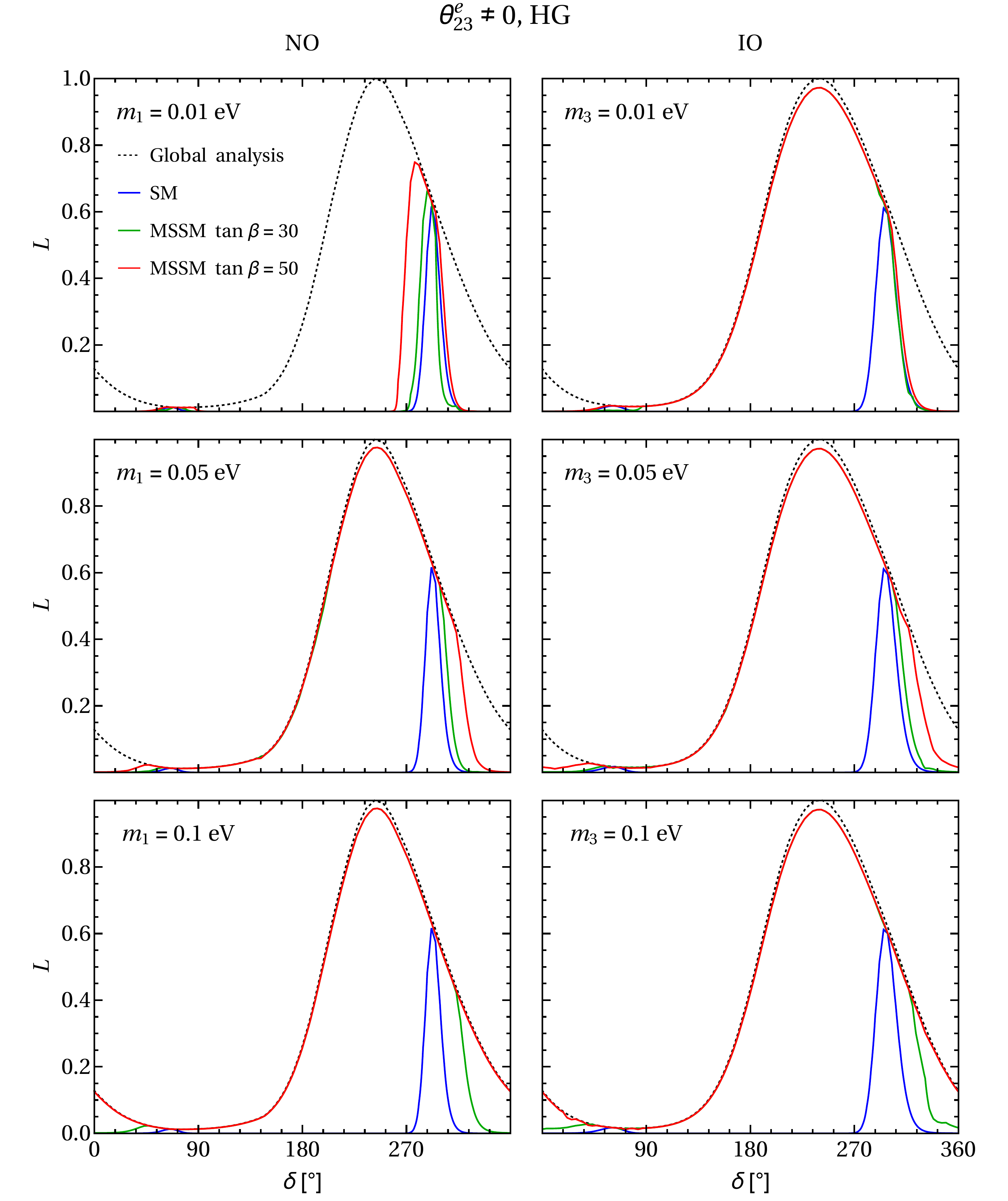}
\caption{Likelihood function vs. $\delta$ 
in the case of non-zero $\theta^e_{23}$ for the HG symmetry form of 
the matrix $\tilde U_\nu$ in all the setups considered. 
The dotted line stands for likelihood extracted from the global analysis
in \cite{Capozzi:2016rtj}.
The blue line is for the SM running;
the green and red lines are for the running within MSSM with 
$\tan\beta = 30$ and $\tan\beta = 50$, respectively.}
\label{fig:likelihoodHG}
\end{figure}

\begin{figure}
\centering
\includegraphics[width=\textwidth]{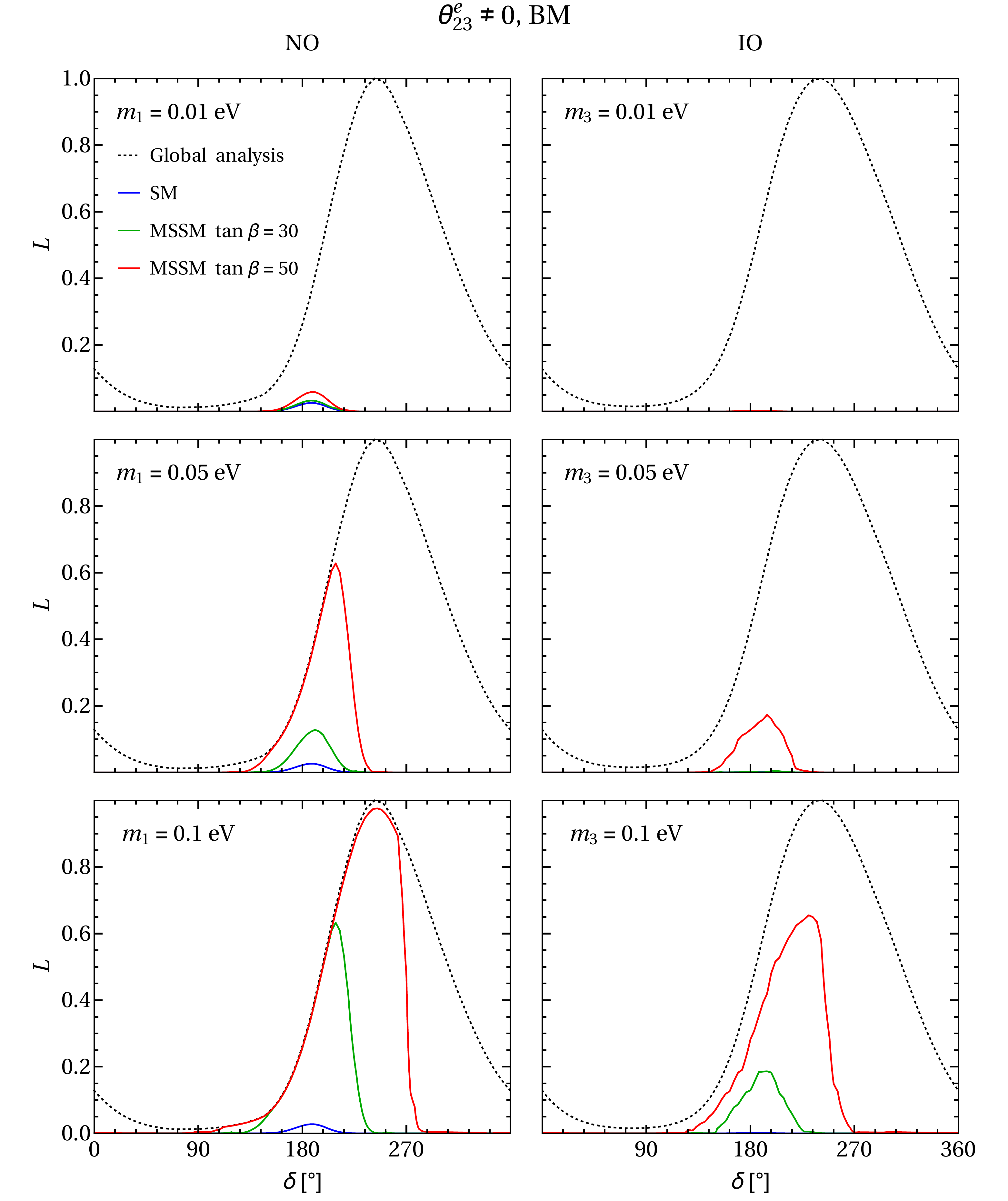}
\caption{Likelihood function vs. $\delta$ 
in the case of non-zero $\theta^e_{23}$ for the BM symmetry form of 
the matrix $\tilde U_\nu$ in all the setups considered. 
The dotted line stands for likelihood extracted from the global analysis
in \cite{Capozzi:2016rtj}.
The blue line is for the SM running, while 
the green and red lines are for the running within MSSM with 
$\tan\beta = 30$ and $\tan\beta = 50$, respectively.}
\label{fig:likelihoodBM}
\end{figure}
%%%%%%%%%%%%%%%%%%%%%%%%%%%%%%%%%
%

We begin our discussion of the numerical results with 
the case of non-zero $\theta^e_{23}$.
In Figs.~\ref{fig:likelihoodTBM}\,--\,\ref{fig:likelihoodHG} we 
show the likelihood functions 
versus $\delta$ for the TBM, GRA, GRB and HG symmetry forms 
of the matrix $\tilde U_\nu$ in all setups. 
The blue line in these figures represents the SM running result, 
the green and red lines are for the MSSM running with 
$\tan\beta = 30$ and $\tan\beta = 50$, respectively.
The SM line practically coincides with the 
line corresponding to the result without running, as expected. 
For this reason we do not show the latter in the plots.
The dotted black line stands for the likelihood extracted from 
the global analysis \cite{Capozzi:2016rtj} which corresponds to the
likelihood for $\delta$ without imposing any sum rule. 
We note that the whole procedure is numerically very demanding and hence
there are some tiny wiggles in the likelihoods which do not
have any physical meaning.
Note also that the mixing sum rule has two solutions but the
solution $\delta\approx 90^\circ$ has a small likelihood and is therefore
barely visible in the plots.

As we have already indicated, the SM results 
are very similar to the  results obtained in 
\cite{Girardi:2014faa} without running.
This implies that, as was concluded in 
\cite{Girardi:2014faa} (see also \cite{Petcov:2014laa}),
 using the data on neutrino mixing angles 
and a sufficiently precise measurement of $\cos\delta$ 
it will be possible to distinguish between 
the three groups of schemes:
the TBM and GRB group, the GRA and HG group, 
and the BM scheme.
Distinguishing between the GRA and HG schemes is 
experimentally very demanding, but not impossible, 
while distinguishing between the TBM and GRB
seems practically extremely difficult (if not impossible) 
to achieve 
(see \cite{Girardi:2014faa,Girardi:2015zva} for further details).

In the MSSM, the results depend on the value of the 
lightest neutrino mass, 
the type of spectrum~--~NO or IO~--~the neutrino masses 
obey, on the value of $\tan\beta$ as well as 
on the uncertainties in the measured values 
of the neutrino oscillation parameters.
As expected, for increasing $\tan\beta$ and increasing 
absolute neutrino mass scale, the difference with the predictions 
without running increases.
The allowed regions for $\delta$ 
start to broaden and, e.g.,  
for the largest value of $\tan\beta = 50$
and $m_1 = 0.05$ eV and 0.10 eV 
($m_3 = 0.01$ eV, 0.05 eV and 0.10 eV) 
in the case of NO (IO) spectrum,
the likelihood profile in the cases of the 
TBM, GRA, GRB and HG mixing schemes 
practically coincides with 
the likelihood 
for $\delta$ obtained without 
imposing the sum rule constraint, 
the difference between the two profiles 
being noticeable only for values of 
$\delta$ lying approximately in the interval 
$\delta \sim (270^\circ - 360^\circ)$.
As already discussed in the previous section, 
the running of $\cos\delta$ in the TBM, GRA, GRB and HG mixing schemes
is mainly influenced by the 
running of $\theta_{12}$ which has a fixed negative sign and hence
has a tendency to shift $\delta$ to values smaller than $270^\circ$. 
For NO spectrum, $m_1 \leq 0.01$ eV
and $\tan\beta = 30$, a measured value of 
$\delta \ltap 260^\circ$ 
would favour the TBM and GRB schemes. 
For $m_1 = 0.05$ eV (or $m_1 = 0.01$ eV)
and the same value of 
$\tan\beta = 30$, a measurement of 
$\delta \gtap 290^\circ$ 
would make the GRA and HG schemes 
more probable.
For $\tan\beta = 50$, $m_1 = 0.05$ eV (or $m_1 = 0.10$ eV),
and given the current uncertainties 
in the measured values of the 
neutrino oscillation parameters,
the  TBM, GRA, GRB and HG  schemes
lead to very similar predictions for 
$\delta$.

For the IO spectrum the RG effects are larger and 
therefore the broadening happens in the four schemes
under discussion~--~TBM, GRA, GRB and HG~--~already
for the ``small'' neutrino mass scale.
Since the likelihood profiles are so broad and nearly identical even for 
the ``small'' and ``medium'' mass scales, 
except for certain differences in the interval 
$\delta \cong (270^\circ - 360^\circ)$, and
given the current uncertainties in the measured values of 
the neutrino oscillation parameters, 
it will be difficult in the MSSM with $\tan\beta \gtap 30$ 
to distinguish between any of the four schemes considered 
using only a determination of $\delta$.

For the BM mixing scheme the results are very different. 
This scheme is strongly disfavoured for the currently 
allowed ranges of the mixing parameters without 
considering RG effects. 
Therefore, the maximal value of the likelihood  
in the SM running case is relatively small.
In the MSSM the running  increases the value of 
$\cos\delta$ to physical values,
as explained in the previous section. 
In addition both 
the maximal value of the likelihood function increases 
and the position of the likelihood maximum shifts
from $\delta \cong 180^\circ$ towards $\delta = 270^\circ$ 
(see Fig.~\ref{fig:likelihoodBM}).
Again the likelihood profile broadens with increasing 
of the absolute neutrino mass scale and
$\tan\beta$ and
at $\delta \ltap 270^\circ$ for NO spectrum  
tends to  approach the likelihood function 
for $\delta$ obtained without imposing the sum rule.
In the case of IO spectrum, the BM scheme is strongly 
disfavoured for $m_3 \ltap 0.05$~eV even 
for $\tan\beta = 50$.

\subsection{Results for Different Mixing Schemes 
in the Case of Zero $\boldsymbol{\theta^e_{23}}$}

%%%%%%%%%%%%%%%%%%%%%%%%%%%%%%%%%%%%%%%%
\begin{figure}
\centering
\includegraphics[width=\textwidth]{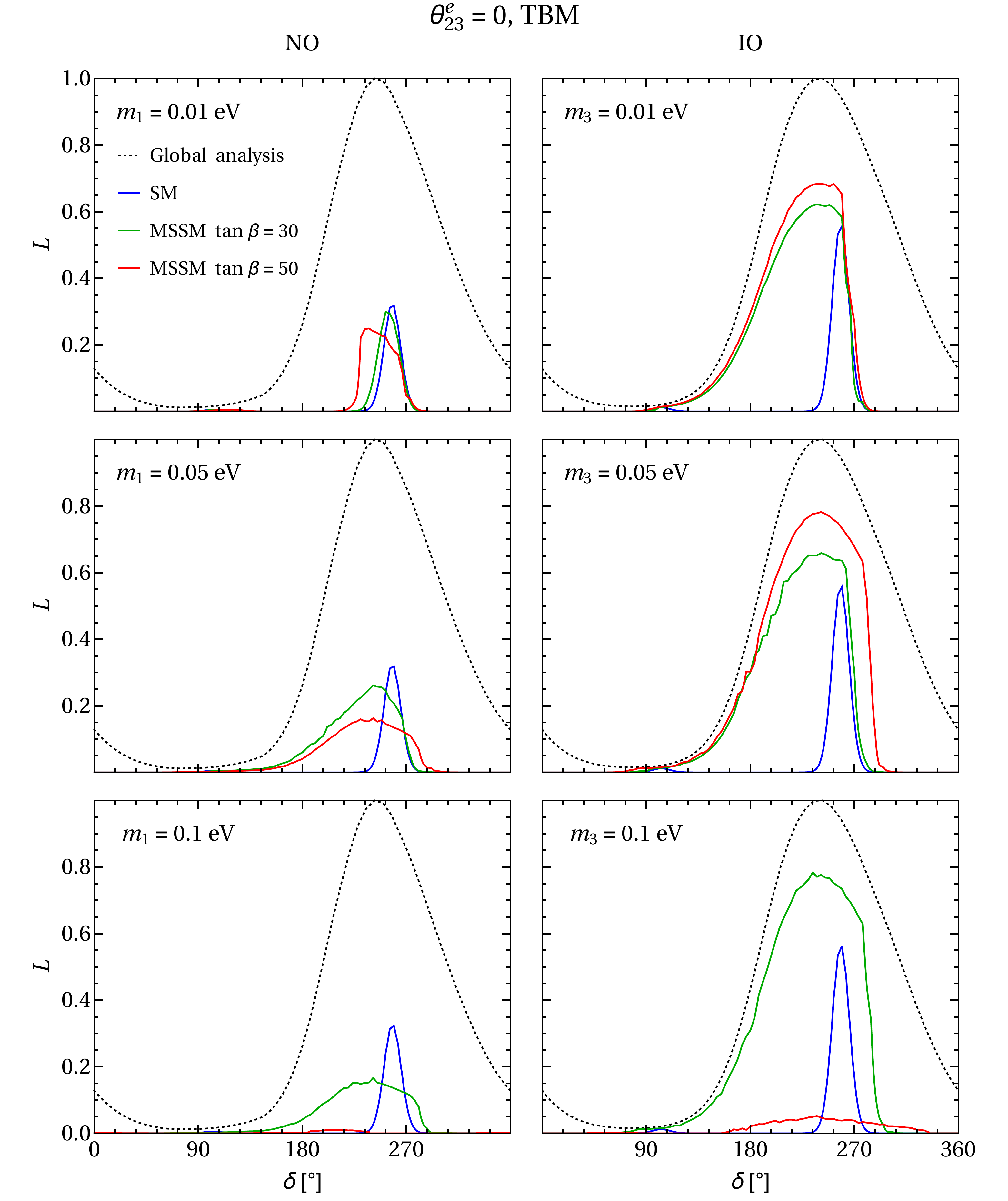}
\caption{Likelihood function vs. $\delta$ in the case of zero 
$\theta^e_{23}$ for the TBM symmetry form of 
the matrix $\tilde U_\nu$ in all the setups considered. 
The dotted line stands for likelihood extracted from the global analysis
in \cite{Capozzi:2016rtj}.
The blue line is for the SM running. 
Finally, the green and red lines are for the running within MSSM with 
$\tan\beta = 30$ and $\tan\beta = 50$, respectively.}
\label{fig:likelihoodTBM2}
\end{figure}

\begin{figure}
\centering
\includegraphics[width=\textwidth]{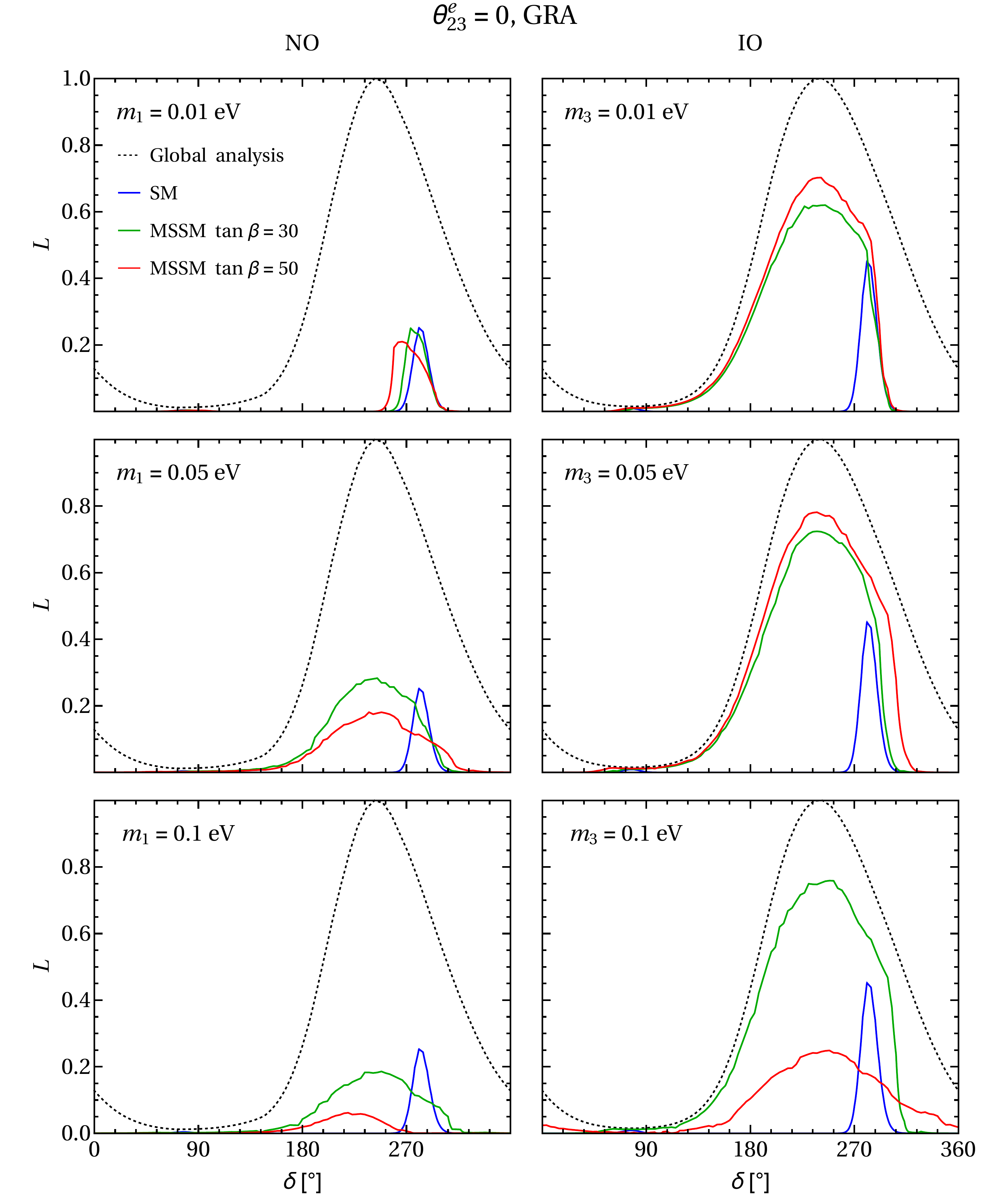}
\caption{Likelihood function vs. $\delta$ in the case of zero 
$\theta^e_{23}$ for the GRA symmetry form of 
the matrix $\tilde U_\nu$ in all the setups considered. 
The dotted line stands for likelihood extracted from the global analysis
in \cite{Capozzi:2016rtj}.
The blue line is for the SM running. 
Finally, the green and red lines are for the running within MSSM with 
$\tan\beta = 30$ and $\tan\beta = 50$, respectively.}
\label{fig:likelihoodGRA2}
\end{figure}

\begin{figure}
\centering
\includegraphics[width=\textwidth]{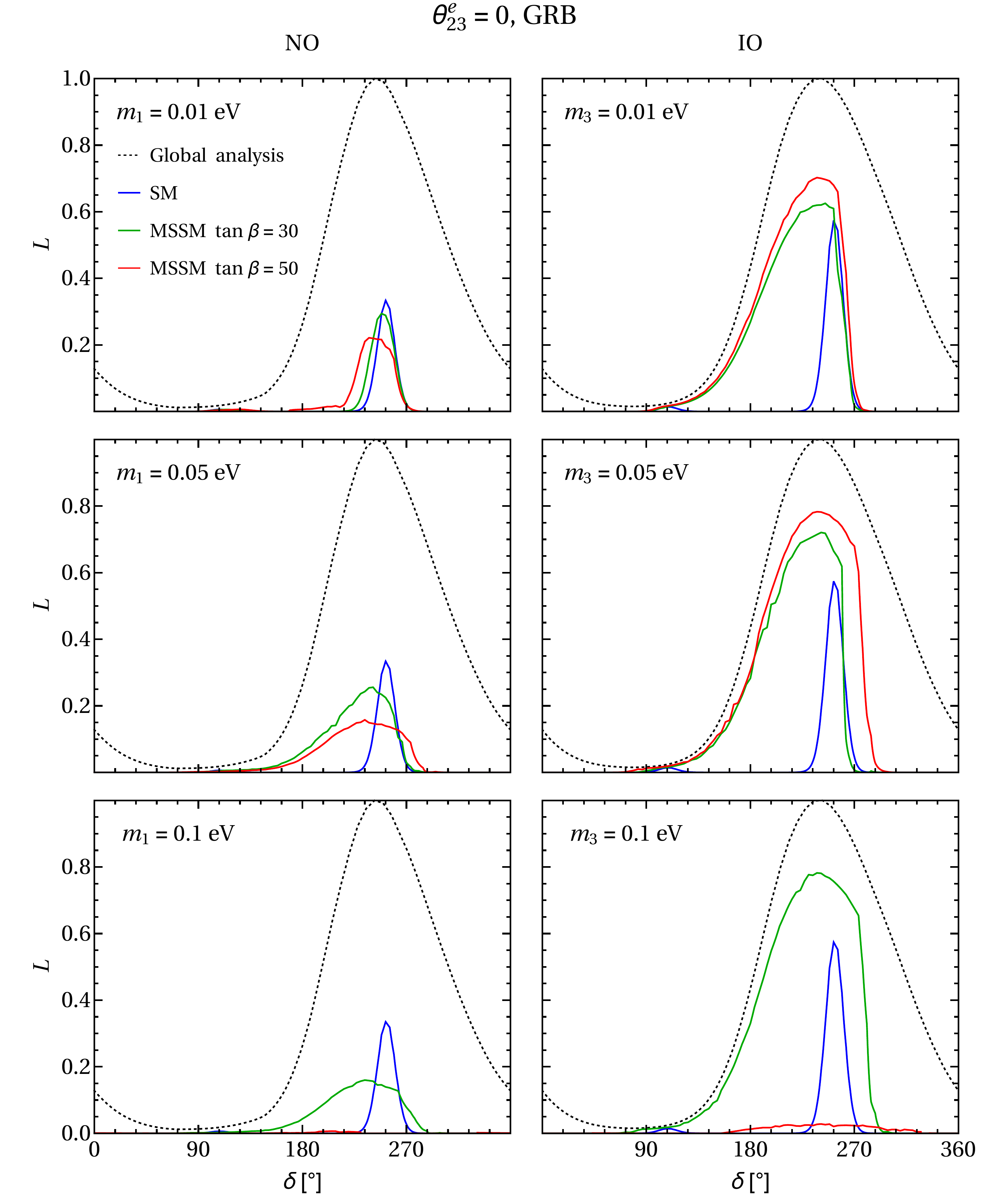}
\caption{Likelihood function vs. $\delta$ in the case of zero 
$\theta^e_{23}$ for the GRB symmetry form of 
the matrix $\tilde U_\nu$ in all the setups considered. 
The dotted line stands for likelihood extracted from the global analysis
in \cite{Capozzi:2016rtj}.
The blue line is for the SM running. 
Finally, the green and red lines are for the running within MSSM with 
$\tan\beta = 30$ and $\tan\beta = 50$, respectively.}
\label{fig:likelihoodGRB2}
\end{figure}

\begin{figure}
\centering
\includegraphics[width=\textwidth]{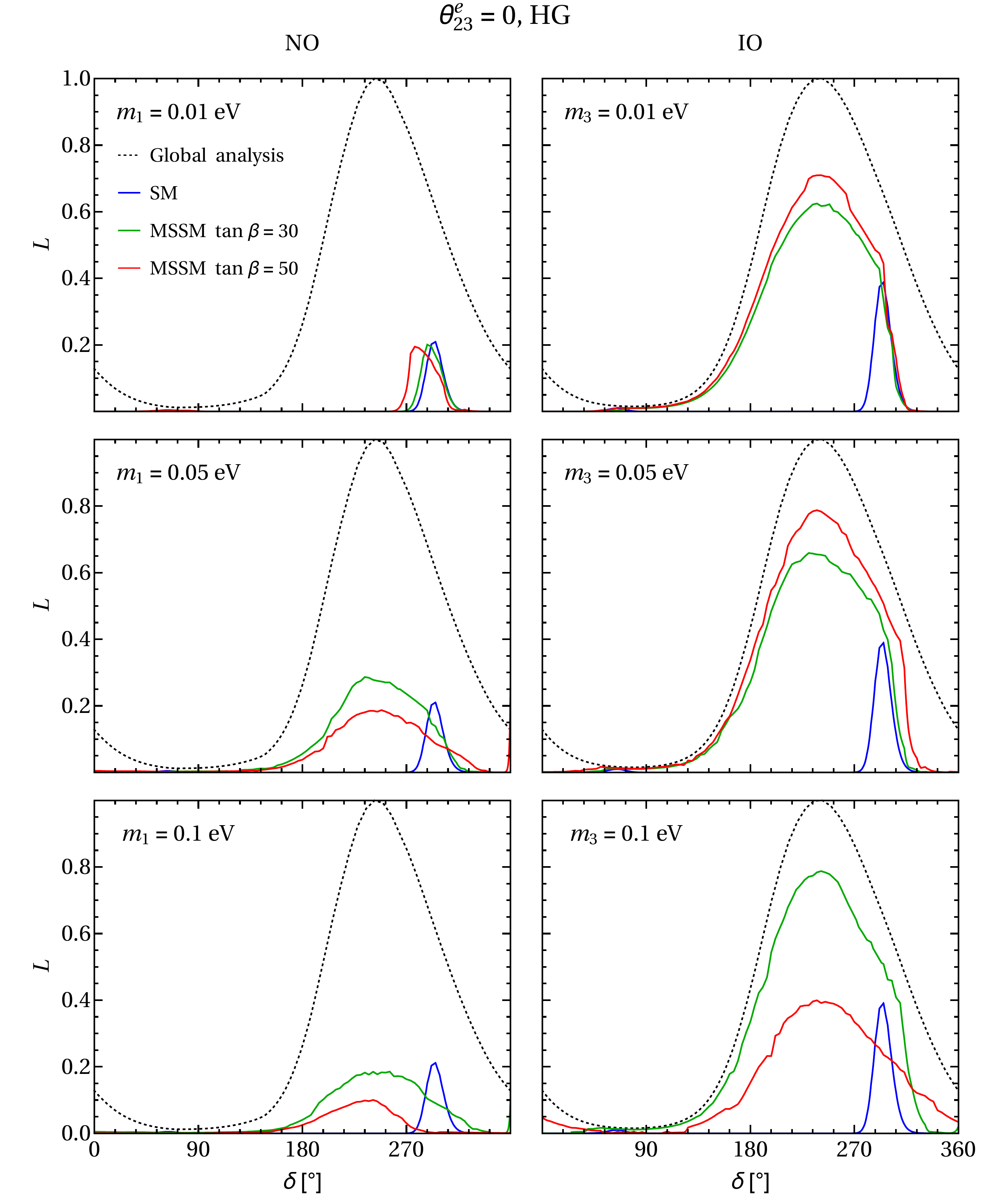}
\caption{Likelihood function vs. $\delta$ in the case of zero 
$\theta^e_{23}$ for the HG symmetry form of 
the matrix $\tilde U_\nu$ in all the setups considered. 
The dotted line stands for likelihood extracted from the global analysis
in \cite{Capozzi:2016rtj}.
The blue line is for the SM running. 
Finally, the green and red lines are for the running within MSSM with 
$\tan\beta = 30$ and $\tan\beta = 50$, respectively.}
\label{fig:likelihoodHG2}
\end{figure}
%%%%%%%%%%%%%%%%%%%%%%%%%%%%%%%%%%%%%%

In Figs.~\ref{fig:likelihoodTBM2}\,--\,\ref{fig:likelihoodHG2} we present 
the results in the case of $\theta^e_{23} = 0$.
Again, the blue line in these figures represents the SM running result, 
the green and red lines are for the MSSM running with 
$\tan\beta = 30$ and $\tan\beta = 50$, respectively.
The dotted black line stands for the likelihood extracted from 
the global analysis \cite{Capozzi:2016rtj} which corresponds to the
likelihood for $\delta$ without imposing any sum rule. 
Similar to the case of non-zero $\theta_{23}^e$, the SM line practically
coincides with the line corresponding to the result without running, as
expected. Therefore we do not show the latter in the plots.
Note again that the small wiggles in the likelihoods are of numerical origin
and not physical.

 For the TBM, GRA, GRB and HG mixing schemes we observe similar to the case
of non-zero $\theta_{23}^e$ broadening of the likelihood with
increasing $\tan\beta$ and increasing absolute neutrino mass scale. 
But in contrast to the case of $\theta^e_{23}\neq 0$,  
the likelihood does not reach the likelihood for $\delta$ without imposing 
the sum rule considered.
The major difference with respect to the results obtained 
in the case of $\theta_{23}^e \neq0$ is that due to the 
constraint on $\theta_{23}$ from eq.~\eqref{eq:ssth23} at the high scale, 
the low-scale mixing parameters are more severely constrained and 
not necessarily close to their respective best fit values.
 
As Figs.~\ref{fig:likelihoodTBM2}\,--\,\ref{fig:likelihoodHG2}
show, for the values of ${\rm min}(m_j)$ and $\tan\beta$ considered, 
the NO spectrum is less favoured (i.e., has a smaller likelihood 
for any given $\delta$ 
and smaller maximum likelihood)  
than the IO spectrum.
The sum rule, eq.~\eqref{eq:ssth23}, restricts $\theta_{23}$ to 
be slightly smaller than $45^\circ$ at the high scale. 
Since the running of this angle
has a fixed negative sign for NO spectrum, 
its low-scale value is larger than its high scale
value and pushed outside of the NO 1$\sigma$ region. 
On the other hand, for IO spectrum 
the low-scale value of $\theta_{23}$ is always smaller
than $45^\circ$ due to the running and the sum rule. 
However, in this case there is a second
1$\sigma$ region below maximal mixing besides the region 
around the best fit value which is larger than $45^\circ$.

In the case of the TBM and GRB schemes, the case of 
${\rm min}(m_j) =0.10$ eV and $\tan\beta = 50$ is 
strongly disfavoured for both NO and IO spectra, while for the 
GRA and HG schemes it is less favoured than the 
${\rm min}(m_j) =0.10$ eV and $\tan\beta = 30$
case.

  As explained in subsection~\ref{sec:th23ecase}, in order to satisfy the sum rule
eq.~\eqref{eq:delta12e}  for zero $\theta_{23}^e$, $\theta_{12}$ is not allowed
to run strongly. This leads to 
the relatively small likelihood for $\tan\beta=50$ and
$m_{\text{lightest}}=0.1$~eV seen in 
Figs.~\ref{fig:likelihoodTBM2}\,--\,\ref{fig:likelihoodHG2}. 
For TBM and GRB mixing the constraint on the
running of $\theta_{12}$ is even more 
severe than for GRA and HG mixing and
the likelihood in these schemes is hence 
even smaller for $\tan\beta=50$ and $m_{\text{lightest}}=0.1$~eV.

 For BM mixing our analytical estimates have indicated that this scheme is not
valid due to the severe constraint on the running of $\theta_{12}$. In
our extensive numerical scans we did not find any valid, 
i.e., physically acceptable,  
parameter points as well.

\section{Summary and Conclusions}
\label{sec:summary}

We presented a systematic study of the effects of RG corrections 
on sum rules for the Dirac CPV phase, eqs. (\ref{eq:delta}) 
and (\ref{eq:delta12e}).  
These corrections are present in every high-energy model, 
when running down to the low scale where experiments take place. 
We answered the question how stable the predictions from 
the sum rules are 
in the cases of charged lepton corrections characterised by 
i)~$\theta_{12}^e\neq0,~\theta_{23}^e\neq0,~\theta_{13}^e=0$ and 
ii)~$\theta_{12}^e\neq0, ~\theta_{23}^e=0,~\theta_{13}^e=0$ 
to TBM, BM, GRA, GRB or HG mixing in the neutrino sector.

To this aim we first reviewed the framework in which 
we obtain the mixing sum rules. 
Then we presented  analytical estimates of the allowed parameter space 
if we take RG corrections into account. 
These estimates were subsequently verified numerically.
To obtain the numerical results for the allowed ranges of $\delta$ 
we used as  three benchmark cases  the SM running 
(where the running effects are small) and 
the MSSM running with $\tan \beta=30$ and $\tan \beta=50$ 
(where  the running effects become larger with increasing $\tan \beta$).
Furthermore, we considered three mass scales: 
a ``small'' mass scale ($m_{\text{lightest}}=0.01$~eV), 
a ``medium'' mass scale ($m_{\text{lightest}}=0.05$~eV) 
and a ``large'' mass scale ($m_{\text{lightest}}=0.1$~eV), 
where  the RG effects increase with the mass scale. 
We presented the results in terms of the likelihood functions 
for each case (SM or MSSM with a given $\tan\beta$, and a given 
mass scale). 
Our numerical results are obtained using the current 
best fit values and uncertainties on the neutrino oscillation parameters 
derived in the global analysis of the neutrino oscillation data 
performed in \cite{Capozzi:2016rtj}.

Our results have shown that the RG effects can 
change significantly 
the allowed low-energy ranges for $\delta$, 
especially when we employ the MSSM running with the 
``medium'' and ``large'' 
mass scales.  In the case of $\theta^e_{23} \neq 0$
the allowed regions for $\delta$ 
broaden and the likelihood profiles approach 
the likelihood for $\delta$ extracted from the global analysis 
(without imposing the sum rules considered).
For the TBM, GRA, GRB and HG symmetry forms 
we found the allowed ranges of values of $\delta$ to be shifted 
from values close to (somewhat larger than) 
$270^\circ$ to values somewhat smaller than (close to) $270^\circ$.
For BM mixing, which is strongly disfavoured by the current data 
without taking into account the running of the neutrino parameters,
we found that the RG corrections partially reconstitute 
compatibility of this symmetry form with the data.
With the increasing of ${\rm min}(m_j)$ and $\tan\beta$,  
the values of $\delta$ in this case shift from $\delta \sim 180^\circ$ 
towards $270^\circ$.
 In the case of $\theta^e_{23} = 0$ and 
for the TBM, GRA, GRB and HG mixing schemes the likelihood profiles 
broaden with increasing $\tan\beta$ and increasing mass scale, 
similarly to the case of non-zero $\theta^e_{23}$.
The main difference is that now they
do not reach the likelihood for $\delta$
obtained without imposing the sum rule. 
The reason for that is the constraint on $\theta_{23}$ from 
eq.~\eqref{eq:ssth23} at the high scale, 
due to which the low-scale mixing parameters are more severely constrained 
and not necessarily close to their respective best fit values.
Finally, we found that in this case the RG corrections 
are not sufficient to restore even partial 
compatibility of BM mixing with the current data.

 In conclusion, our results show that the RG effects on 
the mixing sum rules in SUSY models 
with ${\rm min}(m_j)\gtap 0.01$ eV and 
$\tan\beta \gtap 30$
have to be taken into account to realistically probe 
the predictions from the sum rules in concrete models.

\section*{Acknowledgements}

JG and MS would like to thank Christoph Wiegand for 
helping us to parallelize our numerics
more efficiently. MS is  supported  by  BMBF  under  contract no.\ 05H12VKF and would like to thank
LIPI and KEkini for kind hospitality during which parts of this project were done. 
AVT would like to thank F. Capozzi, E. Lisi, A. Marrone, D. Montanino 
and A. Palazzo for kindly sharing with us the data files for 
one-dimensional $\chi^2$ projections. 
This work was supported in part 
by the INFN program on Theoretical Astroparticle Physics (TASP),
by the research grant  2012CPPYP7 ({\sl Theoretical Astroparticle Physics})
under the program  PRIN 2012 funded by the Italian Ministry 
of Education, University and Research (MIUR), 
by the European Union FP7 ITN INVISIBLES 
(Marie Curie Actions, PITN-GA-2011-289442-INVISIBLES), and 
by the World Premier International Research Center
Initiative (WPI Initiative), MEXT, Japan (STP).

\appendix
\section{Likelihood Functions for $\boldsymbol{\cos\delta}$}

\begin{figure}
\centering
\includegraphics[width=\textwidth]{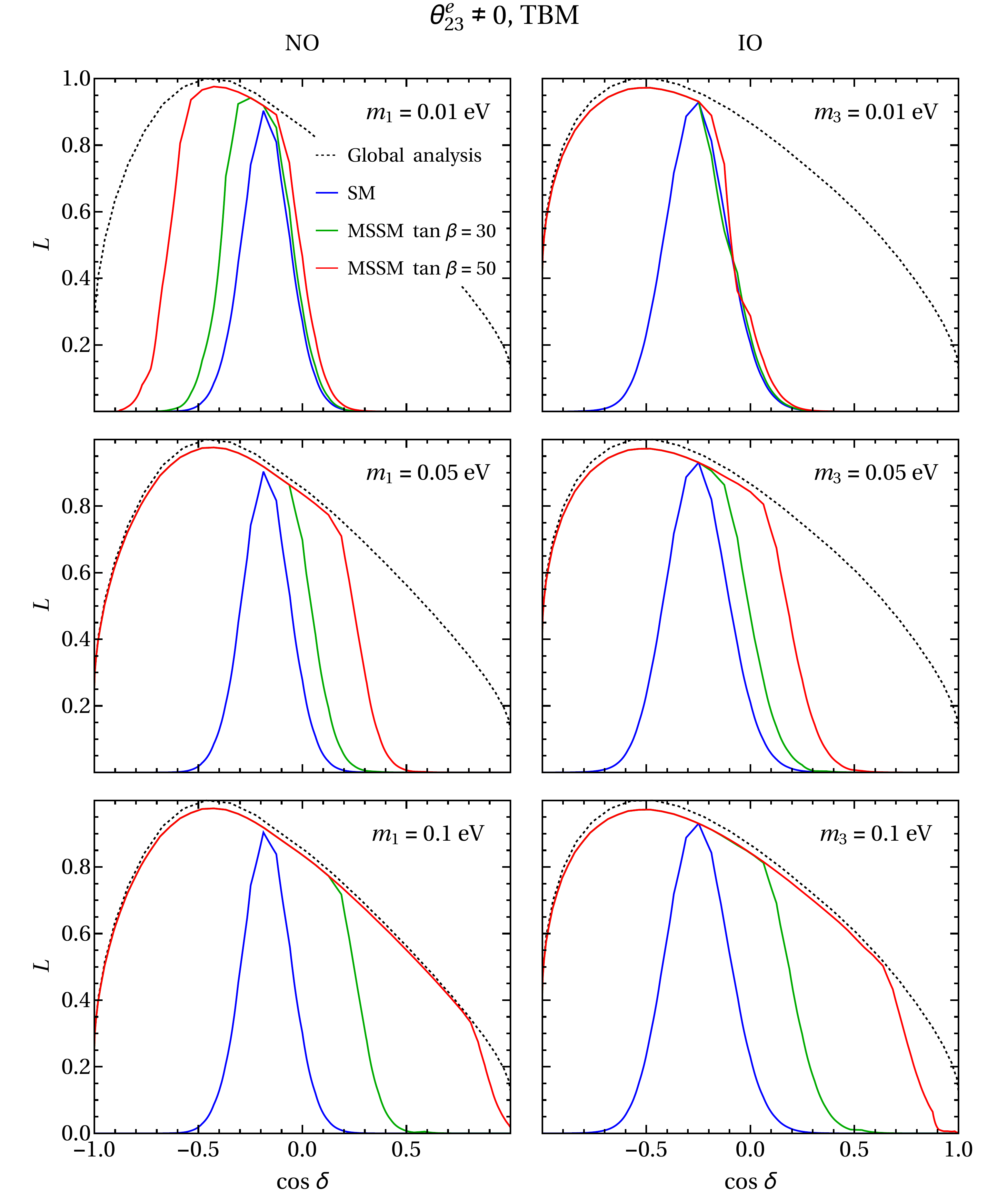}
\caption{Likelihood function vs. $\cos\delta$ 
in the case of non-zero $\theta^e_{23}$ for the TBM symmetry form of 
the matrix $\tilde U_\nu$ in all the setups considered. 
The dotted line stands for likelihood extracted from the global analysis
in \cite{Capozzi:2016rtj}.
The blue line is for the SM running. 
Finally, the green and red lines are for the running within MSSM with 
$\tan\beta = 30$ and $\tan\beta = 50$, respectively.}
\label{fig:likelihoodTBMdelta}
\end{figure}

\begin{figure}
\centering
\includegraphics[width=\textwidth]{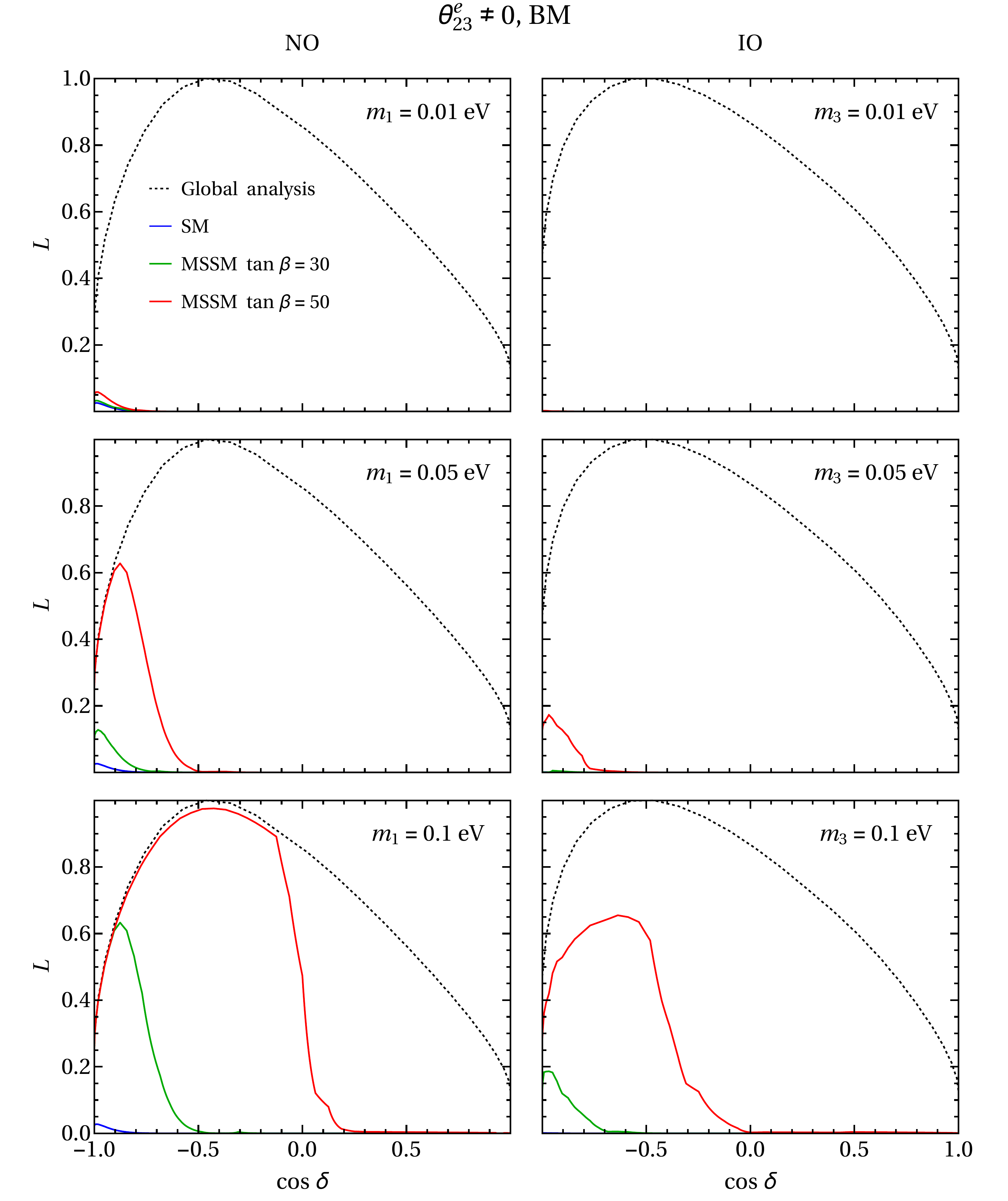}
\caption{Likelihood function vs. $\cos\delta$ 
in the case of non-zero $\theta^e_{23}$ for the BM symmetry form of 
the matrix $\tilde U_\nu$ in all the setups considered. 
The dotted line stands for likelihood extracted from the global analysis
in \cite{Capozzi:2016rtj}.
The blue line is for the SM running. 
Finally, the green and red lines are for the running within MSSM with 
$\tan\beta = 30$ and $\tan\beta = 50$, respectively.}
\label{fig:likelihoodBMdelta}
\end{figure}

\begin{figure}
\centering
\includegraphics[width=\textwidth]{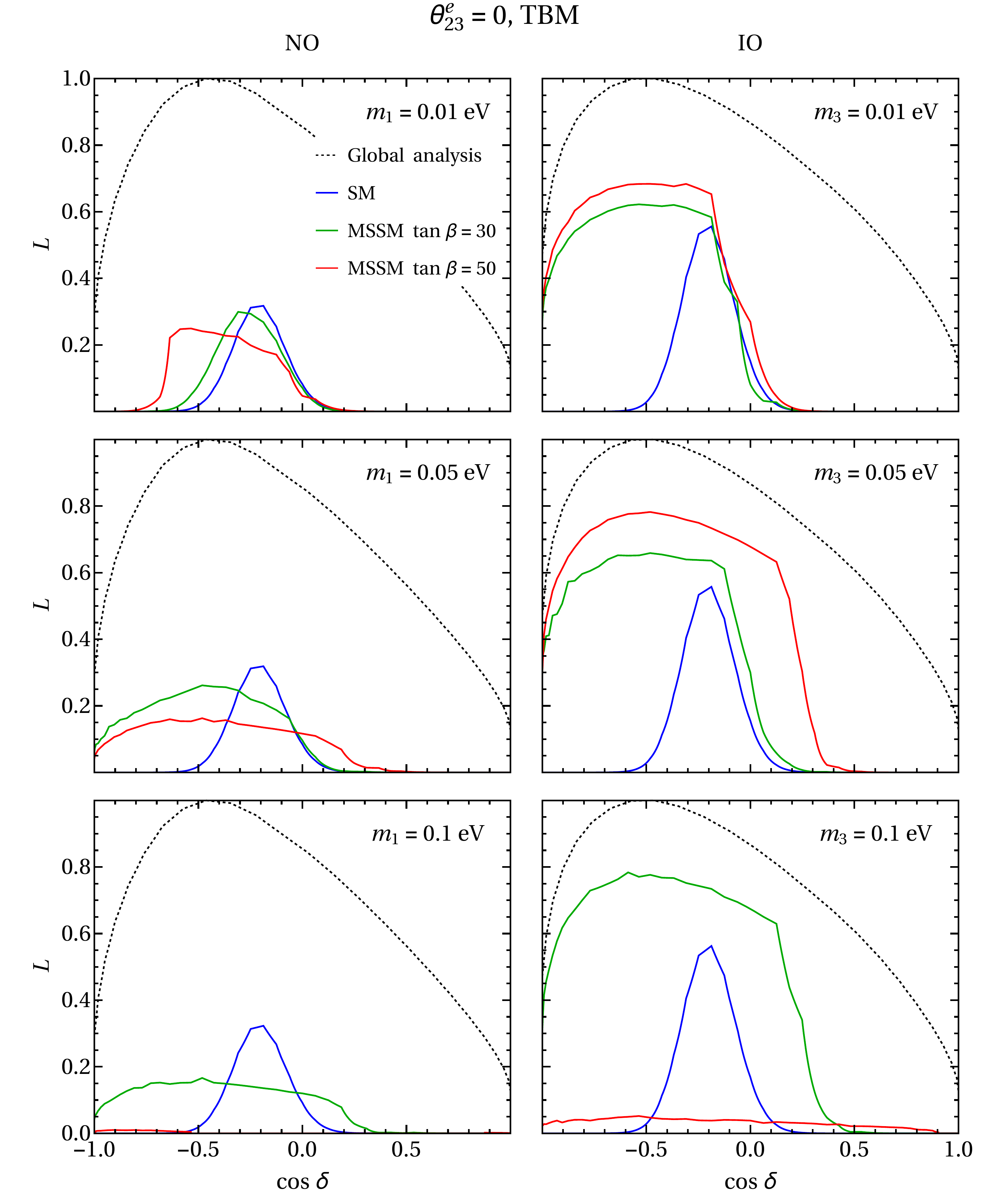}
\caption{Likelihood function vs. $\cos\delta$ in the case of zero 
$\theta^e_{23}$ for the TBM symmetry form of 
the matrix $\tilde U_\nu$ in all the setups considered. 
The dotted line stands for likelihood extracted from the global analysis
in \cite{Capozzi:2016rtj}.
The blue line is for the SM running. 
Finally, the green and red lines are for the running within MSSM with 
$\tan\beta = 30$ and $\tan\beta = 50$, respectively.}
\label{fig:likelihoodTBM2delta}
\end{figure}

\begin{figure}
\centering
\includegraphics[width=\textwidth]{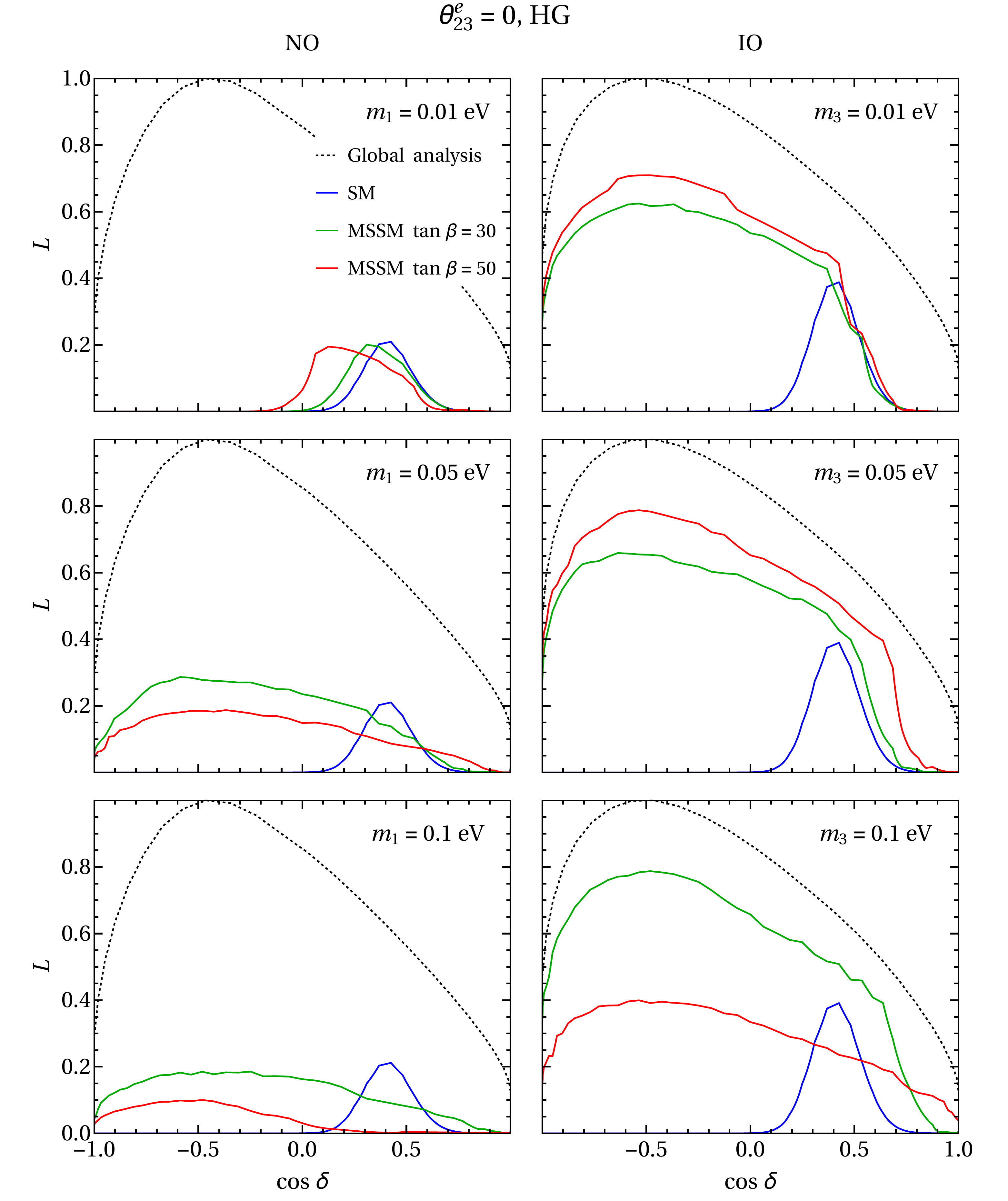}
\caption{Likelihood function vs. $\cos\delta$ in the case of zero 
$\theta^e_{23}$ for the HG symmetry form of 
the matrix $\tilde U_\nu$ in all the setups considered. 
The dotted line stands for likelihood extracted from the global analysis
in \cite{Capozzi:2016rtj}.
The blue line is for the SM running. 
Finally, the green and red lines are for the running within MSSM with 
$\tan\beta = 30$ and $\tan\beta = 50$, respectively.}
\label{fig:likelihoodHG2delta}
\end{figure}

In the past there have been already extensive studies on the
likelihoods for the Dirac CPV phase derived from
mixing sum rules. 
In \cite{Girardi:2014faa,Girardi:2015zva,Girardi:2015vha,Ballett:2014dua}, 
in particular,  results for the TBM, GRA, GRB, HG and BM mixing schemes
were presented neglecting the RG corrections.
However, in the indicated publications the likelihoods for
$\cos \delta$ and not for $\delta$ have been 
derived. For better comparison with these results 
we include in the present Appendix 
Figs.~\ref{fig:likelihoodTBMdelta}\,--\,\ref{fig:likelihoodBMdelta} 
(Figs.~\ref{fig:likelihoodTBM2delta}\,--\,\ref{fig:likelihoodHG2delta})
with the likelihood functions for $\cos \delta$ 
in the case of  $\theta^e_{23} \neq 0$ ($\theta^e_{23} = 0)$.

\begin{figure}
\centering
\includegraphics[width=\textwidth]{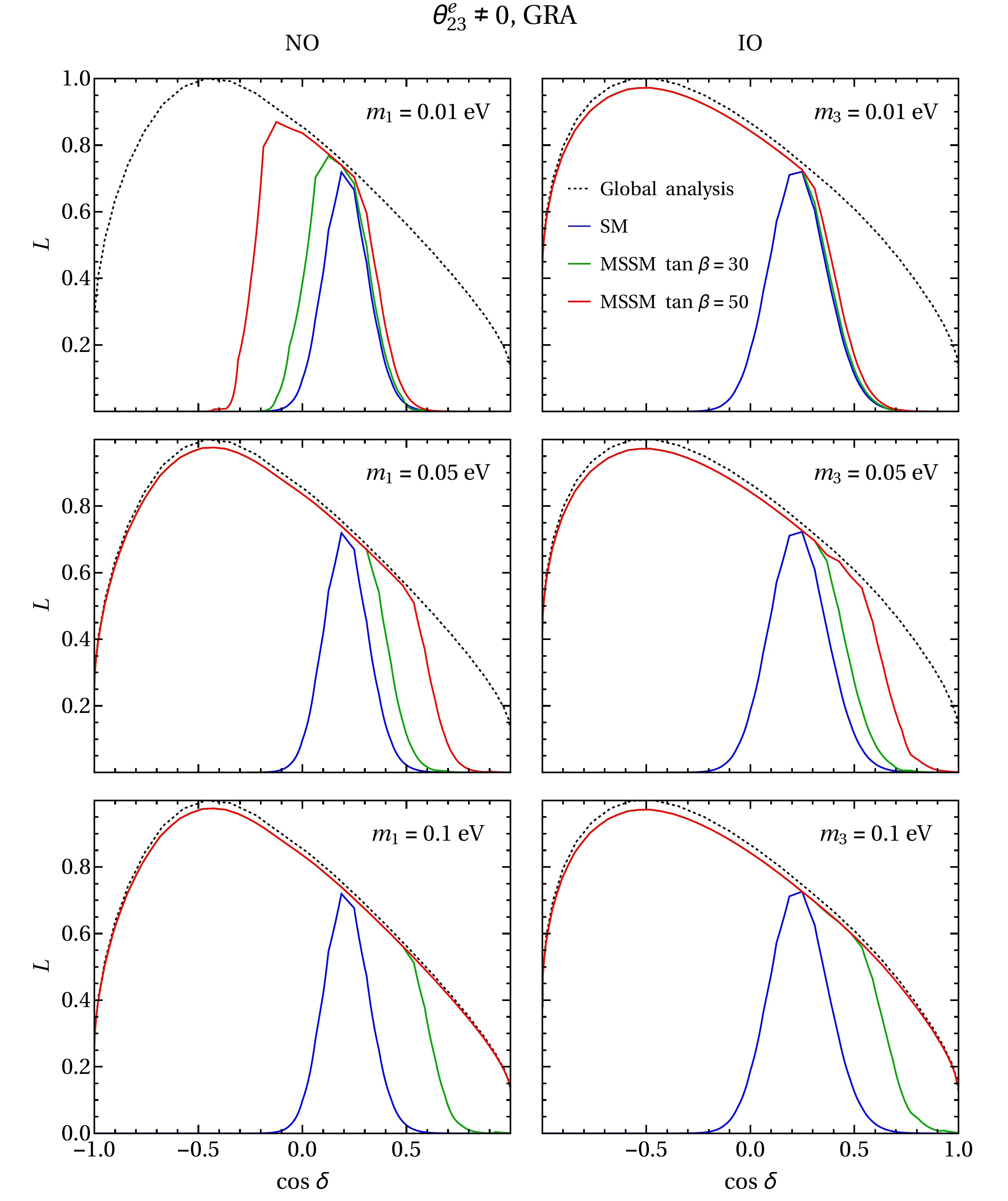}
\caption{Likelihood function vs. $\cos\delta$ 
in the case of non-zero $\theta^e_{23}$ for the GRA symmetry form of 
the matrix $\tilde U_\nu$ in all the setups considered. 
The dotted line stands for likelihood extracted from the global analysis
in \cite{Capozzi:2016rtj}.
The blue line is for the SM running. 
Finally, the green and red lines are for the running within MSSM with 
$\tan\beta = 30$ and $\tan\beta = 50$, respectively.}
\label{fig:likelihoodGRAdelta}
\end{figure}

\begin{figure}
\centering
\includegraphics[width=\textwidth]{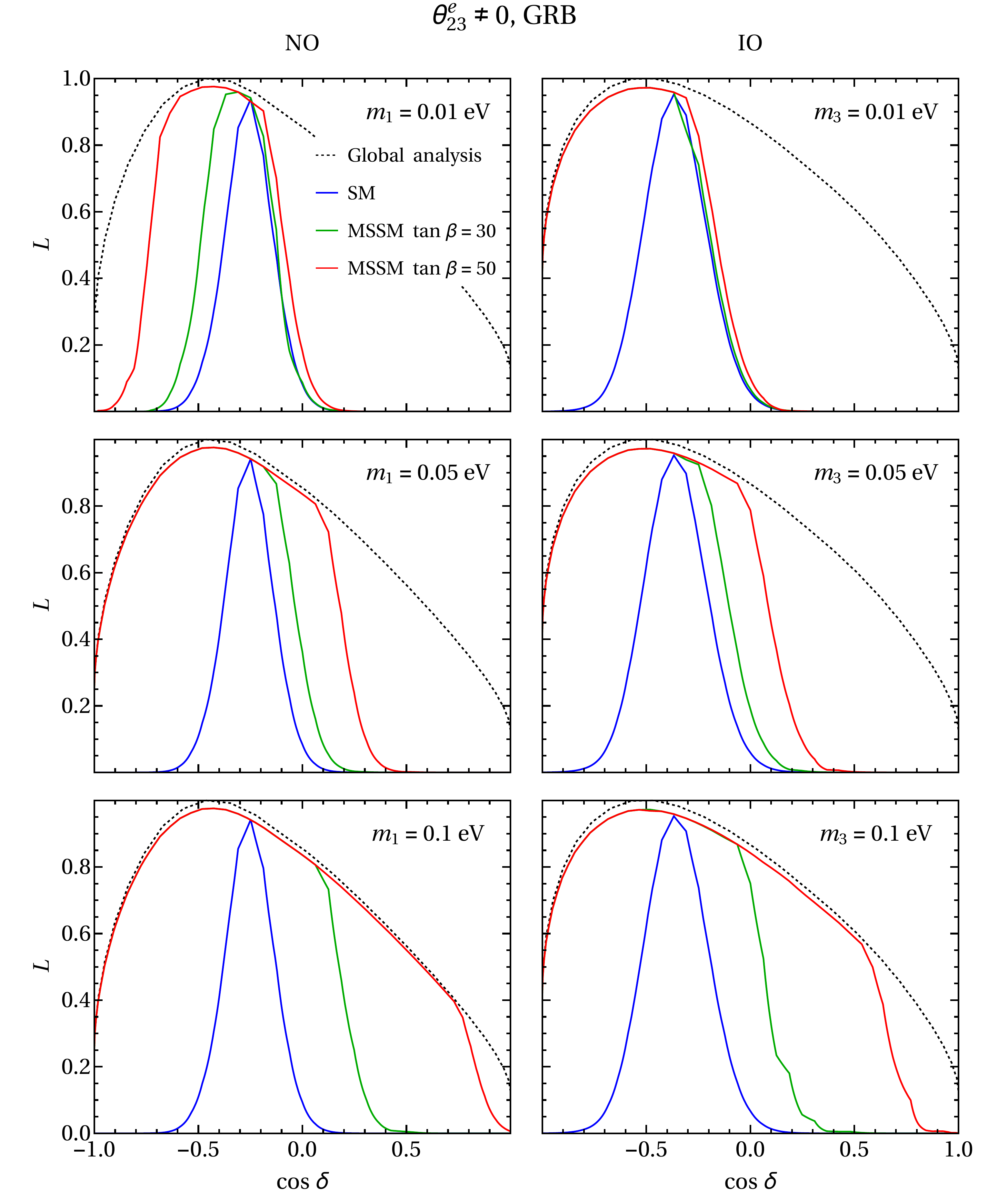}
\caption{Likelihood function vs. $\cos\delta$ 
in the case of non-zero $\theta^e_{23}$ for the GRB symmetry form of 
the matrix $\tilde U_\nu$ in all the setups considered. 
The dotted line stands for likelihood extracted from the global analysis
in \cite{Capozzi:2016rtj}.
The blue line is for the SM running. 
Finally, the green and red lines are for the running within MSSM with 
$\tan\beta = 30$ and $\tan\beta = 50$, respectively.}
\label{fig:likelihoodGRBdelta}
\end{figure}

\begin{figure}
\centering
\includegraphics[width=\textwidth]{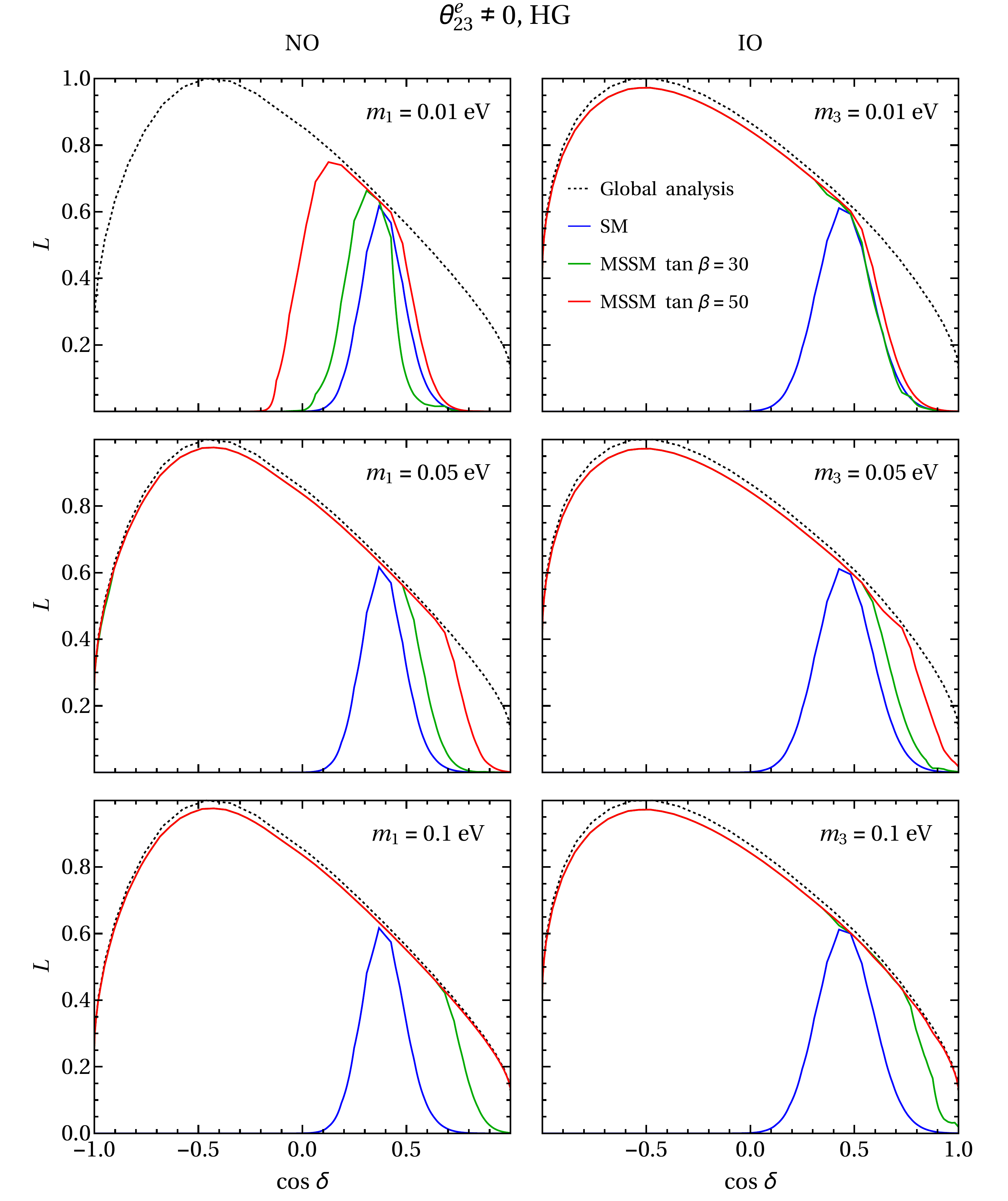}
\caption{Likelihood function vs. $\cos\delta$ 
in the case of non-zero $\theta^e_{23}$ for the HG symmetry form of 
the matrix $\tilde U_\nu$ in all the setups considered. 
The dotted line stands for likelihood extracted from the global analysis
in \cite{Capozzi:2016rtj}.
The blue line is for the SM running. 
Finally, the green and red lines are for the running within MSSM with 
$\tan\beta = 30$ and $\tan\beta = 50$, respectively.}
\label{fig:likelihoodHGdelta}
\end{figure}

%%%%%%%%%%%%%%%%%%%%%%%%%%%%%%%%%

\begin{figure}
\centering
\includegraphics[width=\textwidth]{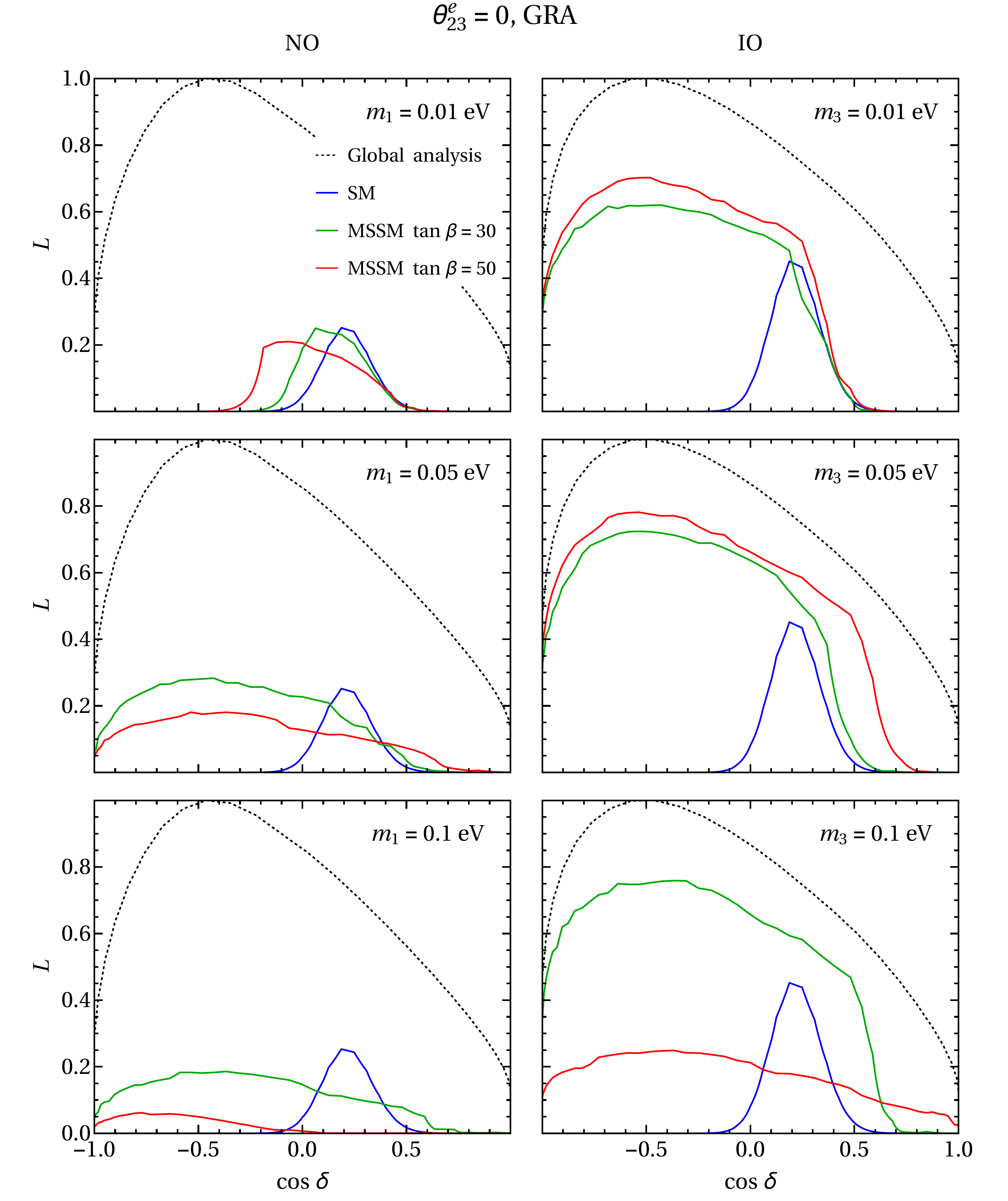}
\caption{Likelihood function vs. $\cos\delta$ in the case of zero 
$\theta^e_{23}$ for the GRA symmetry form of 
the matrix $\tilde U_\nu$ in all the setups considered. 
The dotted line stands for likelihood extracted from the global analysis
in \cite{Capozzi:2016rtj}.
The blue line is for the SM running. 
Finally, the green and red lines are for the running within MSSM with 
$\tan\beta = 30$ and $\tan\beta = 50$, respectively.}
\label{fig:likelihoodGRA2delta}
\end{figure}

\begin{figure}
\centering
\includegraphics[width=\textwidth]{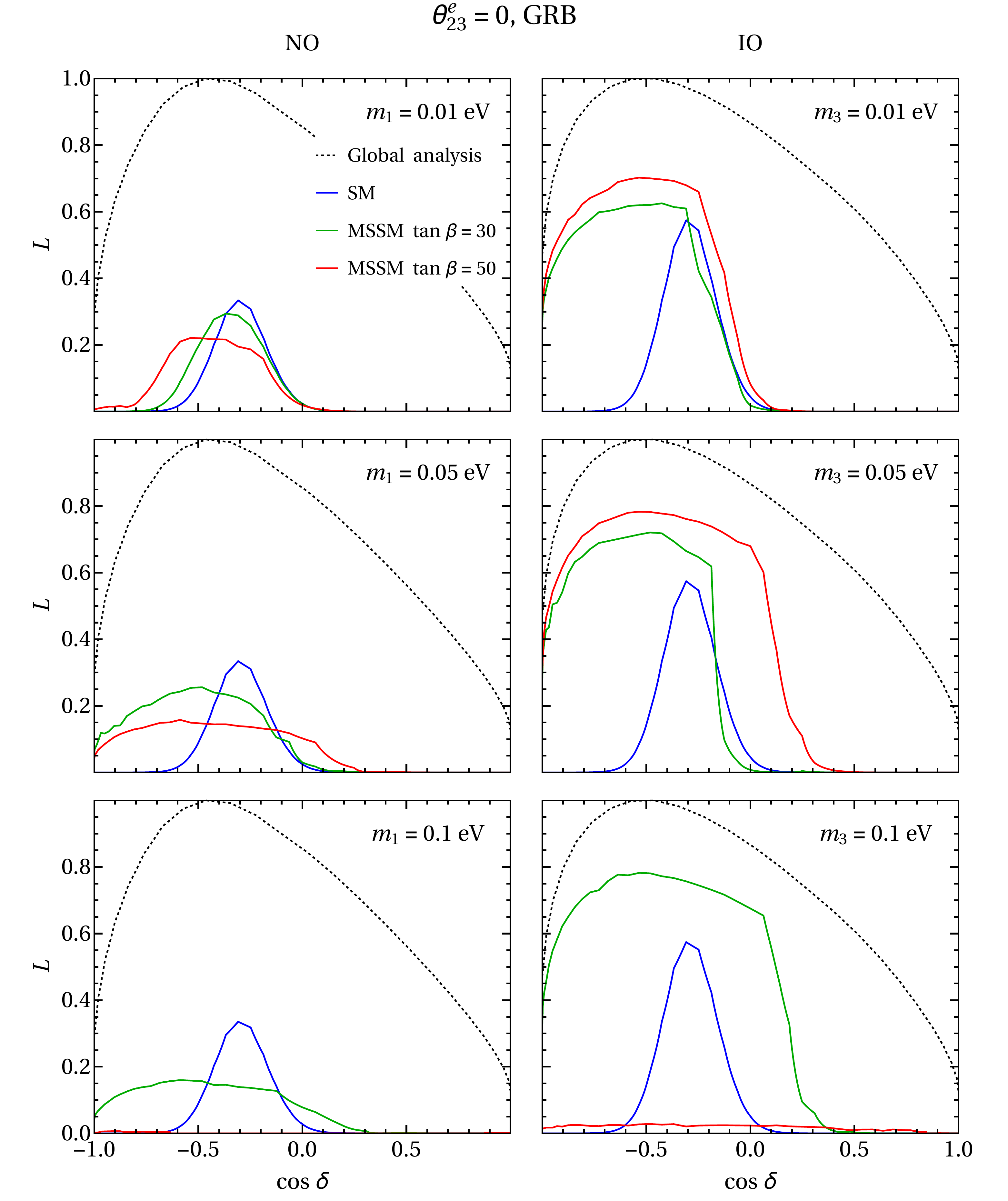}
\caption{Likelihood function vs. $\cos\delta$ in the case of zero 
$\theta^e_{23}$ for the GRB symmetry form of 
the matrix $\tilde U_\nu$ in all the setups considered. 
The dotted line stands for likelihood extracted from the global analysis
in \cite{Capozzi:2016rtj}.
The blue line is for the SM running. 
Finally, the green and red lines are for the running within MSSM with 
$\tan\beta = 30$ and $\tan\beta = 50$, respectively.}
\label{fig:likelihoodGRB2delta}
\end{figure}

\end{document}